\documentclass[traditabstract]{aa}

\usepackage[T1]{fontenc}
\usepackage[latin1]{inputenc}
\usepackage{wasysym}
\usepackage{graphicx}
\usepackage{natbib}
\usepackage{txfonts}

\newcommand{\eq}[1]{\begin{equation}  #1 \end{equation}}

\newcommand{\eqa}[1]{\begin{eqnarray}   #1 \end{eqnarray}}

\newcommand{\br}[1]{\left( #1 \right)}
\newcommand{\bc}[1]{\left\{ #1 \right\}}
\newcommand{\bb}[1]{\left[ #1 \right]}
\newcommand{\ba}[1]{\left\langle #1 \right\rangle}

\newcommand{\nn}{\nonumber}

\newcommand{\dd}{{\rm d}}
\newcommand{\expo}[1]{~{\rm e}^{ #1 }}
\newcommand{\vek}[1]{\mbox{\boldmath $#1$}}
\newcommand{\svek}[1]{\mbox{\boldmath \scriptsize $#1$}}  
\newcommand{\ic}{{\rm i}}

\newcommand{\hmpc}{$\,h^{-1}$Mpc}

\begin{document}

\title{Constraints on intrinsic alignment contamination of weak lensing surveys using the MegaZ-LRG sample}

\author{B. Joachimi\inst{1,2} \and R. Mandelbaum\inst{3} \and F. B. Abdalla\inst{2} \and S. L. Bridle\inst{2}}

\institute{
Argelander-Institut f\"ur Astronomie (AIfA), Universit\"at Bonn, Auf dem H\"ugel 71, 53121 Bonn, Germany\\ \email{joachimi@astro.uni-bonn.de}
\and
Department of Physics and Astronomy, University College London, London WC1E 6BT, UK
\and
Department of Astrophysical Sciences, Princeton University, Peyton Hall, Princeton, NJ 08544, USA
}

\date{Received 20 August 2010 / Accepted 15 December 2010}

\abstract{
Correlations between the intrinsic shapes of galaxies and the large-scale galaxy density field provide an important tool to investigate galaxy intrinsic alignments, which constitute the major potential astrophysical systematic in cosmological weak lensing (cosmic shear) surveys, but also yield insight into the formation and evolution of galaxies. We measure galaxy position-shape correlations in the MegaZ-LRG sample for more than 800{,}000 luminous red galaxies for comoving transverse separations of $0.3 < r_p < 60$\hmpc, making the first such measurement with a photometric redshift sample. In combination with a re-analysis of several spectroscopic SDSS samples, we constrain an intrinsic alignment model for early-type galaxies over long baselines in redshift ($z \lesssim 0.7$) and luminosity ($4\,$mag) with high statistical precision.

We develop and test the formalism to incorporate photometric redshift scatter in the modelling of these observations. For $r_p > 6$\hmpc, the fits to galaxy position-shape correlation functions are consistent with the scaling with $r_p$ and redshift of a revised, nonlinear version of the linear alignment model (Hirata \& Seljak 2004) for all samples. An extra redshift dependence $\propto (1+z)^{\eta_{\rm other}}$ is constrained to $\eta_{\rm other} = -0.3 \pm 0.8\, (1\sigma)$. To obtain consistent amplitudes for all data, an additional dependence on galaxy luminosity $\propto L^\beta$ with $\beta=1.1^{+0.3}_{-0.2}$ is required. The normalisation of the intrinsic alignment power spectrum is found to be $(0.077 \pm 0.008)\, \rho_{\rm cr}^{-1}$ for galaxies at redshift 0.3 and $r$ band magnitude of $-22$ ($k$- and evolution-corrected to $z=0$).

Assuming zero intrinsic alignments for blue galaxies, we assess the bias on cosmological parameters for a tomographic CFHTLS-like lensing survey given our new constraints on the intrinsic alignment model parameter space. Both the resulting mean bias and its uncertainty are smaller than the $1\sigma$ statistical errors when using the constraints from all samples combined. The addition of MegaZ-LRG data is critical to achieving constraints this strong, reducing the uncertainty in intrinsic alignment bias on cosmological parameters by factors of three to seven.
}


\maketitle

\section{Introduction}
\label{sec:intro}

Within the past decade, cosmic shear, the weak gravitational lensing effect by the large-scale structure of the Universe, has evolved from its first detections \citep{bacon00,kaiser00,vwaer00,wittman00} into a well-established and regularly used cosmological probe (see for instance \citealp{benjamin07,fu07,schrabback09} for recent results). Cosmic shear is considered one of the most promising techniques to unravel the properties of dark matter, dark energy and new gravitational physics \citep[e.g.][]{albrecht06,peacock06,kitching08b,thomas08,daniel10} and acts as a major science driver for ongoing, upcoming and future ambitious surveys like Pan-STARRS\footnote{Panoramic Survey Telescope \& Rapid Response System, \texttt{http://pan-starrs.ifa.hawaii.edu}}, DES\footnote{Dark Energy Survey, \texttt{https://www.darkenergysurvey.org}}, LSST\footnote{Large Synoptic Survey Telescope, \texttt{http://www.lsst.org}}, JDEM\footnote{Joint Dark Energy Mission, \texttt{http://jdem.gdfc.nasa.gov}}, and Euclid\footnote{\texttt{http://sci.esa.int/science-e/www/area/\\index.cfm?fareaid=102}}.

The strength of weak lensing lies in its sensitivity to both the geometry of the Universe and the growth of structure, as well as its \lq clean\rq\ relation to the underlying theory which is, except for the smallest scales, solely determined by gravitational interaction. Hence, most efforts in the past years to prepare the analysis of cosmic shear data for high-precision measurements have concentrated on technical issues, in particular the unbiased extraction of shapes from galaxy images \citep[e.g.][]{bridle09,bernstein10,cypriano10,voigt10} and the estimation of galaxy redshifts from multi-colour photometry \citep[e.g.][]{abdalla07,abdalla08,zhang09,bernstein09}. Most astrophysical sources of systematic errors such as the limited accuracy of N-body simulations \citep[e.g.][]{hilbert09} and baryonic effects on structure growth \citep[e.g.][]{white04} are currently subdominant and can, like the technical systematics, largely be reduced by more powerful simulations, specifically designed instrumentation, and optimised analysis techniques.

However, this statement does not apply to the intrinsic alignment of galaxies, which is a major astrophysical systematic affecting all two-point and higher-order weak lensing statistics. The estimation of gravitational shear correlations from the observed ellipticities of galaxy images is simple if one assumes that the intrinsic ellipticity of one galaxy is not correlated with either the intrinsic ellipticity of or the gravitational shear acting on another galaxy. Both of these assumptions are incorrect in the presence of intrinsic alignments which are e.g. caused by tidal torquing and stretching of galaxies by the large-scale gravitational field. These effects depend intricately on galaxy formation and evolution including baryonic physics, thus hampering modelling via analytical calculations or N-body simulations. The bias on cosmological parameters when ignoring intrinsic alignments can be severe; e.g. for the dark energy equation of state parameter, \citet{bridle07} find a deviation of $50\,\%$ from a fiducial value of $-1$ for a Euclid-like survey if intrinsic alignment contamination is not accounted for in the analysis.

Correlations between the intrinsic ellipticities of two galaxies (hereafter II) can occur if the galaxies are subject to the tidal forces of the same local or large-scale matter structures. Thus, a pair of galaxies with II correlations must be physically close, i.e. at small separation both in redshift and on the sky, which allows for a relatively straightforward removal of the II signal from cosmic shear data if photometric redshifts are sufficiently accurate \citep{king02,king03,heymans03,takada04b}. The mutual alignment of halo shapes and spins has been studied extensively in N-body simulations \citep{splinter97,onuora00,faltenbacher02,hopkins05,faltenbacher07,faltenbacher08,lee08}, to a limited degree also including the effect of baryonic physics \citep{bosch02,bett10,hahn10}. Models for II correlations have been developed analytically or via fits to simulations \citep{croft00,heavens00,catelan01,crittenden01,jing02,mackey02,hirata04,schneiderm09}, with widely varying results. The II signal can either be observed using galaxy ellipticity correlations at very low redshift where cosmic shear is negligible (\citealp{brown02} in SuperCOSMOS) or by selecting galaxy pairs with small physical separation (\citealp{mandelbaum06,hirata07,okumura08,brainerd09} in various SDSS samples).

\citet{hirata04} identified gravitational shear-intrinsic ellipticity cross-correlations (hereafter GI) as a further, potentially more serious contaminant of cosmic shear surveys. GI correlations are generated by matter that aligns a nearby foreground galaxy and at the same time contributes to the gravitational lensing signal of a background galaxy. Thus, GI correlations are not restricted to close (along the line-of-sight) pairs. This type of intrinsic alignment has a redshift dependence that is very similar to that of lensing, so that GI correlations could be particularly prominent in the deep data of future cosmic shear surveys. The underlying correlations between halo shapes or orientations and the surrounding matter structure have been detected on a large range of mass scales in simulations \citep{bailin05,altay06,heymans06,kuhlen07}. However, as for II correlations, modelling attempts \citep{hui02,hirata04,heymans06,schneiderm09} do not currently yield robust predictions.

In the absence of a compelling model for GI correlations, methods with a weak dependence on the intrinsic alignment model are required to eliminate biases on cosmology in cosmic shear analyses. \citet{joachimi08b,joachimi09} made use of the characteristic redshift dependence of the GI signal to null it in a fully model-independent way, while \citet{king05}, \citet{bridle07}, \citet{bernstein08}, and \citet{joachimi10} introduced very general parametrisations of the intrinsic alignment contributions containing parameters that are then marginalised over (see \citealp{kirk10} for an application to data). Both approaches feature a considerable loss of information, substantially weakening constraints on cosmological model parameters. Observational constraints on the GI contribution are thus crucial, as they can tighten priors on the intrinsic alignment parameters that need to be marginalised over, and shed light on the underlying physical processes by constraining intrinsic alignment models.

To study the GI signal, one must measure the correlations between the matter distribution and the intrinsic shear, i.e. the correlated part of the intrinsic galaxy shapes \citep[see e.g.][]{hirata04}. Assuming linear biasing, and constructing correlation functions with negligible gravitational lensing contributions, this measurement involves cross-correlating galaxy number densities and ellipticities \citep{mandelbaum06,hirata07,okumura09,mandelbaum09}. These galaxy number density-intrinsic shear correlations were detected by \citet{mandelbaum06} in SDSS spectroscopic data at low redshift ($z\sim 0.1$). \citet{hirata07} extended the analysis to SDSS Luminous Red Galaxies (LRGs) at slightly higher redshift ($0.15<z<0.35$), finding evidence for a dependence on galaxy type and for an increase of the intrinsic alignment signal of early-type galaxies with luminosity. They also considered a small sample of galaxies at intermediate redshifts ($z \sim 0.5$) from the 2dF-SDSS LRG and Quasar Survey (2SLAQ, \citealp{cannon06}), but only marginally detected a signal in a bright subsample due to poor statistics. Recently, \citet{mandelbaum09} increased the redshift baseline for position-shape correlation measurements of blue galaxies out to $z \sim 0.7$ by incorporating a large number of spectroscopic redshifts from the WiggleZ Survey \citep{drinkwater10}, reporting a null detection for all subsamples. 

Analogous to \citet{mandelbaum09}, this work presents observational constraints for early-type galaxies (out to $z \lesssim 0.7$), using SDSS shape measurements together with photometric redshift information from the MegaZ-LRG catalogue \citep{collister07,abdalla08}. In combination with previously analysed red galaxy samples, we cover a wide range of redshifts and galaxy luminosities with high statistical precision, which allows us to narrow down the redshift and luminosity evolution of galaxy number density-intrinsic shear correlations. The longer baselines in redshift and luminosity enable a meaningful extrapolation to typical parameter values found for galaxies in cosmic shear surveys, so that we can estimate the contamination due to intrinsic alignments for present-day surveys. For the first time, we include a galaxy sample that only has photometric redshift information into the analysis, and we develop the corresponding formalism. In particular, we account for the spread of the number density-intrinsic shear correlations along the line of sight due to photometric redshift scatter, and determine the importance of other signals such as galaxy-galaxy lensing in the observables.

This paper is structured as follows: In Sect.$\,$\ref{sec:data} we provide an overview of the galaxy samples used in the analysis. Section \ref{sec:cfmeasurement} contains the methodology for the correlation function measurement. We develop the modelling of photometric redshift number density-intrinsic shear correlations in Sect.$\,$\ref{sec:modelling}, deferring the technical aspects of the derivation, as well as a revision of the redshift dependence of the linear alignment model, to two appendices. In Sect.$\,$\ref{sec:results} the results of our analysis are presented, including constraints on an intrinsic alignment model and a discussion of systematic tests. These findings are then applied in Sect.$\,$\ref{sec:impact} to a prediction of the intrinsic alignment contamination of a generic present-day tomographic cosmic shear survey, before we summarise and conclude in Sect.$\,$\ref{sec:conclusions}. 

Where appropriate, we follow the notation of \citet{mandelbaum06}, \citet{hirata07}, and \citet{mandelbaum09}. Since in contrast to these foregoing works we have to consider various contributing signals, we denote the observable by \lq galaxy position-shape\rq\ correlations and otherwise follow the indexing scheme of \citet{joachimi10}. Throughout we will assume as our cosmological model a spatially flat $\Lambda$CDM universe with parameters $\Omega_{\rm m}=0.25$, $\Omega_{\rm b}=0.05$, $\sigma_8=0.8$, $h=0.7$, and $n_{\rm s}=1.0$. While $h=0.7$ is used for e.g. power spectrum calculations, all distances etc. will be given in units that are independent of $h$, i.e. in \hmpc. We use the AB magnitude system and specify luminosities of the galaxy samples under consideration in the SDSS $r$ filter. Absolute magnitudes are consistently given in terms of $h=1$ and typically $k+e$-corrected to $z=0$ unless stated otherwise.

\section{Data}
\label{sec:data}

All data used in this paper come from the Sloan Digital Sky Survey
(SDSS). The SDSS \citep{york00} imaged roughly $\pi$ steradians of the
sky, and followed up approximately one million of the detected objects
spectroscopically \citep{eisenstein01,richards02,strauss02}. The
imaging was carried out by drift-scanning the sky in photometric
conditions \citep{hogg01,ivezic04}, in five bands ($ugriz$)
\citep{fukugita96,smith02} using a specially-designed wide-field
camera \citep{gunn98}. These imaging data were used to create the
galaxy shape measurements that we employ in this paper.  All of the data
were processed by completely automated pipelines that detect and
measure photometric properties of objects, and astrometrically
calibrate the data \citep{lupton01,tucker06}. The SDSS has had seven
major data releases, and is now complete
\citep{stoughton02,abazajian03, abazajian04,abazajian05,finkbeiner04,
adelman06,adelman07,adelman08,abazajian09}.

\subsection{MegaZ-LRG}
\label{sec:megaz}

The MegaZ-LRG sample \citep{collister07} is based on SDSS five-band ($ugriz$) imaging data. It contains more than a million luminous red galaxies at intermediate redshifts between 0.4 and 0.7, i.e. beyond the redshifts of the LRGs already targeted with the SDSS spectrograph ($z \lesssim 0.45$, \citealp{eisenstein01}). While the original catalogue was selected from the 4th SDSS data release, we use the updated version presented in \cite{abdalla08} based on data release 6 \citep{adelman08}. 

The photometry in five bands is used to determine photometric redshifts for the MegaZ-LRG sample. For a subset of the galaxies, spectroscopic redshift information is required for calibration and cross-checking, which is provided by the 2SLAQ survey \citep{cannon06}. Consequently, the selection criteria of MegaZ-LRG have been designed to match those of 2SLAQ, using a series of magnitude and colour cuts \citep[for details see][]{cannon06,collister07}. These criteria have an efficiency of $95\,\%$ in detecting LRGs in the redshift range $0.4 \leq z \leq 0.7$, the failures being almost entirely due to M-type stars. In Sect.$\,$\ref{sec:cfmeasurement} we will describe how we account for this contamination in our analysis. 

\begin{figure}[t]
\centering
\includegraphics[scale=.36,angle=270]{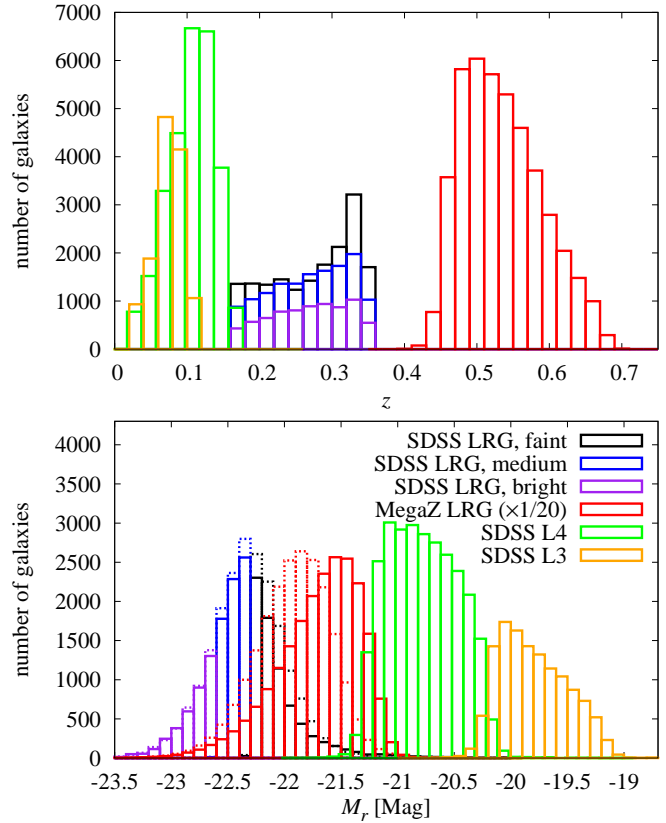}
\caption{\textit{Top panel}: Redshift distributions of the galaxy
  samples analysed. Shown are the histograms for the SDSS LRG samples
  in black (faint, $M_r > -22.3$), blue (medium, $-22.6 < M_r <
  -22.3$), and purple (bright, $M_r < -22.6$), for the full MegaZ-LRG
  sample in red, and for the SDSS Main samples in green (L4) and
  orange (L3). Note that both SDSS LRG and MegaZ-LRG samples are split
  into two redshift bins each, the SDSS LRG samples at $z=0.27$ and
  the MegaZ-LRG sample at $z=0.529$. \textit{Bottom panel}:
  Distribution of rest-frame absolute magnitudes $M_r$. The colour
  coding of the histograms is the same as in the top panel. In the
  case of the SDSS LRG and MegaZ-LRG samples solid lines correspond to
  the low redshift bin, dotted lines to the high redshift bin,
  respectively. Note that the MegaZ-LRG histograms rely on photometric
  redshift estimates, and that they have been downscaled by a factor
  of 20 to facilitate the comparison with the other samples.}
\label{fig:distributions}
\end{figure}

The 2SLAQ selection criteria fluctuated a little at the beginning of the survey.  Specifically, the faint limit of the $i$ band magnitude $i_{\rm deV}$ (the total magnitude estimated using a de Vaucouleurs profile), and the minimum value of $d_{\perp} = (r-i) - (g-r)/8$ (a colour variable used to select LRGs), were varied slightly.  For the majority of the 2SLAQ survey, the criteria $i_{\rm deV} \le 19.8$ and $d_{\perp} \ge 0.55$ were used. For further details on this see \cite{cannon06}. However, for the full MegaZ-LRG sample described in \cite{collister07}, the flux limit is $i_{\rm deV} \le 20$, which means that roughly $1/3$ of the sample is fainter than the 2SLAQ flux limit. Details about the photometric redshift estimation are provided in Sect.$\,$\ref{sec:photoz}.

\begin{table*}[t]
\centering
\caption{Overview of the galaxy samples analysed.}
\begin{tabular}[t]{lccccc}
\hline\hline
sample & area $[{\rm deg}^2]$ & $N_{\rm gal}\,$(density) & $N_{\rm gal}\,$(shape) & $\ba{z}$ & $\ba{L}/L_0$\\[0.5ex]
\hline\\[-1ex]
MegaZ-LRG, all z             & 5363 & 863813 & 427604 & 0.54 & 0.96 \\
MegaZ-LRG, $z < 0.529$       &      & 434321 & 214660 & 0.49 & 0.87 \\
MegaZ-LRG, $z > 0.529$       &      & 429492 & 212944 & 0.59 & 1.05 \\
SDSS LRG, faint, $z < 0.27$  & 4314 &  16701 &   7030 & 0.21 & 1.06 \\
SDSS LRG, faint, $z > 0.27$  &      &  19397 &   9038 & 0.32 & 1.07 \\
SDSS LRG, medium, $z < 0.27$ &      &  16701 &   6139 & 0.22 & 1.50 \\
SDSS LRG, medium, $z > 0.27$ &      &  19397 &   6700 & 0.31 & 1.50 \\
SDSS LRG, bright, $z < 0.27$ &      &  16701 &   3532 & 0.22 & 2.13 \\
SDSS LRG, bright, $z > 0.27$ &      &  19397 &   3659 & 0.31 & 2.12 \\
SDSS Main L4 red             & 3423 & 288863 &  26872 & 0.10 & 0.33 \\
SDSS Main L3 red             &      & 288863 &  10867 & 0.07 & 0.14 \\
\hline
\end{tabular}
\tablefoot{The columns give the sample name, the survey area covered, the number of galaxies $N_{\rm gal}$ used to compute both the galaxy number densities and the shapes, the mean redshift $\ba{z}$, and the mean luminosity $\ba{L}/L_0$ in terms of a fiducial luminosity $L_0$ that corresponds to a $k+e$-corrected (to $z=0$) absolute $r$ band magnitude $M_0=-22$.}
\label{tab:samples}
\end{table*}

As summarised in Table \ref{tab:samples}, we use about 860{,}000 galaxies with a mean redshift of 0.54 in the full MegaZ-LRG sample to compute galaxy number densities. The total number of galaxies is less than that of the full MegaZ-LRG catalogue by \citet{collister07} because a fraction of the area comes from imaging data that was not yet processed by the shape measurement pipeline used to estimate galaxy shapes. As will be discussed in Sect.$\,$\ref{sec:shapemeasurements}, accurate shape measurements could be obtained for about $50\,\%$ of the MegaZ-LRG galaxies, which are then used to trace the intrinsic shear field. 

We also compute $r$ band luminosities, taking into account galactic dust extinction (using reddening maps from \citealp{schlegel98} and extinction-to-reddening ratios from \citealp{stoughton02}), and $k\!+\!e$-correcting the model magnitudes to $z=0$ using the same templates as in \citet{wake06}. Luminosities are given in terms of a fiducial luminosity $L_0$, corresponding to a $k\!+\!e$-corrected absolute $r$ band magnitude of $M_0=-22\,$mag. Due to this procedure, the corrected luminosity acts as a tracer for the total stellar mass, and consequently also for the total mass, of the galaxy. For reference, \citet{blanton03b} obtain a rest-frame magnitude at $z=0.1$ of $M_r^*=-20.44$ for a purely flux-limited sample of SDSS spectroscopic galaxies at $z \lesssim 0.2$. Using \citet{wake06} to $k+e$-correct this value to $z=0$, we find $M_r^*=-20.32$ or $L^*=0.21 L_0$.

Due to a magnitude-dependent redshift success rate and a 0.2 magnitude difference in the $i_{\rm deV}$ cuts (see above), the MegaZ-LRG sample is somewhat fainter in absolute magnitude than the 2SLAQ sample.  Thus, we define the $z=0$ absolute magnitudes using the MegaZ-LRG photometric redshift values, but we also multiply them by a correction factor that is derived from 2SLAQ.  In 2SLAQ, we find that if we use the photometric redshift to define the luminosity rather than the spectroscopic redshift, the mean sample luminosity appears to be 5\% too low. Possible sources of this bias are discussed in Sect.$\,$\ref{sec:photoz}. Thus, we correct the mean $\ba{L}/L_0$ for the MegaZ-LRG sample, as defined using photometric redshifts, upwards by 5\% to account for the impact of photometric redshift errors, the corrected values being given in Table \ref{tab:samples}.

In addition, with a cut at photometric redshift $z=0.529$, we split the sample into two redshift bins, each containing roughly the same number of galaxies. We show the MegaZ-LRG photometric redshift distribution, as well as the luminosity distributions of both MegaZ-LRG subsamples in Fig.$\,$\ref{fig:distributions}, where the latter are also based on photometric redshift estimates. As is evident from the figure, the redshift cut for the MegaZ-LRG sample also segregates the galaxies in luminosity since the sample is magnitude limited.

\subsection{Spectroscopic LRGs}
\label{sec:sdsslrg}

We also use the SDSS DR4 spectroscopic LRG sample
\citep{eisenstein01}, for which measurements of galaxy position-shape
correlations were presented in \citet{hirata07}.  No new measurements
of this sample are made for the current work; instead, we combine the
previous measurements with the new MegaZ-LRG data to provide
constraints over a longer redshift baseline, see
Fig.$\,$\ref{fig:distributions} and Table \ref{tab:samples}.

The SDSS spectroscopic LRG sample has a flux limit of $r<19.2$ and
colour cuts to isolate LRGs.  We include these galaxies in the
redshift range $0.16<z<0.35$, for which the sample is approximately
volume-limited, and includes 36~278 galaxies total.

In order to study variation within this sample, we use cuts on several
parameters.  First, we construct luminosities using the $r$ band model
magnitudes in the same way as for the MegaZ-LRG sample
(Sect.$\,$\ref{sec:megaz}), and define three luminosity subsamples
with $M_r < -22.6$ (\lq bright\rq), $-22.6 \leq M_r < -22.3$ (\lq
medium\rq), and $M_r \geq -22.3$ (\lq faint\rq), see Table
\ref{tab:samples}.  The absolute magnitude cuts are defined in terms
of $h=H_0/(100\,$km$\,$s$^{-1}\,$Mpc$^{-1})$ such that one can
implement the cuts without specifying the value of $H_0$. The
magnitudes have been corrected for galactic extinction, and are
$k+e$-corrected to $z=0$ using the same templates as in
\citet{wake06}. Each luminosity subsample is in addition split at
$z=0.27$ into a high- and a low-redshift bin.

\subsection{Main sample}
\label{sec:sdssmain}

Furthermore we consider galaxies at a more typical luminosity, the
same subsamples of the DR4 SDSS Main spectroscopic sample as in
\citet{mandelbaum06}, divided by luminosity and other properties.  For
this work, we use the red galaxies in L3 (from roughly 1.5 to 0.5
magnitudes fainter than typical $L_*$ galaxies) and L4 (from 0.5
magnitudes fainter to 0.5 magnitudes brighter than $L_*$)\footnote{We
refrain from including the L5 and L6 samples as well because of their
significant overlap with the spectroscopic LRG sample.}.  The sample
properties were described in full in that paper; for this work, we
mention only that the luminosities described to select the sample are
Petrosian magnitudes, extinction-corrected using reddening maps from
\citet{schlegel98} with the extinction-to-reddening ratios given in
\citet{stoughton02}, and $k$-corrected to $z=0.1$ using {\sc kcorrect
v3\_2} software as described by \citet{blanton03}.  For consistency
with previous work, L3 and L4 were initially selected with respect to
this type of absolute magnitude.  However, in order to compare this
sample with the others, we must also compute absolute magnitudes in
the same way as for those (model instead of Petrosian magnitudes,
$k+e$-corrected to $z=0$ instead of $0.1$ using specifically red
galaxy templates).  To use the $k+e$-corrections as in the previous
sections, we must first ensure that we are selecting a properly red
galaxy sample for which they are appropriate.

\citet{hirata07} used an empirically-determined redshift-dependent
colour separator of $u-r = 2.1 + 4.2z$, which uses observer-frame
rather than rest-frame colours; within these luminosity bins, the
fractions of red galaxies were 0.40 (L3), 0.52 (L4), 0.64 (L5), and
0.80 (L6). For the following analysis, we want to define a
red-sequence sample that is comparable to the higher-redshift samples
defined previously (which are typically more strictly defined, though
see \citealp{wake06} and Sect.$\,$\ref{sec:colorcomparison} for a
discussion of the differences between MegaZ-LRG and SDSS LRGs).  Thus,
for this work we use a different colour separator and recompute the
GI correlations for the new \lq red\rq\ samples.

To generate the new colour separator, we first $k$-correct the
extinction-corrected model magnitudes using {\sc kcorrect} to $z=0$.
The reason for beginning in this fashion, rather than using the
\cite{wake06} templates as for the SDSS LRG and MegaZ-LRG samples, is
that those $k+e$-corrections are not actually applicable to the
majority of the Main sample since it is flux-limited and therefore has
a significant fraction of blue galaxies.  Thus, we begin with {\sc
kcorrect} which imposes a template-dependent $k$-correction.  Then, we
place a cut in the distribution of rest-frame $g-r$, to strictly
isolate the reddest galaxies.  The cut value was motivated by
\citet{padmanabhan04}, which shows typical values for
pressure-supported elliptical galaxies in the Main sample; however the
actual cut value is shifted because the k-correction in that paper is
to $z=0.1$ rather than $0$.  The new red fractions in L3 and L4 are
$22\,\%$ and $26\,\%$, respectively.  For the galaxies satisfying this
cut, we then get absolute magnitudes using the \cite{wake06}
$k+e$-corrections, which we find are typically 0.2 magnitudes brighter
than the original Petrosian $z=0.1$ magnitudes used to define the L3
and L4 samples. 

The detailed issue of consistency between the different samples in
this paper (MegaZ-LRG, spectroscopic LRGs, and Main L3 and L4 red
samples) is irrelevant to any attempts to simply fit each sample to an
intrinsic alignment amplitude.  However, before attempting to combine
the measurements in the different samples, we must address this issue.
Thus, a detailed comparison will be given in
Sect.$\,$\ref{sec:colorcomparison}.

\subsection{Photometric redshifts}
\label{sec:photoz}

The MegaZ-LRG catalogue is selected from the SDSS imaging database
using a series of colour and magnitude cuts \citep{collister07} that
match the default selection criteria of the 2SLAQ survey
\citep{cannon06} except for going 0.2 magnitudes fainter. The
spectroscopic redshifts available from 2SLAQ were used to train and
test the photometric redshift code, which was applied to the entire
set of LRGs selected from the SDSS imaging database.

Around $13{,}000$ spectroscopic objects were obtained in the 2SLAQ
survey, of which about $8000$ were used to train a neural network
\citep{collister04} to produce photometric redshifts, leaving
approximately 5100 galaxies to verify the estimates
\citep{collister07,abdalla08}.  The quality of the photometric
redshifts is very good because of the large 4000 Angstrom break present in
luminous red galaxies. Accurate photometric redshifts are paramount for
our analysis in order to obtain good signal-to-noise, and to ensure that the
matter-intrinsic correlations dominate the observed signals, see
Sect.$\,$\ref{sec:othersignals}.

\begin{figure}[t]
\centering
\includegraphics[scale=.36,angle=270]{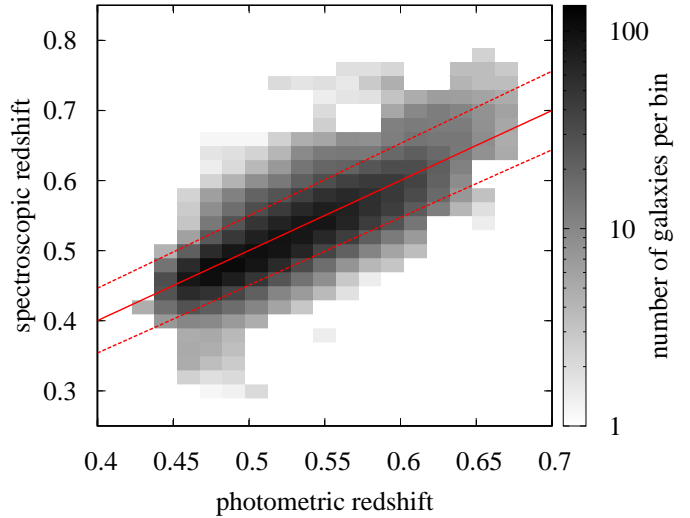}
\caption{Binned histogram of spectroscopic redshifts from 2SLAQ and photometric redshift estimates from the MegaZ-LRG catalogue. Note that the shading of the bins is logarithmic. The solid line indicates a one-to-one relation between spectroscopic and photometric redshifts, coinciding with the mean trend to high accuracy. The dotted lines correspond to the $\pm 1\sigma$ scatter.}
\label{fig:2SLAQ-histogram}
\end{figure}

We plot in Fig.$\,$\ref{fig:2SLAQ-histogram} the quality of the
photometric redshifts that can be tested with the verification sample
of $\sim$ 5000 2SLAQ spectra.  We find that the distribution of the
differences between photometric redshift estimate and spectroscopic
redshift has a mean of zero. As a consequence, the mean trend of the
photometric redshift distribution, given a spectroscopic redshift, is
indistinguishable from a one-to-one relation.  The number of galaxies
with a difference between photometric redshift $\bar{z}$ and
spectroscopic redshift $z$ largely exceeding the typical scatter
($5\sigma$ or more) is always less than $3\,\%$ for any photometric
redshift bin in the range $\bar{z} < 0.65$. The distribution of
differences between photometric and spectroscopic redshift is well fit
by a Gaussian with width $0.024(1+z)$ (corresponding to the scatter
shown in the figure), in good agreement with the results by
\citet{collister07} who find a very similar scatter in the range $0.45
< \bar{z} < 0.50$ in which most galaxies of our sample reside. Their
scatter increases by up to $50\,\%$ for higher photometric redshifts
in the range $0.60 < \bar{z} < 0.65$. 

While the distribution of spectroscopic redshifts given a photometric redshift, i.e. $p(z|\bar{z})$ is unbiased on average, the distributions $p(\bar{z}|z)$ have significant systematic offsets, as is evident from Fig.$\,$\ref{fig:2SLAQ-histogram}. Photometric redshift estimates for galaxies at low redshift are biased high, those for galaxies at high redshift are biased low, leading to a distribution of photometric redshifts that is more compact than the corresponding spectroscopic distribution. Since faint galaxies are preferentially found at low redshift, and bright galaxies at high redshift, the luminosity distribution of MegaZ-LRG galaxies based on photometric redshifts as displayed in Fig.$\,$\ref{fig:distributions} also appears more compact than the true distribution. We are not directly affected by this change in the shape of the distribution because we base our analysis on mean sample luminosities. However, as was discussed in Sect.$\,$\ref{sec:megaz}, the re-distribution of galaxies due to photometric redshifts also leads to a small change in the mean luminosity of the MegaZ-LRG samples which we correct for via the photometric versus spectroscopic redshift relation from 2SLAQ. 

Note that we employ the photometric versus spectroscopic redshift histogram of Fig.$\,$\ref{fig:2SLAQ-histogram} directly in the correlation function models as an approximation to $p(z|\bar{z})$, despite the $0.2$ magnitudes fainter catalogue used in this work. We will show in Sect.$\,$\ref{sec:gicorr_piscaling} that this is not a significant cause of systematic error in our analysis, as expected because the photometric redshift properties are only slightly extrapolated to $i_{\rm deV}=20$. We note furthermore that, due to the small field of the 2SLAQ survey, cosmic variance limits the accuracy of using this photometric versus spectroscopic redshift relation for MegaZ-LRG as well.

\subsection{Galaxy shape measurements}
\label{sec:shapemeasurements}

To measure the intrinsic shears, we use PSF-corrected shape
measurements from SDSS, specifically the galaxy ellipticity measurements by
\citet{mandelbaum05}, who obtained shapes for more than 30
million galaxies in the SDSS imaging data down to extinction-corrected
model magnitude $r=21.8$ using the {\sc Reglens} pipeline.  We refer the
interested reader to \citet{hirata03} for an outline of the
PSF correction technique (re-Gaussianisation) and to
\citet{mandelbaum05} for all details of the shape measurement.
The two main criteria for the shape measurement to be considered high
quality are that galaxies must (a) have extinction-corrected $r$ band
model magnitude $r<21.8$, and (b) be well-resolved compared to the
PSF size in both $r$ and $i$ bands.

\begin{figure}[t]
\centering
\includegraphics[scale=.42,angle=270]{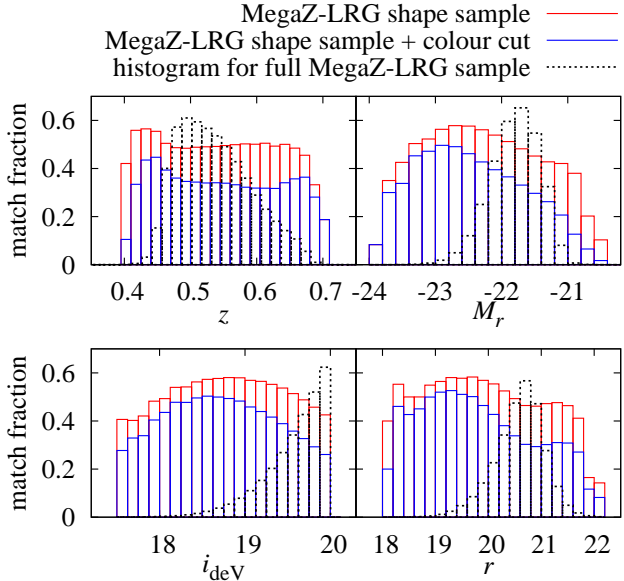}
\caption{Fraction of galaxies in the MegaZ-LRG sample with
  high-quality shape measurements, as a function of photometric
  redshift (top left panel), rest-frame absolute $r$ band magnitude
  $M_r$ (top right panel), $i$ band de Vaucouleurs magnitude $i_{\rm
  deV}$ which was used as a selection criterion for the MegaZ-LRG
  catalogue (bottom left panel), and apparent observer-frame $r$ band
  magnitude used to impose a magnitude limit on the shape catalogue
  (bottom right panel). The red histograms show the match fraction for
  the full MegaZ-LRG shape sample, and the blue histograms for shape
  sample with the additional colour cut that will be discussed in
  Sect.$\,$\ref{sec:colorcomparison}. For reference we have added to
  each panel the histogram of the full MegaZ-LRG sample with arbitrary
  normalisation as black dotted lines. Note that the fraction of
  galaxies with shape information does not depend strongly on redshift
  and $r$ band magnitude.}
\label{fig:matchfrac}
\end{figure}

The fraction of MegaZ-LRG galaxies with high-quality shape measurements is $50\,\%$, nearly independent of photometric redshift and $r$ band magnitude, see Fig.$\,$\ref{fig:matchfrac}. We also plot this match fraction as a function of the $i$ band de Vaucouleurs magnitude, where $i_{\rm deV} \leq 20$ was used as the magnitude cut for the MegaZ-LRG catalogue, and do not find a significant evolution either. It is interesting to note that the observer-frame $r-i_{\rm deV}$ strongly increases with redshift, exceeding unity around $z=0.5$, i.e. approximately at the peak of the MegaZ-LRG redshift distribution. Therefore all galaxies close to the $r$ band magnitude limit have to be located at high redshift, given the limit on $i_{\rm deV}$ for MegaZ-LRG, so that any evolution of the match fraction with redshift should imply an evolution with $r$. Since both are roughly constant in spite of the size cuts in galaxy shape measurements, the decrease in apparent galaxy size due to the larger distance at higher redshift has to be balanced by an increase of the physical dimensions of these galaxies. This can be explained by the positive correlation between galaxy size and absolute magnitude, where for a flux-limited survey the galaxies at higher redshift are intrinsically brighter on average.

The match fraction as a function of absolute rest-frame $r$ band magnitude $M_r$ displays a moderate decrease towards intrinsically fainter galaxies. Since in this case the effects of different galaxy distances have been removed, the galaxy size-luminosity relation causes the more luminous galaxies to have generally a higher share of good shape measurements. Note that in Fig.$\,$\ref{fig:matchfrac} we also show match fractions for the MegaZ-LRG sample with an additional colour cut that removes the bluest galaxies from the sample, see Sect.$\,$\ref{sec:colorcomparison}. This cut predominantly removes galaxies which are faint in the $r$  and $i_{\rm deV}$ bands, without a significant dependence on redshift.

Moreover we find that the fraction of galaxies with high-quality shape measurements exhibits a considerable dependence on observing conditions due to the resolution cut. We have generated mock catalogues that account for all of these variations in the ability to measure galaxy shapes with magnitude and observing conditions.

\section{Measurement of correlation functions}
\label{sec:cfmeasurement}

The software for the computation of galaxy position-shape correlation functions is one of the codes used in \citet{mandelbaum06},  \citet{hirata07}, and \citet{mandelbaum09}. Here we summarise the methodology, and refer the reader to \citet{mandelbaum09} for details. This software finds pairs of galaxies (with one belonging to the shape-selected sample used to trace the intrinsic shear field, and the other belonging to the full sample used to trace the density field) using the SDSSpix package\footnote{\tt http://lahmu.phyast.pitt.edu/\~{}scranton/SDSSPix/}. We measure the correlations as a function of comoving transverse separation $r_p$ and comoving line-of-sight separation $\Pi$ of the galaxy pairs, over the complete range of redshifts covered by a sample. The correlation functions must be computed using a large range of $\Pi$ because of photometric redshift error (see Sect.$\,$\ref{sec:modelling}); we divide this range into bins of size $\Delta\Pi = 10$\hmpc. 

We adopt a variant of the estimator presented in \citet{mandelbaum06}, which is given by 
\eq{
\label{eq:LSgp}
\hat{\xi}_{\rm g+}(r_p,\Pi) = \frac{S_+D-S_+R}{D_S R}\;,
}
where $S_+D$ stands for the correlation between all galaxies in the full MegaZ-LRG catalogue, tracing the density field, and those from the subset with shape information, 
\eq{
\label{eq:defsplusd}
S_+D = \sum_{i \neq j | r_p,\Pi} \frac{e_+(j|i)}{2{\cal R}}\;.
}
Here, $e_+(j|i)$ denotes the radial component of the ellipticity of galaxy $j$ measured with respect to the direction towards galaxy $i$ out of the number density sample. A similar equation holds for $S_+R$, but in this case the galaxies of the number density sample are taken from a random catalogue compliant with the selection criteria of the MegaZ-LRG data. The subtraction of $S_+R$ is meant to remove any spurious shear component, as described in previous works.  The shear responsitivity ${\cal R}$ represents the response of our particular ellipticity definition to a shear, and for an ensemble with rms ellipticity per component of $e_\mathrm{rms}$, is roughly $1-e_\mathrm{rms}^2 \approx 0.85$.  For more details on the random catalogue generation and treatment and the shear responsitivity ${\cal R}$ see \citet{mandelbaum06}.   

The normalisation of (\ref{eq:LSgp}) is given by the number of pairs $D_S R$ with one real galaxy from the shape subsample $D_S$ and one random galaxy from the full random catalogue $R$. All these pair counts are done for galaxies with transverse comoving separation $r_p$ and comoving line-of-sight separation $\Pi$, for all redshifts $z$ in the sample. 

To deduce matter-intrinsic shear correlations from $\xi_{\rm g+}$, we also measure galaxy clustering for MegaZ-LRG via 
\eq{
\label{eq:LSgg}
\hat{\xi}_{\rm gg}(r_p,\Pi) = \frac{D_S D}{D_S R} - 1\;,
}
where $D_S D$ denotes the number of galaxy pairs between the full MegaZ-LRG catalogue (used to trace the density field) and the intrinsic shear sample, and $D_S R$ is the number of pairs with one real galaxy in the sample used to trace the intrinsic shear and one in the (full) random catalogue representing the density field sample. Again, all these galaxies are selected from bins in $r_p$ and $\Pi$ over all redshifts in the sample. By cross-correlating the MegaZ-LRG full and shape samples in (\ref{eq:LSgg}), we intend to mitigate the effect of a residual star contamination $f_{\rm contam}=0.05$ in the sample (see Sect.$\,$\ref{sec:megaz}) which should enter (\ref{eq:LSgg}) only linearly, as will be detailed in Sect.$\,$\ref{sec:bias}. Determining galaxy clustering from auto-correlations of the full galaxy sample would result in a contamination which is quadratic in $f_{\rm contam}$ to leading order.

The projected correlation function is then computed by summation of the correlation functions in (\ref{eq:LSgp}) and (\ref{eq:LSgg}) along the line of sight, multiplied by $\Delta\Pi$. This calculation is done in $N_\mathrm{bin}=10$ logarithmically spaced bins in transverse separation in the range $0.3<r_p<60$\hmpc. We re-bin the existing correlation functions in $r_p$ for the SDSS LRG and Main samples accordingly. For the spectroscopic samples, bins in line-of-sight separation have a width of $\Delta \Pi=4$\hmpc. The cut-off in this stacking process is $\Pi_{\rm max}=60$\hmpc\ for the spectroscopic data sets, capturing virtually all of the signal \citep{mandelbaum06,hirata07,padmanabhan07,mandelbaum09}. MegaZ-LRG correlation functions are computed for $\Pi_{\rm max}=90$\hmpc\ and $\Pi_{\rm max}=180$\hmpc, where we will investigate the effect of these truncations in more detail in Sect.$\,$\ref{sec:gicorr_piscaling}. Note that these values of $\Pi_{\rm max}$ were chosen to roughly coincide with the $1\sigma$ and $2\sigma$ photometric redshift scatter.

Covariance matrices for MegaZ-LRG are determined using a jackknife with 256 regions, in order to account properly for shape noise, shape measurement errors, and cosmic variance. For the total survey area probed by MegaZ-LRG the maximum comoving transverse separations contained in each jackknife region are well above $100$\hmpc\ at $z=0.5$, and thus considerably larger than the maximum $r_p$ used for the analysis, so that the jackknife samples are independent. 50 jackknife regions were used to obtain covariances for the SDSS LRG and Main samples \citep{hirata07}, where the smaller number of regions is a consequence of the lower mean redshift and the smaller survey area covered. We address the issue of noise in the covariance matrices below in Sect.$\,$\ref{sec:fittingmethod}.

To compute correlation functions as a function of the comoving separations $r_p$ and $\Pi$, one needs to assume a fiducial cosmology to transform the observable galaxy redshifts and angular separations. The cosmological model for this conversion differs from our fiducial cosmology in the value $\Omega_{\rm m}=0.3$ for all samples under consideration, in order to maintain consistency with the signals computed for the SDSS LRG sample from \cite{hirata07}. Since we use our default value of $\Omega_{\rm m}=0.25$ for all model calculations, this discrepancy could bias our results. We evaluate the effect of the difference in $\Omega_{\rm m}$ for the MegaZ-LRG high-redshift bin which is the sample at the highest redshift and should thus be affected most. The change in $\Pi$ can be safely neglected while $r_p$ changes by $2\,\%$ at $z=0.59$. Converting the transverse separation of the observed correlation functions from $\Omega_{\rm m}=0.3$ to $\Omega_{\rm m}=0.25$, we find changes in the fit results for the galaxy bias and the intrinsic alignment model amplitude below $1\,\%$ each, and thus conclude that the discrepancy in $\Omega_{\rm m}$ can be neglected in our analysis.

\section{Modelling}
\label{sec:modelling}

While the methodology for the analysis of spectroscopic galaxy samples is already well established \citep{mandelbaum06,hirata07}, we consider for the first time a sample with only photometric redshift information, obtained from the MegaZ-LRG catalogue. In this section and in Appendix \ref{sec:appgicorr} we derive the models that are later compared to the observational data, incorporating photometric redshift uncertainty. The photometric redshift scatter has two major effects on our measurements: First, the truncation of the observed correlation function at large line-of-sight separations $\Pi$ has to be taken into account explicitly in the model. Second, we must assess the importance of contributions other than galaxy number density-intrinsic shear correlations to the observations. For reasons of optimum signal-to-noise and a simplified physical interpretation, the observations are expressed as line-of-sight projected correlation functions as a function of comoving transverse separation between galaxy pairs, and we transform the model accordingly.

\subsection{The number density-intrinsic shear correlation function}
\label{sec:gIcorrelationfct}

As a first step, we compute the correlation function between galaxy number density and intrinsic shear (hereafter gI), $\xi_{\rm gI}^{\rm phot}(\bar{r}_p,\bar{\Pi},\bar{z}_{\rm m})$. As before, $\bar{r}_p$ denotes the comoving transverse separation, and $\bar{\Pi}$ the comoving line-of-sight separation of galaxy pairs which are located at a mean redshift $\bar{z}_{\rm m}$. Note that we assign a bar to all quantities derived from photometric redshift estimates. In the following we will refer to correlations of the form $\xi(r_p,\Pi,z)$ as the three-dimensional correlation function. We follow the notation of \citet{joachimi10} in denoting the different signals that contribute to the observations\footnote{Note that, as a consequence, our notation differs from the one in \citet{mandelbaum06}, \citet{hirata07}, and \citet{mandelbaum09} in that these works used the term \lq GI\rq\ for both the measured galaxy number density-ellipticity cross-correlations (gI in this paper) and the derived GI intrinsic alignment term.}. As a reminder, for each of the galaxy samples considered, the galaxy pairs used to compute $\xi_{\rm gI}^{\rm phot}(\bar{r}_p,\bar{\Pi},\bar{z}_{\rm m})$ consist of one galaxy out of the density sample tracing the galaxy distribution and one galaxy out of the subsample with shape information used to trace the intrinsic shear.

In Appendix \ref{sec:appgicorr} we derive an approximate procedure to model $\xi_{\rm gI}^{\rm phot}(\bar{r}_p,\bar{\Pi},\bar{z}_{\rm m})$ to good accuracy. As we show there, $\xi_{\rm gI}^{\rm phot}(\bar{r}_p,\bar{\Pi},\bar{z}_{\rm m})$ can be obtained via a simple coordinate transformation, see (\ref{eq:trafo1}), from the angular correlation function
\eq{
\label{eq:gIhankel}
\xi^{\rm ang}_{\rm gI} \br{\theta; \bar{z}_1, \bar{z}_2} = - \int_0^\infty \frac{\dd \ell\; \ell}{2\,\pi}\; J_2\br{\ell \theta}\; C_{\rm gI} \br{\ell; \bar{z}_1, \bar{z}_2}\;,
}
where the angular gI power spectrum $C_{\rm gI}$ is given in terms of the underlying three-dimensional power spectrum $P_{\rm gI}$ by the Limber equation
\eqa{
\label{eq:gIlimber}
C_{\rm gI} \br{\ell; \bar{z}_1, \bar{z}_2} &=& \int_0^{\chi_{\rm hor}}
\dd \chi'\; \\ \nn && \hspace*{0.0cm} \times\; \frac{p_n \br{\chi' |
\chi(\bar{z}_1)}\; p_\epsilon \br{\chi' | \chi(\bar{z}_2)}}{\chi'^2}\;
P_{\rm gI}\br{\frac{\ell}{\chi'},z(\chi')}\;.  }  Here, $\chi$ denotes
comoving distance, integrated up to the comoving horizon distance
$\chi_{\rm hor}$. We have introduced the probability distributions of
comoving distances $p_n$ for the galaxy density sample and
$p_\epsilon$ for the galaxy shape sample. They are related via $p_x
\br{\chi | \chi(\bar{z}_i)} = p_x \br{z | \bar{z}_i} \dd z / \dd \chi$
to the probability of a redshift $z$ given the photometric redshift
estimate $\bar{z}_i$.  The latter can be extracted from the histogram
in Fig.$\,$\ref{fig:2SLAQ-histogram} by a vertical section at
$\bar{z}_i$. Note that we use these distributions, extracted from the
2SLAQ photometric versus spectroscopic relation and linearly
interpolated, for both the shape-selected and the full number density
sample as we find that their redshift distributions agree to good
accuracy. This agreement is not obvious because the images of the
galaxies selected for shape measurement need to have a certain minimum
angular size. We trace the agreement of the two redshift distributions
for the MegaZ-LRG sample back to a rough balance between the effect
that galaxies at higher redshift, i.e. at larger distance, appear
smaller due to the larger angular diameter distance, and the
counteracting effect that for a given range of apparent magnitudes,
galaxies at higher redshift are on average intrinsically brighter and
thus have larger intrinsic physical sizes, see also
Sect.$\,$\ref{sec:shapemeasurements}.

Throughout, we will assume that the galaxy bias $b_{\rm g}$ is scale-independent. Then $P_{\rm gI}$ can be related to the matter-intrinsic power spectrum via $P_{\rm gI}\br{k,z} = b_{\rm g}\, P_{\delta {\rm I}}\br{k,z}$, where we calculate $P_{\delta {\rm I}}$ according to the linear alignment model \citep{catelan01,hirata04}
\eq{
\label{eq:GIlinalign}
P_{\delta {\rm I}}\br{k,z} = - A\; C_1\; \rho_{\rm cr}\;\frac{\Omega_{\rm m}}{D(z)}\; P_\delta\br{k,z}\;, 
}
where $D(z)$ denotes the linear growth factor, normalised to unity
today. Equation (\ref{eq:GIlinalign}) differs from the result derived
by \citet{hirata04} in the dependence on redshift, with the latter
expression featuring an additional term $(1+z)^2$. In Appendix
\ref{sec:larederive} we derive the correct scaling with redshift shown
in (\ref{eq:GIlinalign}); see also \citet{hirata10}.  We absorb the
dimensions of the normalisation into the constant $C_1$ which is set
to $C_1\, \rho_{\rm cr} \approx 0.0134$, following \citet{hirata04}
and \citet{bridle07} who matched the amplitude of the linear alignment
model to SuperCOSMOS observations at low redshift. This choice is
common but of no particular relevance for our study, and the
normalisation is in principle arbitrary. Thus, we introduce a
dimensionless amplitude parameter $A$ which is free to vary. If not
stated otherwise, we use $A=1$ to demonstrate our model in the
following subsections.

\begin{figure}[t!!!]
\centering
\includegraphics[scale=.33,angle=270]{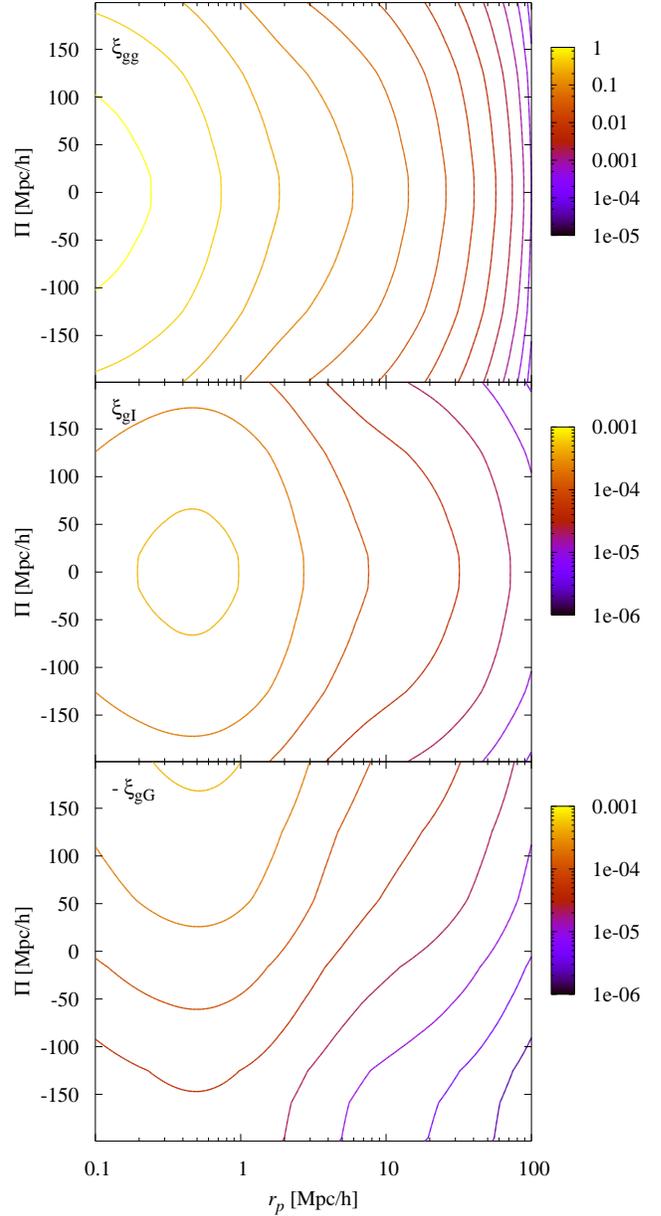}
\caption{Three-dimensional correlation function models of a sample with the MegaZ-LRG photometric redshift error as a function of comoving line-of-sight separation $\Pi$ and comoving transverse separation $r_p$ at $z_{\rm m} \approx 0.5$. The galaxy bias has been set to 1.9 in all panels. \textit{Top panel}: Galaxy clustering correlation (gg). Contours are logarithmically spaced between 1 (yellow shading) and $10^{-5}$ (violet shading) with three lines per decade. \textit{Centre panel}: Galaxy number density-intrinsic shear correlations (gI). Contours are logarithmically spaced between $10^{-3}$ (yellow shading) and $10^{-6}$ (violet shading) with three lines per decade. \textit{Bottom panel}: Galaxy-galaxy lensing (gG). For ease of direct comparison, the contours are encoded exactly like in the centre panel. Note that the galaxy-galaxy lensing signal is not symmetric around $\Pi=0$, in contrast to the gg and gI terms. Also, it is negative, so that the modulus is plotted. For an illustration of the effect of photometric redshift scatter see Fig.$\,$\ref{fig:gicorr_photozeffect}.}
\label{fig:gicorr_signals}
\end{figure}

The original derivation of the intrinsic alignment model requires the linear matter power spectrum in (\ref{eq:GIlinalign}), see e.g. Appendix \ref{sec:larederive}, but following \citet{bridle07} we use the full $P_\delta$ with nonlinear corrections, which provides satisfactory fits to existing data. Thus we refer to (\ref{eq:GIlinalign}) as the nonlinear version of the linear alignment (NLA) model henceforth. The matter power spectrum is computed for our default cosmological model, a spatially flat $\Lambda$CDM universe with parameters $\Omega_{\rm m}=0.25$, $\sigma_8=0.8$, $h=0.7$, and $n_{\rm s}=1.0$. The transfer function is calculated according to \citet{eisenstein98} using $\Omega_{\rm b}=0.05$, and nonlinear corrections are included following \citet{smith03}.

For illustration, in Fig.$\,$\ref{fig:gicorr_signals}, centre panel, the predicted gI correlation function for the MegaZ-LRG sample (assuming alignments consistent with those in SuperCOSMOS, and including photometric redshift errors) is plotted, for $z_{\rm m} \approx 0.5$ and assuming $b_{\rm g}=1.9$ which is roughly in agreement with the results obtained for the full MegaZ-LRG sample by \citet{blake07} and close to the value we determine, see Sect.$\,$\ref{sec:bias}. The signal is strongest around $\Pi=0$, but extends far out along the line-of-sight direction due to the photometric redshift scatter. The correlations have a maximum at $r_p \sim 0.5$\hmpc\ and decrease for larger $r_p$ due to the diminishing physical interaction between galaxies at large separation, and for small $r_p$ since the separation vector between pairs of galaxies gets close to the line-of-sight direction, see also Fig.$\,$\ref{fig:gicorr_photozeffect}. 

Note that throughout this work we do not include redshift-space distortions in our modelling. Since for both spectroscopic and photometric data we integrate the correlation functions over the line-of-sight separation out to at least $60$\hmpc\ and $90$\hmpc, respectively, redshift-space distortions should have a negligible influence on the integrated signals \citep[see also the discussions in][]{padmanabhan07,blake07,mandelbaum09}.  In addition, in the latter case, the redshift space distortions should be subdominant compared to the size of the photometric redshift errors.

\subsection{Contribution of other signals}
\label{sec:othersignals}

Due to the photometric redshift scatter, contributions to galaxy position-shape correlations other than the gI term may become important. In the weak lensing limit, the measured ellipticity of a galaxy image is the sum of the intrinsic ellipticity and the gravitational shear, while the number density is determined by an intrinsic term (whose two-point correlation is the usual galaxy clustering) plus modifications by lensing magnification effects. Hence, in terms of angular power spectra one can write \citep[for details see][]{bernstein08,joachimi10}
\eqa{
\label{eq:contributions}
C_{n \epsilon}(\ell; z_1,z_2) &=& C_{\rm gI}(\ell; z_1,z_2) + C_{\rm gG}(\ell; z_1,z_2)\\ \nn
&& \hspace*{1cm} + C_{\rm mG}(\ell; z_1,z_2) + C_{\rm mI}(\ell; z_1,z_2)\;,
}
for each set of galaxy samples that is correlated. Apart from the gI signal, contributions from galaxy-galaxy lensing (gG), magnification-shear correlations (mG), and magnification-intrinsic correlations (mI) occur. 

If $z_1 \approx z_2$, the gI term is expected to dominate\footnote{It is instructive to bear in mind that the gI term, which we use to constrain the power spectrum $P_{\delta {\rm I}}$, is restricted to pairs of galaxies at similar redshifts while the GI signal, which is generated by $P_{\delta {\rm I}}$, is not limited to close pairs of galaxies and is actually strongest if galaxies are widely separated in redshift. The reason for these opposing redshift dependencies are the different weightings in the Limber equations, see (\ref{eq:gIlimber}) and (\ref{eq:limberGI}) below.}, whereas galaxy-galaxy lensing is the dominant term in $C_{n \epsilon}(\ell; z_1,z_2)$ if a number density and a shape sample at significantly different $z_1<z_2$ are correlated. In addition, correlations between lensing magnification and gravitational shear can have a contribution, e.g. if a matter overdensity causes both tangential shear alignment and an apparent boost in the number density of background galaxies. Likewise this overdensity could tidally align surrounding galaxies and thus create correlations between magnification and the intrinsic galaxy shapes.

All these additional signals are related to the three-dimensional correlation function via relations of the form (\ref{eq:gIhankel}), so that, to assess the importance of their contributions, it is sufficient to compare the angular power spectra. We restrict the consideration to power spectra that correlate galaxy shapes and number densities at the same redshift (redshift auto-correlations), and that hence are representative of correlation functions at small line-of-sight separations. Note that for larger values of $|\Pi|$, the scaling of the different signals with redshift becomes important, so that e.g. the amplitudes of the gG and mG signals relative to the gI term increase if bigger values for $\Pi_{\rm max}$ are chosen.

The corresponding Limber equations of the additional signals read \citep[e.g.][]{joachimi10}
\eqa{
\label{eq:othersignals}
C_{\rm gG} \br{\ell; \bar{z}_1, \bar{z}_2} &=& b_{\rm g}\; \int_0^{\chi_{\rm hor}} \dd \chi'\\ \nn
&& \hspace*{-1cm} \times\; \frac{p_n \br{\chi' | \chi(\bar{z}_1)}\; q_\epsilon \br{\chi',\, \chi(\bar{z}_2)}}{\chi'^2}\; P_\delta \br{\frac{\ell}{\chi'},z(\chi')}\;;\\ \nn
C_{\rm mG} \br{\ell; \bar{z}_1, \bar{z}_2} &=& 2 (\alpha -1)\; \int_0^{\chi_{\rm hor}} \dd \chi'\\ \nn
&& \hspace*{-1cm} \times\; \frac{q_n \br{\chi',\, \chi(\bar{z}_1)}\; q_\epsilon \br{\chi',\, \chi(\bar{z}_2)}}{\chi'^2}\; P_\delta \br{\frac{\ell}{\chi'},z(\chi')}\;;\\ \nn
C_{\rm mI} \br{\ell; \bar{z}_1, \bar{z}_2} &=& 2 (\alpha -1)\; \int_0^{\chi_{\rm hor}} \dd \chi'\\ \nn
&& \hspace*{-1cm} \times\; \frac{q_n \br{\chi',\, \chi(\bar{z}_1)}\; p_\epsilon \br{\chi' | \chi(\bar{z}_2)}}{\chi'^2}\; P_{\delta {\rm I}} \br{\frac{\ell}{\chi'},z(\chi')}\;,
}
where $\alpha = - \dd \bb{\log N(>L)} / \dd \bb{\log L}$ is the logarithmic slope of the cumulative galaxy luminosity function of the density sample \citep[e.g.][]{narayan89,bartelmann01}. Moreover we have defined the lensing weight function
\eq{
\label{eq:weightlensing}
q_x(\chi,\, \chi_1) = \frac{3 H_0^2 \Omega_{\rm m}}{2\, c^2} \frac{\chi}{a(\chi)} \int_{\chi}^{\chi_{\rm hor}} \dd \chi'\; p_x(\chi' | \chi_1)\; \frac{\chi' - \chi}{\chi'}\;
}
for $x=\bc{n,\epsilon}$, where $a$ denotes the cosmic scale factor.

To determine $\alpha$, we calculate the cumulative distribution of $i$ band de Vaucouleurs magnitudes used for selection of the MegaZ-LRG samples, and fit the slopes $s=\dd \bb{\log N(<i_{\rm deV})}/\dd i_{\rm deV}$ at the faint end of each distribution. The slope of the cumulative galaxy luminosity function is then given by $\alpha=2.5\,s$ \citep[e.g.][]{vwaer09}. We find $\alpha=2.26$ for the full MegaZ-LRG sample, $\alpha=1.29$ for the low-redshift sample, and $\alpha=3.04$ for the high-redshift sample.

\begin{figure}[t]
\centering
\includegraphics[scale=.67,angle=0]{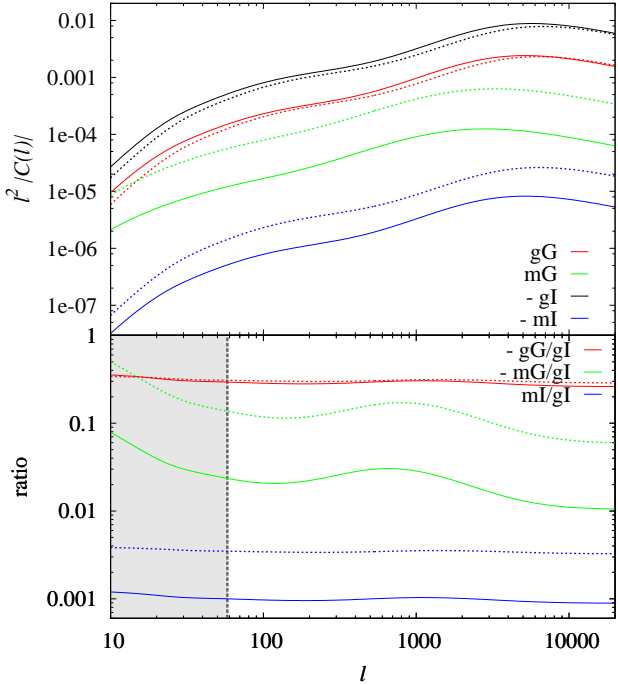}
\caption{\textit{Top panel}: Modulus of the angular power spectra of the different signals contributing to galaxy position-shape correlations. Number density-intrinsic correlations (gI) are shown in black, galaxy-galaxy lensing (gG) in red, magnification-shear correlations (mG) in green, and magnification-intrinsic correlations (mI) in blue. Solid curves correspond to the redshift auto-correlations at $z=0.5$, i.e. close to the mean redshift of the low-redshift MegaZ-LRG sample, and dotted curves to the mean redshift of the high-redshift sample at $z=0.59$. \textit{Bottom panel}: Ratio of the aforementioned signals over the gI correlations, with the same coding of the curves as above. The grey region covers angular scales that do not contribute significantly to the correlation functions. Galaxy-galaxy lensing and possibly mG correlations yield a relevant contribution to number density-shape correlations besides the gI signal.}
\label{fig:tomops}
\end{figure}

In this comparison, we assume that the intrinsic alignment signal follows the corrected NLA model with the normalisation from SuperCOSMOS. Since the strength of the intrinsic alignment signals for LRGs are expected to be significantly higher, see e.g. the results for SDSS LRGs by \citet{hirata07}, this should be a conservative assumption. Note that the SuperCOSMOS normalisation employed in foregoing work was based on the redshift scaling of the II term given in \citet{hirata04}. Since SuperCOSMOS has a mean redshift of $0.1$, the amplitudes of the II signals in the original and corrected version of the linear alignment model should differ by about a factor of $1.1^4$ (see also Appendix \ref{sec:larederive}). Thus we retain the normalisation to SuperCOSMOS for the corrected version by setting $A=1.1^2=1.21$. In addition, we choose a galaxy bias $b_{\rm g}=1.9$, where the latter value is close to the actual fit results for MegaZ-LRG, see Sect.$\,$\ref{sec:bias}. The resulting angular power spectra for all four contributions to (\ref{eq:contributions}) are shown in Fig.$\,$\ref{fig:tomops}. We plot the redshift auto-correlation power spectra, using $\bar{z}_1=\bar{z}_2=0.5$ and $\bar{z}_1=\bar{z}_2=0.59$ corresponding approximately to the mean redshifts of the two MegaZ-LRG redshift-binned samples.

For the photometric redshift accuracy of the MegaZ-LRG sample, the gI signal still clearly dominates the position-shape correlations. It has a slightly lower amplitude at $\bar{z}=0.59$ than at $\bar{z}=0.5$ due to the broader redshift distribution at the higher photometric redshift. To verify that it is indeed the width of the distribution, and not the shift of its mean redshift, that causes the depletion, we shift the redshift distribution at $\bar{z}=0.4525$ to a mean of $0.6$ and re-compute the gI signal which then has a similar, slightly higher amplitude compared to the gI correlations at $\bar{z}=0.4525$. The other signals are less affected by the width of the contributing redshift distributions since they depend on lensing and thus have a much broader kernel in the line-of-sight integration, see (\ref{eq:othersignals}).

The mI signal never attains more than a few per mil of the gI term and is hence irrelevant for our purposes. Due to the steep slopes of the galaxy luminosity functions, magnification-shear correlations can contribute up to 20 per cent of the gI term over a wide range of angular frequencies, and even become the dominant contamination of the gI signal at small $\ell$. However, due to the Bessel function $J_2$ in the kernel of (\ref{eq:gIhankel}), contributions from small $\ell$ are largely suppressed in $\xi_{\rm gI}^{\rm phot}(\bar{r}_p,\bar{\Pi},\bar{z}_{\rm m})$. The first maximum of $J_2$ is at $\ell \theta \sim 3$, and the maximum angle probed by our analysis is $3.1\,{\rm deg}$ corresponding to $r_p=60$\hmpc\ at $z=0.4$, so that only angular frequencies $\ell \gtrsim 60$ are important, as indicated by the grey region in Fig.$\,$\ref{fig:tomops}. Still, the mG term could add of the order $10\,\%$ to the total signal under pessimistic assumptions, which is in the regime of the expected model parameter errors, so that we include the mG signal into our modelling. We note that this contribution is likely to be even more relevant for future intrinsic alignment analyses of surveys with higher statistical power and/or less accurate photometric redshifts, provided that the faint-end slope of the luminosity function has similar steepness.

Since we assume the NLA model, for which $P_{\delta {\rm I}}$ has the same angular dependence as the matter power spectrum, the differences between the signals shown in Fig.$\,$\ref{fig:tomops} can only be due to the weights in the Limber equations (\ref{eq:othersignals}). The gI, mI, and gG signals all have at least one term $p(\chi)$ in the kernel which is thus relatively compact. Only the weight of the mG correlations features a product of lensing efficiencies (\ref{eq:weightlensing}) which smears out the information over comoving distance due to its width and in addition shifts the features in the projected power spectrum because $q(\chi)$ peaks at half the source distance, see (\ref{eq:weightlensing}). Therefore the mG signal has a different angular dependence, causing the waviness in the ratio mG/gI.

Galaxy-galaxy lensing has a scale dependence that is similar to the gI term, thereby yielding a nearly constant contribution of about $30\,\%$. Therefore we need to incorporate the gG term into our model, mainly affecting the amplitude of the model correlation function due to the almost constant ratio gG/gI. Note that contrary to the usual approach to galaxy-galaxy lensing studies, we have defined the correlation function such that radial alignment produces a positive signal. Hence, the inclusion of the gG term into the model will increase the measured gI amplitude. The modulus of the three-dimensional gG correlation function is shown in Fig.$\,$\ref{fig:gicorr_signals}, bottom panel. Due to the lensing contribution, the gG correlation is not symmetric with respect to $\Pi=0$, even if the redshift distributions of the galaxy shape and density samples are identical. We note in passing that if the measurements included correlations between galaxies at largely different redshifts to which galaxy-galaxy lensing would be the dominant contribution, it would be possible to simultaneously measure the gG and gI signals, see e.g. \citet{joachimi10}. However, such joint analyses are beyond the scope of the present work.

The above statements hold only if the amplitude of the intrinsic alignment signal is of the order found in the SuperCOSMOS survey, because we have used $A \sim 1$ in (\ref{eq:GIlinalign}). If the contribution by intrinsic alignments were weaker, the importance of the gG and mG signals would further increase. However, \citet{hirata07} have demonstrated that LRGs show a strong intrinsic alignment signal at $z \sim 0.3$, so unless we find a strong decline of intrinsic alignments with redshift, $A \sim 1$ should be a conservative assumption.

We also consider galaxy clustering (hereafter gg) which will be used to determine the galaxy bias of the different samples. Since in the case of MegaZ-LRG the gg signal is affected in the same way by the photometric redshift scatter as number density-shape cross-correlations, we proceed in exact analogy and compute the three-dimensional correlation function $\xi_{\rm gg}^{\rm phot}(\bar{r}_p,\bar{\Pi},\bar{z}_{\rm m})$ from \citep[e.g.][]{hu03}
\eq{
\label{eq:gghankel}
\xi^{\rm ang}_{\rm gg} \br{\theta; \bar{z}_1, \bar{z}_2} = \int_0^\infty \frac{\dd \ell\; \ell}{2\,\pi}\; J_0\br{\ell \theta}\; C_{\rm gg} \br{\ell; \bar{z}_1, \bar{z}_2}\;,
}
again by means of (\ref{eq:trafo2}), where the angular power spectrum is related to the matter power spectrum via
\eqa{
\label{eq:gglimber}
C_{\rm gg} \br{\ell; \bar{z}_1, \bar{z}_2} &=& b_{\rm g}^2\; \int_0^{\chi_{\rm hor}} \dd \chi'\; \\ \nn
&& \hspace*{-1cm} \times\; \frac{p_n \br{\chi' | \chi(\bar{z}_1)}\; p_\epsilon \br{\chi' | \chi(\bar{z}_2)}}{\chi'^2}\; P_\delta \br{\frac{\ell}{\chi'},z(\chi')}\;.
}
Note that one of the redshift probability distributions is determined from the shape sample because we use cross-correlations between the full and shape samples to measure galaxy clustering in this analysis, see Sect.$\,$\ref{sec:cfmeasurement}. We show the three-dimensional correlation function of galaxy clustering in the top panel of Fig.$\,$\ref{fig:gicorr_signals}. The strong spread of the gg signal along the line of sight demonstrates that in the case of the MegaZ-LRG sample, photometric redshift scatter and the corresponding effect of a truncation at large $\Pi$ when computing the projected correlation function has to be modelled with similar care as for the gI term. Since galaxy clustering produces a strong signal, we can safely neglect potential contributions by lensing magnification effects in this case.

\subsection{Projection along the line of sight}
\label{sec:projection}

As in the spectroscopic case, the quantity that is actually compared to the data is the projected galaxy position-shape correlation function $w_{\rm g+}$, obtained by integrating the three-dimensional correlation function $\xi_{\rm gI}^{\rm phot} (\bar{r}_p,\bar{\Pi},\bar{z}_{\rm m})$ over $\bar{\Pi}$. In addition we take the average over a range of photometric redshifts $\bar{z}_{\rm m}$ which e.g. corresponds to the two redshift bins defined for the MegaZ-LRG sample, resulting in
\eq{
\label{eq:wobs}
w_{\rm g+} (\bar{r}_p) = \int_{-\bar{\Pi}_{\rm max}}^{\bar{\Pi}_{\rm max}} \dd \bar{\Pi} \int \dd \bar{z}_{\rm m}\; {\cal W}(\bar{z}_{\rm m})\; \xi_{\rm gI}^{\rm phot} (\bar{r}_p,\bar{\Pi},\bar{z}_{\rm m})\;,
}
where the truncation at $\Pi_{\rm max}$, taken to be the same as for the data, has now been written explicitly. 

The average over $\bar{z}_{\rm m}$ contains the weighting ${\cal W}(z)$ over redshifts as derived by \citet{mandelbaum09}, which is given by
\eq{
\label{eq:zaverage}
{\cal W}(z) = \frac{p^2(z)}{\chi^2(z)\; \chi'(z)}\; \bb{\int \dd z\; \frac{p^2(z)}{\chi^2(z)\; \chi'(z)}}^{-1}\;
}
where $p(z)$ is in this case the unconditional probability distribution of photometric redshifts determined from the MegaZ-LRG sample\footnote{Note that using the photometric redshift distribution obtained from the 2SLAQ verification sample instead has no significant influence on the models.} (or its redshift-binned subsamples). Here, $\chi'(z)$ denotes the derivative of comoving distance with respect to redshift. Note that the denominator $\chi^2(z)\, \chi'(z)$ in (\ref{eq:zaverage}) is proportional to the derivative of the comoving volume $V_{\rm com}$ with respect to redshift. Equation (\ref{eq:zaverage}) can be illustrated by considering a volume-limited sample for which $p(z) = \dd V_{\rm com}/\dd z$ holds. Then ${\cal W}(z) = p(z)$, as expected for a simple average over redshift in (\ref{eq:wobs}). A flux-limited sample like MegaZ-LRG misses faint galaxies at high redshifts compared to a volume-complete sample, so these redshifts are downweighted accordingly by (\ref{eq:zaverage}) in the averaging process. Note that this procedure of computing the three-dimensional correlation function for each redshift and subsequently averaging over redshift with the weighting (\ref{eq:zaverage}) is equivalent to our treatment of the data, where the correlation function was computed as a function of $r_p$ and $\Pi$ for \emph{all} redshifts, thereby obtaining an average over the full range of redshifts covered by the sample.

In the case of galaxy clustering, the projected correlation function $w_{\rm gg}(\bar{r}_p)$ is determined in exact analogy to (\ref{eq:wobs}). In principle the integral over $\bar{\Pi}$ in (\ref{eq:wobs}) should extend to infinity, see (\ref{eq:defcrosspower}), but it can be truncated at some maximum value $\bar{\Pi}_{\rm max}$ because correlations between pairs of galaxies do not extend to infinite separation. However, the photometric redshift scatter for the MegaZ-LRG sample causes significant contributions to the line-of-sight integral in (\ref{eq:wobs}) from $\Pi \gtrsim 300$\hmpc, although signal-to-noise requirements enforce relatively small values of $\bar{\Pi}_{\rm max}$. 

For a consistency check on the modelling of the line-of-sight truncation of the correlation functions in case of the MegaZ-LRG sample, we compute the model and the observed correlation functions for $\bar{\Pi}_{\rm max}=90$\hmpc\ and $\bar{\Pi}_{\rm max}=180$\hmpc. For the number density-shape correlations as well as the galaxy clustering signal, we compare the ratios of the correlation functions with these two cut-offs in Sect.$\,$\ref{sec:gicorr_piscaling}, finding good agreement between model and observational data. We use cut-offs in the signal integration along the line-of-sight at either $180$\hmpc\ or $90$\hmpc\ for the fits to the full MegaZ-LRG sample and find consistent results for the intrinsic alignment fit parameters with errors of the same order, see Sect.$\,$\ref{sec:iafits}.  The signals for $w_{\rm gg}$ and $w_{\rm g+}$ both have similar signal-to-noise when truncating at these two values of $\bar{\Pi}_{\rm max}$. The correlation functions for the two MegaZ-LRG redshift-binned samples have been cut off at $\bar{\Pi}_{\rm max}=90$\hmpc\ throughout.

In addition to the photometric MegaZ-LRG sample, we will also reconsider spectroscopic samples from SDSS. As discussed before, the line-of-sight truncation can be ignored in the case of spectroscopic data, so that the projected correlation function is simply given by
\eq{
\label{eq:wgispec}
w_{\rm g+}(r_p) = - b_{\rm g} \int \dd z\; {\cal W}(z) \int_0^\infty \frac{\dd k_\perp\, k_\perp}{2\,\pi}\; J_2(k_\perp r_p)\; P_{\delta {\rm I}} (k_\perp,z)\;,
}
where (\ref{eq:defcrosspower}) and the same redshift averaging procedure as in (\ref{eq:wobs}) were used. Similarly, one obtains for the spectroscopic galaxy clustering signal \citep[e.g.][]{hirata07}
\eq{
\label{eq:wgg}
w_{\rm gg}(r_p) = b_{\rm g}^2 \int \dd z\; {\cal W}(z) \int_0^\infty \frac{\dd k_\perp\, k_\perp}{2\,\pi}\; J_0(k_\perp r_p)\; P_\delta (k_\perp,z)\;. 
}

\subsection{Fitting method}
\label{sec:fittingmethod}

\begin{figure*}[t]
\centering
\includegraphics[scale=.38,angle=270]{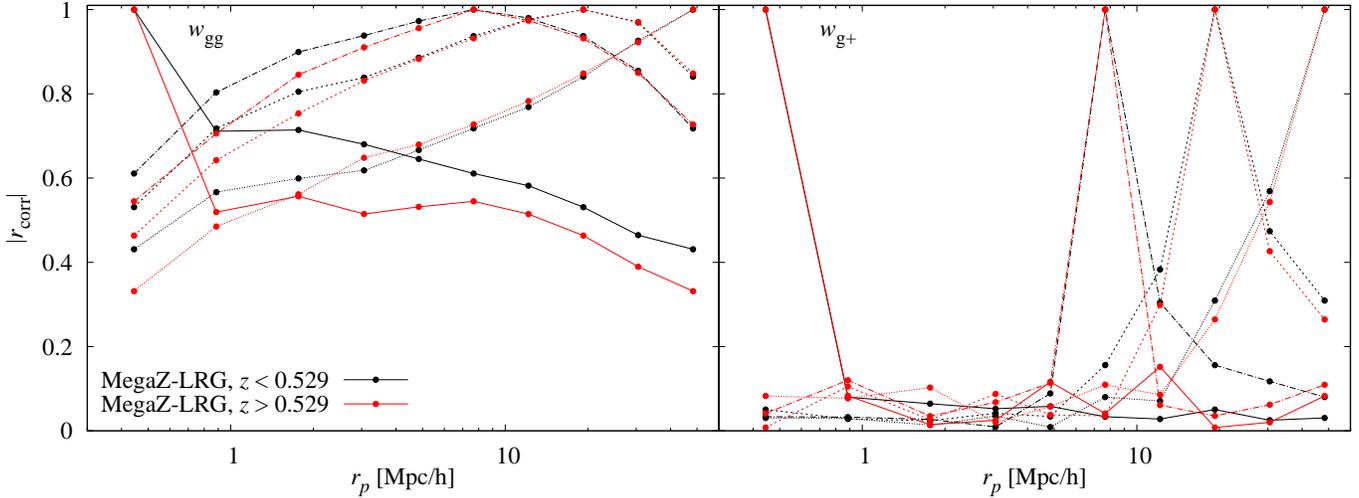}
\caption{Correlation coefficients $r_{\rm corr}$ between different transverse separation bins of the correlation functions $w_{\rm gg}$ (left panel) and $w_{\rm g+}$ (right panel) for the redshift-binned MegaZ-LRG samples. Shown is the modulus of $r_{\rm corr}$ for the 10 $r_p$ bins correlated with the smallest bin considered ($r_p=0.44$\hmpc, solid lines), the smallest bin used for the fits ($r_p=7.67$\hmpc, dot-dashed lines), the centre bin used for the fits ($r_p=19.25$\hmpc, dashed lines), and the largest bin ($r_p=48.30$\hmpc, dotted lines). Black curves correspond to the lower redshift bin, red curves to the higher redshift bin. Note that the black and red lines for the correlations of $w_{\rm g+}$ with $r_p=0.44$\hmpc\ nearly coincide.}
\label{fig:corrcov}
\end{figure*}

We perform the fits to the data via weighted least squares minimisation, using the reduced $\chi^2$ at the minimum to quantify the goodness of fit. The data are compared to signals computed for the NLA model (\ref{eq:GIlinalign}) or a more flexible variant (\ref{eq:iamodel}) introduced in Sect.$\,$\ref{sec:results}. To obtain confidence regions, we compute the likelihood according to $L \propto \exp \br{-\chi^2/2}$, i.e. assuming Gaussianity. The posterior probability in parameter space is computed via Bayes' theorem, using a top-hat prior that is truncated outside the regime where the likelihood deviates substantially from zero, see Sect.$\,$\ref{sec:iafits-combined} for details. When doing three-parameter fits, we marginalise in each case over the hidden parameters. $1\sigma$ and $2\sigma$ confidence contours are then defined by the regions containing $68.3\,\%$ or $95.4\,\%$ of the (marginalised) posterior.

Since we have noisy jackknife covariances obtained from a finite number of realisations, the inverse of these covariances, required for the $\chi^2$, is biased \citep{hirata04b,hartlap06}. We employ the corrected estimator for the inverse covariance presented in \citet{hartlap06},
\eq{
\label{eq:cinvestimator}
\widehat{\br{\rm Cov}^{-1}} = \frac{n-d-2}{n-1}\; \br{\widehat{{\rm Cov}}}^{-1} \equiv {\cal F}\; \br{\widehat{{\rm Cov}}}^{-1}\;,
}
where $d$ is the dimension of the data vector and $n$ the number of realisations used to estimate the covariance. Equation (\ref{eq:cinvestimator}) was derived under the assumption of statistically independent data vectors with Gaussian errors. For the SDSS samples ($d=10$, $n=50$) we find ${\cal F} \approx 0.776$, and for MegaZ-LRG ($d=10$, $n=256$) ${\cal F} \approx 0.957$, the latter result being in excellent agreement with the results obtained from the simulations described in Appendix D of \citet{hirata04b}. 

To study the characteristics of the covariances, we compute the correlation coefficient between different transverse separation bins,
\eq{
\label{eq:corrcoeff}
r_{\rm corr}(r_{p,a},r_{p,b}) = \frac{{\rm Cov}(r_{p,a},r_{p,b})}{\sqrt{{\rm Cov}(r_{p,a},r_{p,a})\;{\rm Cov}(r_{p,b},r_{p,b})}}\;.
}
In Fig.$\,$\ref{fig:corrcov} we have plotted $r_{\rm corr}$ of both $w_{\rm gg}$ and $w_{\rm g+}$ for the two MegaZ-LRG samples at low and high redshift. While $w_{\rm g+}$ decorrelates quickly with only moderate correlation between neighbouring bins on the largest scales, $w_{\rm gg}$ features strong positive, long-range correlation particularly on the larger scales that are used for the fits. The correlation coefficients have similar values for the two redshift bins, with the $z < 0.529$ bin showing slightly higher correlation in most cases. For the SDSS samples we find a similar correlation structure for $w_{\rm g+}$, but much weaker correlations in $w_{\rm gg}$. 

The difference in correlation length between $w_{\rm gg}$ and $w_{\rm g+}$ is caused by the different kernels in the Hankel transformation between power spectrum and correlation function, see (\ref{eq:gIhankel}) and (\ref{eq:gghankel}). Since $J_2(x)$ decreases faster than $J_0(x)$ for increasing $x$, we expect $w_{\rm gg}$ to generally feature stronger correlations. A given transverse separation $r_p$ between galaxies is observed under a smaller angle if these galaxy pairs are located at higher redshift, and it is this angle which enters the argument of the Bessel functions. Therefore the correlation present in $w_{\rm gg}$ is more pronounced in the MegaZ-LRG samples, which are at considerably higher redshift than the other SDSS samples.

\section{Results}
\label{sec:results}

\subsection{Scaling with line-of-sight truncation}
\label{sec:gicorr_piscaling}

To test whether the data and the model show the same behaviour when varying $\bar{\Pi}_{\rm max}$, we compute both $w_{\rm g+}$ and $w_{\rm gg}$ for the MegaZ-LRG sample according to (\ref{eq:wobs}) for $\bar{\Pi}_{\rm max}=90$\hmpc\ and $\bar{\Pi}_{\rm max}=180$\hmpc. Then we compare the ratio of $w_{\rm g+}$ with cut-off $\bar{\Pi}_{\rm max}=90$\hmpc\ over $w_{\rm g+}$ with cut-off $\bar{\Pi}_{\rm max}=180$\hmpc\ (and likewise for $w_{\rm gg}$) obtained from the model to the corresponding ratio computed from the observations. Since the projected correlation functions with the different cut-offs are strongly correlated, we compute the errors on the ratio again via jackknifing. Note that due to these correlations the actual errors on the ratio are significantly smaller than if one assumed them to be independent. Note furthermore that the ratio also inherits a significant correlation between different transverse separations from the individual projected correlation functions, in particular for $w_{\rm gg}$, see Fig.$\,$\ref{fig:corrcov}.

\begin{figure}[t]
\centering
\includegraphics[scale=.35,angle=270]{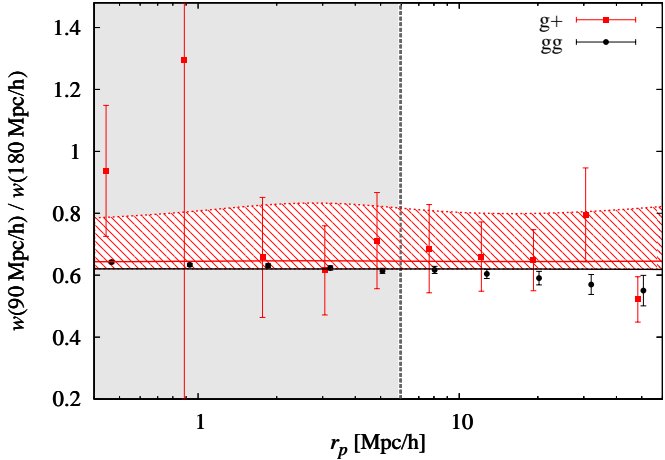}
\caption{Effect of the cut-off in $\Pi$ in the projection of the three-dimensional correlation functions along the line of sight. Shown is the ratio of the projected correlation function computed for $\bar{\Pi}_{\rm max}=90$\hmpc\ over the correlation function obtained with $\bar{\Pi}_{\rm max}=180$\hmpc, for both the galaxy clustering signal (gg, in black) and number density-shape correlations (g+, in red). Points are computed from the MegaZ-LRG data, using the full range in redshifts. Note that the black points have been slightly offset horizontally for clarity. The black line is obtained from the model for $w_{\rm gg}$. The red hatched region comprises the range of ratios for $w_{\rm g+}$ with different relative strengths of the galaxy-galaxy lensing contribution. The red solid line indicates the ratio resulting for the best-fit intrinsic alignment amplitude determined in Sect.$\,$\ref{sec:iafits}. Note that the error bars at different transverse separations are correlated, in particular for $w_{\rm gg}$ at large $r_p$, see Fig.$\,$\ref{fig:corrcov}.}
\label{fig:ratiopicut}
\end{figure}

In Fig.$\,$\ref{fig:ratiopicut} we have plotted the ratios of the projected correlation functions with the different cut-offs. The model prediction for the ratio of the galaxy clustering signals is in fair agreement with the data, yielding a loss of about $40\,\%$ when reducing $\bar{\Pi}_{\rm max}$ from $180$\hmpc\ to $90$\hmpc. On scales where the galaxy bias becomes nonlinear, indicated by the grey shaded region, bias effects may not cancel in the ratio anymore, so that deviations from the model prediction are expected. On scales $r_p \gtrsim 20$\hmpc\ one observes an apparently significant trend in the data to fall below the model prediction. However, note that $w_{\rm gg}$ at large transverse separations features very strong cross-correlations between the data points, and this property is inherited by the ratio. A fit of the model ratio to the observed ratio for the five data points with $r_p \geq 6$\hmpc\ yields a reduced $\chi^2$ of 1.75, corresponding to a $p$-value of 0.12, and hence data and model are still marginally consistent.

The prediction for the ratio of the position-shape correlation functions with cut-offs at $90$\hmpc\ and $180$\hmpc, respectively, depends on the relative strength of the galaxy-galaxy lensing contribution with respect to the gI signal (we neglect the mG contribution whose effect should be very small in this case). We shade the possible range of ratios in Fig.$\,$\ref{fig:ratiopicut} between the lowest ratio resulting for a negligible gG term which almost coincides with the ratio for galaxy clustering, and the highest ratio resulting for the set of parameters we used in Sect.$\,$\ref{sec:othersignals} which we take as a conservative lower limit on the intrinsic alignment signal. We also show the curve obtained with the best-fit intrinsic alignment amplitude from the fits in Sect.$\,$\ref{sec:iafits} below, which is in good agreement with the data, yielding a ratio of $0.64$. In this case the least-squares analysis results in a reduced $\chi^2$ of $0.77$ and a $p$-value of $0.57$.

Note furthermore that both models and data are consistent with the fact that the loss of signal due to the smaller cut-off in $\Pi$ is roughly constant in transverse separation. The general agreement of the observed and modelled ratios confirms that we model the effect of $\Pi_{\rm max}$ on photometric redshift data correctly. Besides, it supports our use of the 2SLAQ photometric redshift error distribution despite the slightly different apparent magnitude limits of the sample, as discussed in Sect.$\,$\ref{sec:photoz}.

\subsection{Galaxy bias}
\label{sec:bias}

To relate the observed galaxy number density-intrinsic correlations (plus the corrections due to galaxy-galaxy lensing and magnification-shear correlations) to the matter-intrinsic correlations that generate intrinsic alignments, the galaxy bias $b_{\rm g}$ for the density tracer sample needs to be measured. As described in Sect.$\,$\ref{sec:cfmeasurement}, we  compute a galaxy clustering signal that represents the cross-correlation between the density tracer sample and the sample used to trace the intrinsic shear.  Given that the latter is a subset of the former with nearly the same properties (redshift and luminosity distributions, see Fig.~\ref{fig:matchfrac}), we assume that the two have the same galaxy bias.  Thus, we compute $b_{\rm g}$ from this galaxy clustering signal, assuming a linear bias model, but using the full matter power spectrum which should extend the validity of the fits into the quasi-linear regime (see also \citealp{hirata07}, who test several methods to determine the galaxy bias in a similar context). Note that all our considerations rely on the hypothesis that we have assumed the correct cosmological model, in particular $\sigma_8=0.8$.

The redshift averaging and the projection along the line of sight of $w_{\rm gg}$ is performed according to (\ref{eq:wobs}) for the photometric MegaZ-LRG samples and following (\ref{eq:wgg}) for the SDSS LRG samples. We do not repeat the bias measurement for the SDSS L3 and L4 samples but adopt the values determined by \citet{hirata07}, rescaled to our value of $\sigma_8$ by employing $b_{\rm g} \propto \sigma_8^{-1}$, which results in $b_{\rm g}=1.04$ and $b_{\rm g}=1.01$, respectively. Note that the assumption of that same bias, despite the use of different colour cuts for the intrinsic shear tracers, is acceptable because the bias we need is that of the density tracer sample, which has not changed.  To all model projected correlation functions $w_{\rm gg}$, we add a constant $C$ as a further fit parameter to account for the undetermined integral constraint on the numerator of the estimator (\ref{eq:LSgg}) due to the unknown mean galaxy number density (\citealp{landy93}; see also \citealp{hirata07}), i.e.
\eq{
w_{\rm gg}^{\rm model}(r_p,b_{\rm g},C) = w_{\rm gg}(r_p,b_{\rm g}) + C\;,
}
where $w_{\rm gg}$ is given by (\ref{eq:wgg}) or the analogue of (\ref{eq:wobs}) in the case of the MegaZ-LRG samples. Note that we have made the dependence on the galaxy bias explicit in the foregoing equation.

For the fit we discard scales $r_p < 6$\hmpc, i.e. the five data
points at the smallest $r_p$ where the assumption of a linear bias is
expected to break down \citep{tasitsiomi04,mandelbaum06}. For the
MegaZ-LRG sample, there is one additional nuisance in this modelling,
which is the $5\,\%$ stellar (M star) contamination fraction.  As
shown in \citet{mandelbaum08}, the imposition of the apparent size cut
that is needed for a robust galaxy shape measurement is sufficient to
remove this contamination to within 1\%. Thus, the galaxy clustering
signal as defined (a {\em cross}-correlation between the shape and
density samples) is diminished by a single power of the contamination
fraction $f_{\rm contam}=0.05$.  The bias determined by the fits is
actually $\sqrt{1-f_{\rm contam}}\, b_{\rm g}$, so we must correct it
upwards to account for that.  Then, since $w_{g+}$ is reduced by a
factor of $1-f_{\rm contam}$, instead of dividing by just $b_{\rm g}$
to get to $w_{\delta +}$, we divide by $(1-f_{\rm contam})\, b_{\rm
g}$.

\begin{table}[t]
\begin{minipage}[t]{\columnwidth}
\centering
\caption{Galaxy bias $b_{\rm g}$ for the different galaxy samples used.}
\begin{tabular}[t]{lccc}
\hline\hline
sample & $b_{\rm g}$ & $\chi^2_{\rm red}$ & $p(>\chi^2)$\\[0.5ex]
\hline\\[-1ex]
MegaZ-LRG, all z, (90)       & $1.94 \pm 0.03$ & 0.21 & 0.89\\
MegaZ-LRG, all z, (180)      & $1.91 \pm 0.03$ & 0.37 & 0.77\\
MegaZ-LRG, $z < 0.529$, (90) & $1.87 \pm 0.04$ & 0.22 & 0.88\\
MegaZ-LRG, $z > 0.529$, (90) & $2.09 \pm 0.04$ & 2.51 & 0.06\\
SDSS LRG, $z < 0.27$         & $1.88 \pm 0.10$ & 0.82 & 0.48\\
SDSS LRG, $z > 0.27$         & $1.89 \pm 0.07$ & 0.97 & 0.41\\
SDSS Main L4 red             & $1.04$ & $-$ & $-$ \\
SDSS Main L3 red             & $1.01$ & $-$ & $-$ \\
\hline
\end{tabular}
\tablefoot{Shown is the $1\sigma$-error on $b_{\rm g}$, marginalised over the additive constant $C$, the reduced $\chi^2$ for 3 degrees of freedom, and the corresponding $p$-value. The values for the SDSS L3 and L4 samples have been derived from the results by \citet{hirata07}. All results assume $\sigma_8=0.8$. The numbers in parentheses indicate $\Pi_{\rm max}$ in units of \hmpc\ for the MegaZ-LRG samples.}
\label{tab:biasresults}
\end{minipage}
\end{table}

\begin{figure}[t]
\centering
\includegraphics[scale=.36,angle=270]{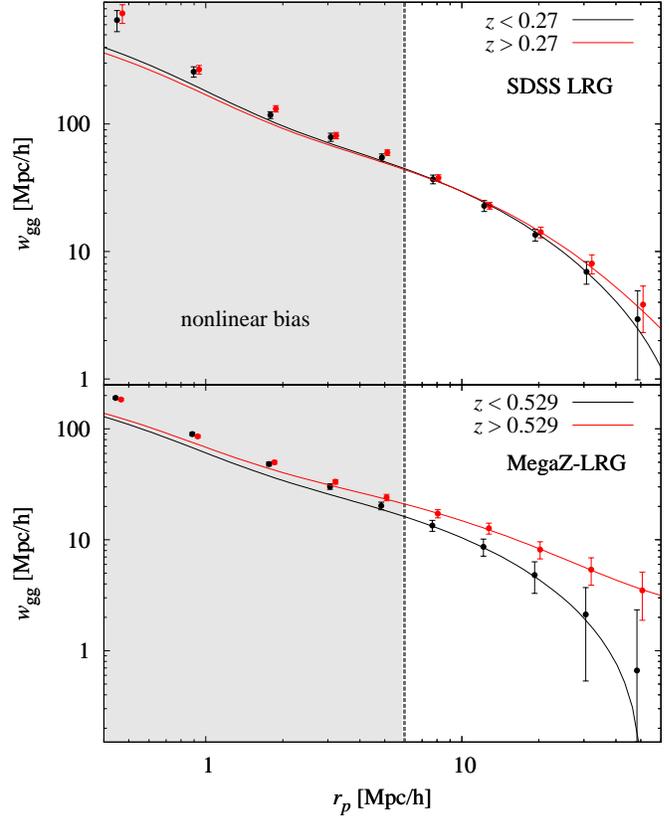}
\caption{Projected correlation function $w_{\rm gg}$ as a function of comoving transverse separation $r_p$. \textit{Top panel}: For the SDSS LRG sample with redshifts smaller than 0.27 (black) and with redshifts larger than 0.27 (red). \textit{Bottom panel}: For the MegaZ-LRG sample with photometric redshifts smaller than 0.529 (black) and with photometric redshifts larger than 0.529 (red). Note that the red points have been slightly offset horizontally for clarity, and that the error bars are strongly correlated. In addition we show the best-fit models as black and red curves, respectively. Only the data points outside the grey region have been used for the fits to avoid the regime of nonlinear bias.}
\label{fig:biasfit}
\end{figure}

In Fig.$\,$\ref{fig:biasfit}, we show the projected correlation functions $w_{\rm gg}$ for the two MegaZ-LRG redshift bins and the two SDSS LRG redshift bins. Note that the SDSS LRG samples have not been split further into luminosity bins because the full SDSS LRG sample, divided into the two redshift bins, is used to trace the galaxy number density field. In each case, we also plot the best-fit models, indicating that the model is a good description of the data on scales where nonlinear bias is not important. At smaller transverse separations (the grey region), which have been excluded from the fits, the data have increasingly larger positive offsets with respect to the model, caused by nonlinear clustering effects.

The best-fit values for $b_{\rm g}$, marginalised over $C$, are summarised in Table \ref{tab:biasresults}. We find good agreement within the $1\sigma$ limits for the best-fit galaxy bias, determined for the different $\Pi_{\rm max}$ in the MegaZ-LRG data, again confirming that we are correctly modelling the truncation in the line-of-sight projection. Splitting the MegaZ-LRG data into two redshift bins at $z=0.529$, we obtain a stronger bias for the bin at higher redshift. This finding is expected, as the bin with $z>0.529$ contains on average significantly more luminous galaxies that are more strongly biased, see Fig.$\,$\ref{fig:distributions}. Only the high-redshift MegaZ-LRG sample yields a reduced $\chi^2$ that significantly exceeds unity which we trace back to the strong correlations between errors as the plot in Fig.$\,$\ref{fig:biasfit} suggests a good fit. Indeed the reduced $\chi^2$ drops well below unity if we repeat the fit ignoring the off-diagonal elements in the covariance. 

We compare our results with the galaxy bias obtained by \citet{blake07} who also studied MegaZ-LRG, albeit with slightly different selection criteria. They used the cuts $i_{\rm deV} \leq 19.8$ and $d_\perp \geq 0.55$ throughout, as well as additional star-galaxy separation criteria that reduced the stellar contamination to $1.5\,\%$. The different selection criteria hinder direct comparison, e.g. the \citet{blake07} criteria (driven mostly by the $i_{\rm deV}$ cut) shift the $r$ band absolute magnitude range $0.15\,$mag brighter. Hence, we expect that the galaxy bias obtained from our analysis should be smaller, and indeed, after rescaling the bias given in Table 2 of \citet{blake07} to $\sigma_8=0.8$, we find $b_{\rm g}=1.89$ and $b_{\rm g}=2.10$ for the two redshift slices roughly coinciding with our MegaZ-LRG low-redshift sample, and $b_{\rm g}=2.18$ and $b_{\rm g}=2.44$ for the two redshift slices closer to our high-redshift sample.

The SDSS LRG samples yield a similar galaxy bias compared to the full MegaZ-LRG sample, with no significant evolution in redshift. Given that the SDSS LRG galaxies have on average a higher luminosity and are located at considerably lower redshift, this finding hints at a stronger bias in the past for galaxies at fixed luminosity. Using again the fact that the bias scales as $b_{\rm g} \propto \sigma_8^{-1}$, our findings for the SDSS LRG samples can be compared to the results for the equivalent bias model in \citet{hirata07} who use $\sigma_8=0.751$. Rescaling the values of Table \ref{tab:biasresults} to this value of $\sigma_8$, we get $b_{\rm g} = 2.00 \pm 0.11$ for the low-redshift sample and $b_{\rm g} = 2.01 \pm 0.07$ for the high-redshift sample. These values agree (within $1\sigma$) with $b_{\rm g} = 2.01 \pm 0.12$ for $z<0.27$ and $b_{\rm g} = 1.97 \pm 0.07$ for $z>0.27$ as found by \citet{hirata07}. Note that the latter analysis used a narrower range in transverse separation with $r_p=7.5-47$\hmpc\ compared to $r_p=6-60$\hmpc\ considered in this work.

\subsection{Intrinsic alignment model fits to individual samples}
\label{sec:iafits}

With the galaxy bias in hand, we can now proceed to fit models of intrinsic alignments to $w_{\rm g+}$. The NLA model features a single free parameter for the amplitude, $A$. Within the physical picture of this model, the amplitude quantifies how the shape of a galaxy responds to the presence of a tidal gravitational field. It is likely that this response depends on the galaxy population under consideration, and thus possibly features an additional evolution with time and hence redshift dependence (on top of the one inherent to the NLA model), and a variation with galaxy luminosity. Therefore we will investigate a more flexible prescription for the gI power spectrum in Sect.$\,$\ref{sec:iafits-combined}.

\begin{figure}[t]
\centering
\includegraphics[scale=.36,angle=270]{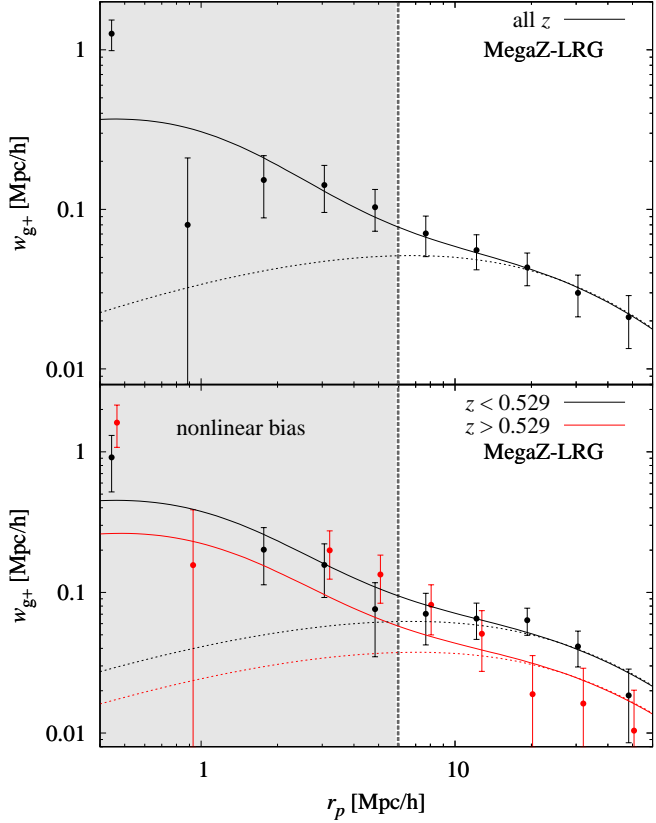}
\caption{Projected correlation function $w_{\rm g+}$ as a function of comoving transverse separation $r_p$ for different MegaZ-LRG subsamples. \textit{Top panel}: Shown is $w_{\rm g+}$ for the MegaZ-LRG sample with the full range in redshifts. The black solid curve corresponds to the best-fit model when only varying the amplitude $A$, without contributions by galaxy-galaxy lensing. The dotted black line is obtained by using the linear matter power spectrum instead of the full power spectrum including non-linear corrections, and identical model parameters otherwise. \textit{Bottom panel}: Same as above, but for the MegaZ-LRG sample split into two photometric redshift bins, where results for $z<0.529$ are shown in black, and for $z>0.529$ in red. Dotted lines again correspond to models computed from the linear matter power spectrum. The point for the $z < 0.529$ subsample at $r_p \approx 1$\hmpc\ is negative and thus not shown. Note that the red points have been slightly offset horizontally for clarity, and that the error bars are correlated. Only the data points outside the grey region have been used for the fits.}
\label{fig:gifit}
\end{figure}

In this section we use (\ref{eq:GIlinalign}) as the intrinsic alignment model, with the single fit parameter $A$. We keep the original SuperCOSMOS normalisation, i.e. $C_1\, \rho_{\rm cr} \approx 0.0134$. To allow for a comparison with foregoing work, we also present some fits with models based on the NLA version with the redshift dependence given in \citet{hirata04}. Note that all intrinsic alignment models applied in this work have a fixed dependence on transverse scales. Since the assumption of a linear bias also enters the model, we again limit the parameter estimation to scales $r_p > 6$\hmpc. Note that we do not explicitly propagate the errors on the galaxy bias determined in the foregoing section through to the uncertainty on intrinsic alignment parameters, as they are marginal compared to the measurement error in $w_{\rm g+}$ (which is dominated by shape noise).

\begin{figure}[t]
\centering
\includegraphics[scale=.36,angle=270]{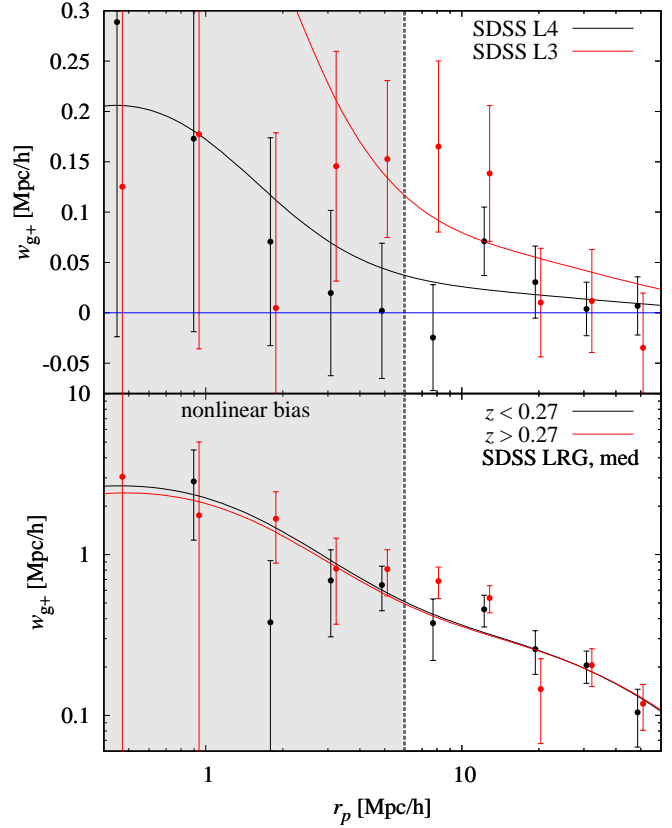}
\caption{Projected correlation function $w_{\rm g+}$ as a function of comoving transverse separation $r_p$ for different SDSS samples. \textit{Top panel}: Shown is $w_{\rm g+}$ for the red SDSS L4 (in black) and L3 (in red) samples. The curves correspond to the best-fit models when only varying the amplitude $A$. \textit{Bottom panel}: Same as above, but for the SDSS LRG medium brightness sample split into two redshift bins, where results for $z<0.27$ are shown in black, and for $z>0.27$ in red. The point for the $z < 0.27$ subsample at the smallest $r_p$ is negative and thus not shown. Note that the red points have been slightly offset horizontally for clarity, and that the error bars are correlated. Only the data points outside the grey region have been used for the fits.}
\label{fig:gifit_sdss}
\end{figure}

In Fig.$\,$\ref{fig:gifit}, the projected correlation functions for the full MegaZ-LRG sample as well as for the two MegaZ-LRG redshift bins, split at $z=0.529$, are plotted. The fit results for $A$ are presented in Table \ref{tab:fitsAonly}. On the scales usable for the fit, the best-fit gI model, which is also plotted in Fig.$\,$\ref{fig:gifit} for each case, traces the data points well with reduced $\chi^2$-values below one, whereas for $r_p \lesssim 1$\hmpc\ points lie several $\sigma$ above and below the model curve, possibly indicating strongly nonlinear effects. The nature of these deviations is unknown, but since they occur on scales near the virial radius of LRGs, one may hypothesise that at these ranges of $r_p$, complicated dependencies on the tidal field or a change in the intrinsic alignment mechanism play a role. Moreover we find very good agreement between the best-fit amplitudes obtained for the full MegaZ-LRG sample with different values of $\Pi_{\rm max}$.

In addition, we show in Fig.$\,$\ref{fig:gifit} models for $w_{\rm g+}$ that have been calculated using the linear matter power spectrum instead of the nonlinear one in (\ref{eq:GIlinalign}), holding all other model parameters fixed. As expected, the signals for linear and nonlinear power spectrum coincide on the largest scales, but already at $r_p \sim 10$\hmpc, $w_{\rm g+}$ computed from linear theory drops below the correlation function that includes nonlinear clustering and yields a worse fit to the data in case of the full and the high-redshift MegaZ-LRG sample. Thus, although our analysis is still restricted to relatively large scales, non-linear effects in the intrinsic alignment of galaxies must be taken into account.

\begin{table*}[t!!!]
\centering
\caption{$1\sigma$ confidence limits on the amplitude $A$ of the gI correlation function for each sample.}
\vspace*{0.1cm}
\begin{tabular}[t]{lcccccc}
\hline\hline
sample & $\ba{z}$ & $\ba{L}/L_0$ & $A$ & $\chi^2_{\rm red}$ & $p(>\chi^2)$ & $A$ (HS04)\\[0.5ex]
\hline\\[-1ex]
MegaZ-LRG, all z, (90)       & 0.54 & 0.96 &  $4.52^{+0.64}_{-0.64}$ & 0.04 & 1.00 & $1.98^{+0.28}_{-0.28}$ \\
MegaZ-LRG, all z, (180)      & 0.54 & 0.96 &  $4.51^{+0.64}_{-0.63}$ & 0.36 & 0.84 & $1.98^{+0.28}_{-0.28}$ \\
MegaZ-LRG, $z < 0.529$, (90) & 0.49 & 0.87 &  $5.31^{+0.86}_{-0.86}$ & 0.47 & 0.76 & $2.40^{+0.39}_{-0.39}$ \\
MegaZ-LRG, $z > 0.529$, (90) & 0.59 & 1.05 &  $3.53^{+0.97}_{-0.96}$ & 0.58 & 0.68 & $1.46^{+0.40}_{-0.40}$ \\
MegaZ-LRG, all z, $g-i$ cut                         & 0.53 & 0.99 &  $5.23^{+0.75}_{-0.75}$ & 0.26 & 0.90 & $2.29^{+0.33}_{-0.33}$ \\
MegaZ-LRG, $z < 0.529$, $g-i$ cut                   & 0.49 & 0.91 &  $5.91^{+0.94}_{-0.94}$ & 0.17 & 0.95 & $2.67^{+0.43}_{-0.42}$ \\
MegaZ-LRG, $z > 0.529$, $g-i$ cut                   & 0.59 & 1.08 &  $4.53^{+1.17}_{-1.16}$ & 0.20 & 0.94 & $1.87^{+0.48}_{-0.48}$ \\
SDSS LRG, $z < 0.27$, faint                        & 0.21 & 1.06 &  $4.51^{+1.70}_{-1.74}$ & 1.30 & 0.27 & $3.03^{+1.18}_{-1.15}$ \\
SDSS LRG, $z > 0.27$, faint                        & 0.32 & 1.07 &  $7.67^{+1.71}_{-1.75}$ & 1.82 & 0.12 & $4.44^{+1.02}_{-1.00}$ \\
SDSS LRG, $z < 0.27$, med                          & 0.22 & 1.50 &  $9.98^{+1.55}_{-1.51}$ & 0.51 & 0.73 & $6.77^{+1.04}_{-1.03}$ \\
SDSS LRG, $z > 0.27$, med                          & 0.31 & 1.50 & $10.03^{+1.49}_{-1.45}$ & 2.03 & 0.09 & $5.84^{+0.85}_{-0.86}$ \\
SDSS LRG, $z < 0.27$, bright                       & 0.22 & 2.13 & $12.93^{+2.14}_{-2.11}$ & 1.15 & 0.33 & $8.77^{+1.08}_{-1.09}$ \\
SDSS LRG, $z > 0.27$, bright                       & 0.31 & 2.12 & $16.09^{+2.75}_{-2.76}$ & 2.07 & 0.08 & $9.35^{+1.55}_{-1.55}$ \\
SDSS L4 red                                        & 0.10 & 0.33 &  $1.20^{+0.90}_{-0.88}$ & 0.80 & 0.52 & $1.00^{+0.75}_{-0.74}$ \\
SDSS L3 red                                        & 0.07 & 0.14 &  $3.61^{+2.06}_{-2.09}$ & 0.83 & 0.51 & $3.16^{+1.83}_{-1.82}$ \\[1ex]
\hline
\end{tabular}\\
\vspace*{0.4cm}
\begin{tabular}[t]{lcccc}
\hline\hline
sample & $A$ (w/o gG+mG)& $\chi^2_{\rm red}$ & $p(>\chi^2)$ & $A$ (w/o mG)\\[0.5ex]
\hline\\[-1ex]
MegaZ-LRG, all z, (90)       & $4.11^{+0.64}_{-0.64}$ & 0.03 & 1.00 & $4.47^{+0.64}_{-0.64}$\\
MegaZ-LRG, all z, (180)      & $3.95^{+0.63}_{-0.64}$ & 0.36 & 0.84 & $4.45^{+0.64}_{-0.64}$\\
MegaZ-LRG, $z < 0.529$, (90) & $4.92^{+0.86}_{-0.86}$ & 0.47 & 0.76 & $5.26^{+0.86}_{-0.86}$\\
MegaZ-LRG, $z > 0.529$, (90) & $3.06^{+0.96}_{-0.97}$ & 0.58 & 0.68 & $3.47^{+0.97}_{-0.96}$\\[1ex]
\hline
\end{tabular}
\tablefoot{\textit{Top section}: Fits results for all samples considered in this work. For reference, we give the means $\ba{z}$ and $\ba{L}/L_0$ for each sample in the second and third column. The fourth to sixth columns contain the best-fit intrinsic alignment amplitude, and the corresponding reduced $\chi^2$ and $p$-values of the fit for 4 degrees of freedom. To facilitate comparisons with preceding work, we also fit the amplitude of the version of the NLA model based on \citet{hirata04}, shown in the last column. Note that these fits produce the same $\chi^2$ as the foregoing ones. We also show the fit results for MegaZ-LRG samples with a colour cut as introduced in Sect.$\,$\ref{sec:colorcomparison}. The numbers in parentheses indicate $\Pi_{\rm max}$ in units of \hmpc\ for the MegaZ-LRG samples. \textit{Bottom section}: Amplitude fits to MegaZ-LRG samples neglecting the contributions by galaxy-galaxy lensing (gG) and magnification-shear cross-correlations (mG). The rightmost column contains best-fit amplitudes when including the gG but not the mG signal. Due to the very similar $r_p$-dependence of the gI, gG, and mG signals, the $\chi^2$-values of these fits are almost identical.}
\label{tab:fitsAonly}
\end{table*}

We also perform the analysis on $w_{\rm g+}$ for the SDSS LRG data, which is divided into three luminosity bins in addition to the two redshift bins split at $z=0.27$, see Table \ref{tab:samples}. As redshifts are determined spectroscopically in this case, we use (\ref{eq:wgispec}) to compute $w_{\rm g+}$. For reasons of simplicity we use the redshift distribution for all SDSS LRG luminosities combined because we find that employing the individual distributions for the faint, medium, and bright subsamples instead leads only to sub-per cent changes in the signal on all scales. The resulting correlation functions and their best-fit models are shown in Fig.$\,$\ref{fig:gifit_sdss}, and the resulting parameter constraints on $A$ listed in Table \ref{tab:fitsAonly}.

In Table \ref{tab:fitsAonly} we additionally present constraints on $A$ using the NLA model with the redshift dependence as derived by \citet{hirata04}. If the redshift distribution of a galaxy sample is sufficiently compact, the amplitudes of the two NLA models considered are approximately related by a factor $\br{1+\ba{z}}^2$, caused by the different redshift dependencies.  Our findings for the SDSS LRG samples are compatible with the results of the power-law fits by \citet{hirata07}, yielding a maximum intrinsic alignment amplitude of close to ten times that found in SuperCOSMOS, for the bright high-redshift SDSS LRG sample. In contrast, the resulting values for $A$ using the SDSS L3 and L4 samples are small for both NLA models, the constraints being consistent with zero at the $2\sigma$-level. 

By default we include both galaxy-galaxy lensing and magnification-shear correlations in the modelling for the photometric redshift MegaZ-LRG data, but the bottom panel of Table \ref{tab:fitsAonly} also lists results for $A$ when dropping either the mG term only or both additional contributions. Since the gG and mG signals yield a negative contribution to $w_{\rm g+}$, a lower amplitude $A$ than in the case including these contributions is needed to get a good fit to the data. Dropping the mG term causes a drop in $A$ for all samples which is below the $2\,\%$ level and hence much smaller than the $1\sigma$ error on the amplitude. Since the intrinsic alignment amplitude is about three to four times higher than assumed in the prediction of Sect.$\,$\ref{sec:othersignals}, which yielded a maximum mG contribution of the order $10\,\%$ on relevant scales, effects at the per-cent level by the mG signal are indeed expected.

The change in amplitude when discarding all additional signals ranges between $7\,\%$ for the low-redshift MegaZ-LRG sample and $13\,\%$ for the high-redshift sample. As Fig.$\,$\ref{fig:tomops} suggests, the relative contribution of the gG to the gI signal is approximately constant as a function of redshift for MegaZ-LRG if the intrinsic alignment amplitude is fixed. Since we obtain a larger $A$ for the low-redshift than for the high-redshift sample, the contribution by galaxy-galaxy lensing is correspondingly smaller for $z<0.529$. In Sect.$\,$\ref{sec:othersignals} we calculated a $30\,\%$ gG/gI ratio for $A = 1.21$, which is again in good agreement with our fit results.

It is interesting to note that our default analysis yields nearly perfect agreement between the best-fit values for $A$ obtained from the full MegaZ-LRG samples with cut-offs at $90$\hmpc\ and $180$\hmpc, respectively, but that, when excluding galaxy-galaxy lensing, one observes a moderate discrepancy in the fit results. The galaxy-galaxy lensing increases for larger line-of-sight separation of the galaxy pairs correlated, whereas the gI term diminishes, so that the subsample with $\Pi_{\rm max}=180$\hmpc\ is affected more strongly. This finding again confirms that we are modelling the effect of photometric redshift scatter and the truncation of signals at large $\Pi$ correctly. Besides, both additional signals that contribute to galaxy position-shape correlations have a dependence on transverse separation that is very similar to the one of the gI term. Thus only the amplitude $A$ is affected while the goodness of fit remains almost unchanged when the gG and mG terms are included.

In general, the dependence on $r_p$ given by the NLA model describes the data reasonably well, yielding reduced $\chi^2$ values of order unity (or sometimes below). Only the high-redshift SDSS LRG samples tend to a $\chi^2$ that significantly exceeds unity, which is caused by an excess signal around $r_p=10$\hmpc\ of unknown origin, as can be seen in Fig.$\,$\ref{fig:gifit_sdss}, bottom panel. The $p$-values remain above, but are close to, the significance level of 0.05.

\begin{figure}[t]
\centering
\includegraphics[scale=.365,angle=270]{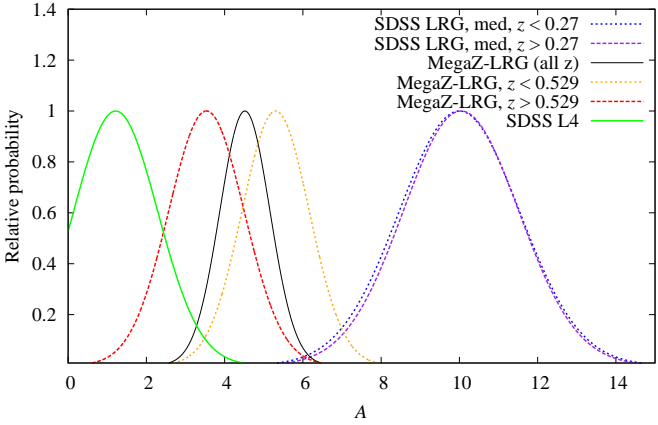}
\caption{Constraints on the amplitude $A$ of the intrinsic alignment model. The black solid curve corresponds to using the full MegaZ-LRG sample. The constraints from the individual MegaZ-LRG redshift bins are shown as red dashed lines ($z > 0.529$) and orange dotted lines ($z < 0.529$). For comparison we also show the constraints on $A$ for the SDSS L4 sample as green solid line and for the SDSS LRG medium luminosity samples with $z<0.27$ (dark blue dotted line) and $z>0.27$ (purple dashed line).}
\label{fig:contours_A}
\end{figure}

In all cases except the two SDSS Main samples the intrinsic alignment amplitude is higher than for the original SuperCOSMOS normalisation, which would correspond to $A=1.21$ in the corrected linear alignment model and $A=1$ when fitting the original version. The SDSS Main samples are at similar redshift to SuperCOSMOS, and consistent with the results by \citet{brown02}, although only at the $2\sigma$ level in the case of the L3 sample which prefers a higher amplitude (but note the possible excess signal at $r_p \sim 10$\hmpc\ in the top panel of Fig.$\,$\ref{fig:gifit_sdss}).

As is obvious from the compact and mutually inconsistent posterior probabilities for $A$ shown in Fig.$\,$\ref{fig:contours_A}, the different galaxy samples are inconsistent with an intrinsic alignment model that has only $A$ as a free parameter. The SDSS LRGs span a very similar and relatively short range in redshifts, so that no strong evolution with redshift is expected in these subsamples. Then it is evident from the fit results in Table \ref{tab:fitsAonly} that the intrinsic alignment amplitude increases with galaxy luminosity, with the brightest sample attaining a high amplitude of $A \approx 16$. Moreover, despite a mean luminosity that is $20\,\%$ higher than that of the low-redshift MegaZ-LRG sample, the high-redshift MegaZ-LRG sample has a smaller amplitude parameter $A$, indicative of a decrease of the intrinsic alignment amplitude with redshift beyond the redshift dependence inherent to the NLA model. However, note that the amplitudes of the two samples are still consistent with each other at about the $1\sigma$ level, so that the NLA model is still in agreement with these findings for MegaZ-LRG.

\subsection{Compatibility of the different samples}
\label{sec:colorcomparison}

As has been shown in \citet{hirata07} and many other papers, red and
blue galaxies behave differently with respect to intrinsic alignments.
Thus, in order to obtain combined constraints from these samples and
to understand their results in some unified way, we now address the
sample definitions as regards the resulting colour properties in
greater detail. 

\begin{figure}[t]
\centering
\includegraphics[scale=.45]{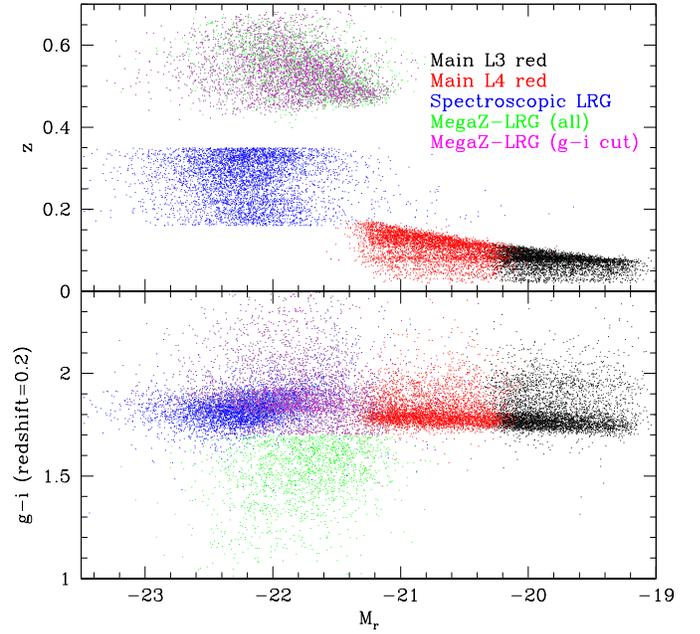}
\caption{Redshift, colour, and magnitude properties of all samples
    used in this paper. \textit{Top panel:} $k+e$-corrected (to $z=0$)
    $M_r$ used to define the luminosities relative to $L_0$, versus
    redshift $z$. Galaxies from MegaZ-LRG are shown in green, those
    from the SDSS LRG samples in blue, and galaxies from the red SDSS
    Main L4 (L3) sample in red (black). To avoid confusion, a subset
    of the points from each sample was used. \textit{Bottom panel:}
    The colour-redshift relation. The colour coding is the same as
    above. As shown, the full MegaZ-LRG sample is the only one with
    significant contributions below $^{0.2}(g-i)=1.7$. Thus, we define
    the cut sample (pink points) using all MegaZ-LRG galaxies that are
    redder than this value, for consistency with the other samples.}
\label{fig:2dtrends}
\end{figure}

\cite{wake06} performed a comparison of the SDSS spectroscopic LRG and
the MegaZ-LRG sample definition, with the goal of creating a subset of
the MegaZ-LRG sample that would pass the SDSS LRG joint
colour-magnitude cuts if it were shifted to redshift $z=0.2$.  The
reason for this choice of comparison redshift is that, with the
MegaZ-LRG sample concentrated at $z=0.55$, this difference in redshift 
corresponds to shifting the SDSS filters over by one, and thus the
$k$-corrections are less prone to systematic error.  As shown in their
comparison, when using the $g-i$ colour shifted to $z=0.2$, only $\sim
30\,\%$ of the MegaZ-LRG galaxies would pass the joint colour-magnitude
cut of the SDSS spectroscopic LRGs.

\begin{figure}[t]
\centering
\includegraphics[scale=.36,angle=270]{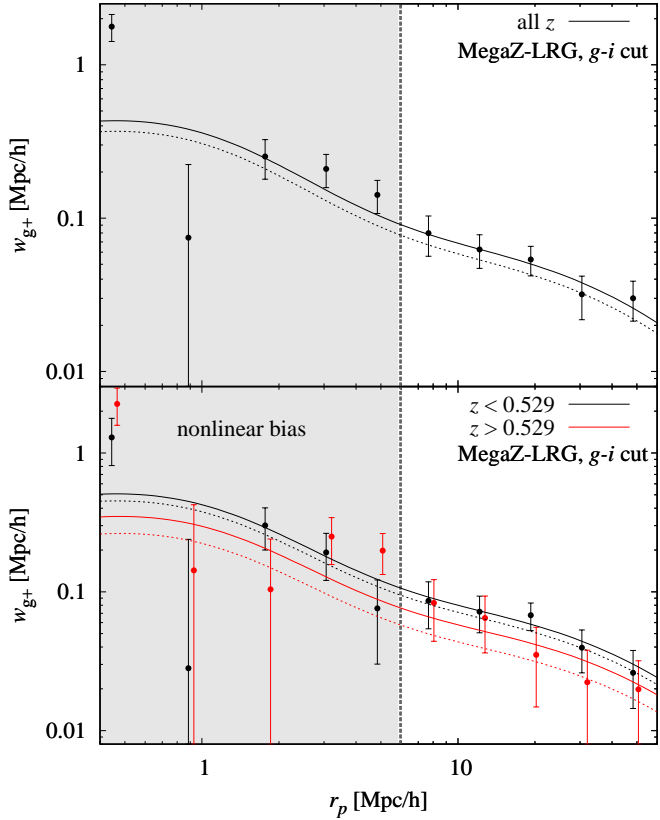}
\caption{Projected correlation function $w_{\rm g+}$ as a function of comoving transverse separation $r_p$ for the MegaZ-LRG subsamples with the cut $^{0.2}(g-i)>1.7$ imposed. \textit{Top panel}: Shown is $w_{\rm g+}$ for the MegaZ-LRG sample with the full range in redshifts. The black solid curve corresponds to the best-fit model when only varying the amplitude $A$, without contributions by galaxy-galaxy lensing and magnification-shear cross-correlations. The dotted line represents the best-fit model to the full sample shown in Fig.$\,$\ref{fig:gifit}. \textit{Bottom panel}: Same as above, but for the cut MegaZ-LRG sample split into the two photometric redshift bins, where results for $z<0.529$ are shown in black, and for $z>0.529$ in red. Dotted lines again indicate the best-fit model for the full MegaZ-LRG samples, respectively. Note that the red points have been slightly offset horizontally for clarity, and that the error bars are correlated. Only the data points outside the grey region have been used for the fits.}
\label{fig:gifit_cc}
\end{figure}

However, this cut is not what we want to impose on our sample.  The
reason for this is that, as shown in \citet{wake06}, Fig.$\,$3, many
of the MegaZ-LRG galaxies that are excluded are at the faint end,
where the SDSS spectroscopic LRG cut starts to exclude more and more
of the red sequence.  In contrast, for this paper, we want to keep all of the red sequence without regard for 
matching the luminosity ranges (indeed, we would like to study
samples on a wide luminosity baseline in order to measure the variation of
intrinsic alignments with luminosity).  Thus, we wish to define a
minimum $^{0.2}(g-i)$ that corresponds roughly to that for the SDSS
spectroscopic LRGs and our revised definition of the SDSS Main L3 and L4
red samples.  The best choice in this context appears to be a cut at
$^{0.2}(g-i)>1.7$, which should ensure consistency with the other
samples within the limits of our uncertainty in
$k+e$-corrections.

To illustrate this cut, we present Fig.$\,$\ref{fig:2dtrends}, which
shows two-dimensional projections of the relationship between
redshift, colour, and absolute rest-frame magnitude of the samples.
As shown, they span a wide range of redshifts ($0.05<z<0.7$) and of
luminosities (four magnitudes), and with the imposition of this new
colour cut, the colour ranges are quite similar.  MegaZ-LRG shows the
largest scatter to redder colours; however, this is expected given
that, as the highest redshift sample, they have the largest
photometric errors which significantly widens the colour distribution
at the red end where the $g$ band flux is often only weakly
detected, especially once the 4000\AA\ break moves from $g$ to $r$ band. The result of this cut is to reduce the MegaZ-LRG sample to
$70\,\%$ of its original size.  The typical sample redshift does not
change, and the mean luminosity increases marginally by $2\,\%$
compared to the values of the full MegaZ-LRG sample, see Table
\ref{tab:fitsAonly}. 

We repeat the intrinsic alignment amplitude fits of Sect.$\,$\ref{sec:iafits} for the cut MegaZ-LRG samples, including the contributions by galaxy-galaxy lensing and magnification-shear correlations. We continue to use the relation between photometric and spectroscopic redshifts from the 2SLAQ verification sample of Sect.$\,$\ref{sec:photoz} because we do not observe any significant effect by the colour cut on this relation, e.g. neither the mean nor the scatter of the distribution of differences between photometric and spectroscopic redshifts change beyond the $1\sigma$ level. The resulting correlation functions with the best-fit models are plotted in Fig.$\,$\ref{fig:gifit_cc}, and the corresponding best-fit values for $A$ listed in Table \ref{tab:fitsAonly}. For comparison we also show the best-fit models to the uncut MegaZ-LRG samples in the figure. Since the cut and uncut samples have the same redshifts and luminosities to high accuracy, we can ascribe any difference in the signals to a dependence of intrinsic alignments on galaxy colour. 

For both the high- and low-redshift MegaZ-LRG samples as well as the full sample we find an increase in $A$ for the cut sample which has higher $^{0.2}(g-i)$, thus being in line with the expectation that redder galaxies have stronger intrinsic alignments. The increase amounts to $11\,\%$ for the low-redshift sample and $28\,\%$ for the high-redshift sample (the corresponding error bars feature a similar increase), suggesting a stronger colour dependence at higher redshift. However, it should be noted that all observed changes due to the colour cut in $g-i$ remain within the $1\sigma$ errors and are therefore not statistically significant.

\subsection{Intrinsic alignment model fits to combined samples}
\label{sec:iafits-combined}

\begin{figure*}[t]
\centering
\includegraphics[scale=.318,angle=270]{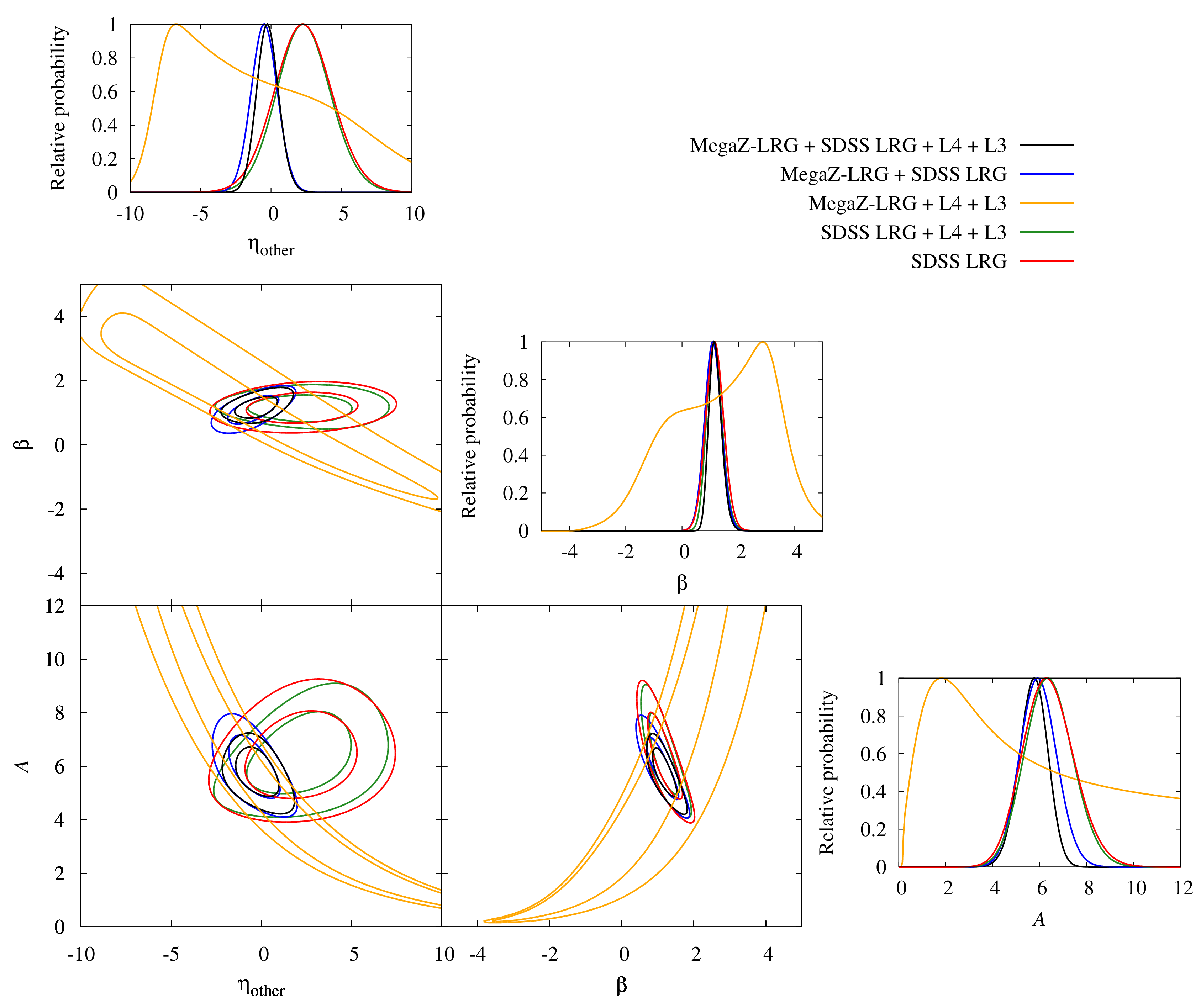}
\caption{Constraints by the joint fit to MegaZ-LRG and SDSS spectroscopic samples on the amplitude $A$ of the intrinsic alignment model, the extra redshift dependence with power-law index ${\eta_{\rm other}}$, and the index $\beta$ of the luminosity dependence. In the lower left panels, the two-dimensional $1\sigma$ and $2\sigma$ confidence contours are given, marginalised in each case over the parameter not shown with flat priors in the range $A \in \bb{0;20}$, $\eta_{\rm other} \in \bb{-10:10}$, and $\beta \in \bb{-5;5}$. The upper right panels display the constraints on $A$, ${\eta_{\rm other}}$, and $\beta$, each marginalised over the two remaining parameters. Red lines are obtained via fits to the six SDSS LRG samples, green lines for the SDSS LRG and Main samples, yellow lines via fits to the MegaZ-LRG and SDSS Main samples, blue lines for MegaZ-LRG and SDSS LRGs combined, and black lines result for the joint fit to the MegaZ-LRG, SDSS LRG, L4, and L3 samples.}
\label{fig:contours_AzL}
\end{figure*}

Having addressed the question of the compatibility of the samples, we repeat the fits to $w_{\rm g+}$ for different combinations of galaxy samples, now allowing for an additional redshift and luminosity dependence according to the extended model
\eq{
\label{eq:iamodel}
P^{\rm model}_{\rm gI}(k,z,L) = A\; b_{\rm g}\; P_{\delta {\rm I}}(k,z)\; \br{\frac{1+z}{1+z_0}}^{\eta_{\rm other}} \br{\frac{L}{L_0}}^\beta\;,
}
where $z_0=0.3$ is an arbitrary pivot redshift, and $L_0$ is the pivot luminosity corresponding to an absolute $r$ band magnitude of $-22$ (passively evolved to $z=0$). The matter-intrinsic power spectrum $P_{\delta {\rm I}}$ is given by the NLA model with the modified redshift dependence discussed in Appendix \ref{sec:larederive}, including the normalisation to SuperCOSMOS. This model contains three free parameters $\bc{A,\beta,{\eta_{\rm other}}}$, and, as before, has a fixed dependence on transverse scales.

\begin{table*}[t]
\centering
\caption{$1\sigma$ marginalised confidence limits on the amplitude $A$, the slope of the additional redshift dependence ${\eta_{\rm other}}$, and the luminosity dependence $\beta$ of the intrinsic alignment model (\ref{eq:iamodel}).}
\begin{tabular}[t]{lcccccc}
\hline\hline
sample combination & $N_{\rm sample}$ & $A$ & ${\eta_{\rm other}}$ & $\beta$ & $\chi^2_{\rm red}$ & $p(>\chi^2)$\\[0.5ex]
\hline\\[-1ex]
SDSS LRG                       & 6 & $6.27^{+1.10}_{-1.01}$ & $2.27^{+2.07}_{-2.06}$ & $1.17^{+0.30}_{-0.32}$ & 1.37 & 0.09\\
SDSS LRG + L4 + L3             & 8 & $6.32^{+1.05}_{-0.97}$ & $2.27^{+1.93}_{-1.89}$ & $1.13^{+0.27}_{-0.27}$ & 1.23 & 0.16\\
MegaZ-LRG + L4 + L3            & 4 & $1.81^{+9.12}_{-1.44}$ & $-6.80^{+9.47}_{-1.52}$ & $2.87^{+0.71}_{-2.93}$ & 0.60 & 0.89\\
MegaZ-LRG + SDSS LRG           & 8 & $5.92^{+0.77}_{-0.75}$ & $-0.47^{+0.93}_{-0.96}$ & $1.10^{+0.29}_{-0.30}$ & 1.15 & 0.24\\
MegaZ-LRG + SDSS LRG + L4 + L3 & 10 & $5.76^{+0.60}_{-0.62}$ & $-0.27^{+0.80}_{-0.79}$ & $1.13^{+0.25}_{-0.20}$ & 1.09 & 0.32\\[1ex]
\hline
\end{tabular}
\tablefoot{Shown are the fit results for $A$, $\eta_{\rm other}$, and $\beta$ for various sample combinations. In addition we list the number of subsamples $N_{\rm sample}$ used for the joint fit and the reduced $\chi^2$ including $p$-value. The number of the degrees of freedom is given by $5\,N_{\rm sample}-3$. We have used flat priors $A \in \bb{0;20}$, $\eta_{\rm other} \in \bb{-10;10}$, and $\beta \in \bb{-5;5}$, except for the case \lq MegaZ-LRG + L4 + L3\rq, where $\eta_{\rm other} \in \bb{-20;20}$ was assumed instead.}
\label{tab:fitscombined}
\end{table*}

The amplitude parameter $A$ and the luminosity term can be taken out of all integrations leading to $w_{\rm g+}$ because they neither depend on redshift nor comoving distance, so that the parameters $A$ and $\beta$ can be varied in the likelihood analysis with low computational cost. The extra redshift term containing ${\eta_{\rm other}}$ depends on the integration variable in (\ref{eq:gIlimber}) though. To facilitate the likelihood analysis for the MegaZ-LRG samples with photometric redshifts, we assume that this term can be taken out of the integration and is evaluated at the mean $\bar{z}_{\rm m}=(\bar{z}_1+\bar{z}_2)/2$ of the two redshifts entering (\ref{eq:gIlimber}). This approximation holds to fair accuracy because the corresponding redshift probability distributions are sufficiently narrow, given the small photometric redshift uncertainty for the MegaZ-LRG sample. 

The additional redshift dependence is then integrated over in the averaging process in (\ref{eq:wobs}) and (\ref{eq:wgispec}) for photometric and spectroscopic samples, respectively. As also the luminosity distributions of the galaxy samples under consideration are compact and narrow, we use the mean luminosity in the luminosity term in (\ref{eq:iamodel}) instead of integrating $\br{L/L_0}^\beta$ over the full distribution. This is a good approximation even for the full MegaZ-LRG sample, which features the broadest luminosity distribution, the deviation being below $2\,\%$ close to the best-fit values for $\beta$ that we determine below.

We consider joint fits to several combinations of the six SDSS LRG subsamples, the two MegaZ-LRG low- and high-redshift samples with the colour cut, and the two SDSS Main L4 and L3 samples. The resulting two-dimensional marginalised confidence contours and marginal one-dimensional posterior distributions for the parameter set $\bc{A,\beta,\eta_{\rm other}}$ are shown in Fig.$\,$\ref{fig:contours_AzL}. The corresponding marginal $1\sigma$ errors on these parameters and the goodness of fit are given in Table \ref{tab:fitscombined}. In the computation of marginalised constraints we assumed by default flat priors in the ranges $A \in \bb{0;20}$, $\eta_{\rm other} \in \bb{-10:10}$, and $\beta \in \bb{-5:5}$. For the combination of the MegaZ-LRG and SDSS Main samples, which yields weak and degenerate constraints, we extend the prior range of $\eta_{\rm other}$ to $\pm 20$. Note that in this case the posterior has not yet decreased very close to zero at $\beta=5$, but still we expect the influence of the $\beta$-prior on the marginal constraints to be negligible.

Combining all SDSS LRG samples we can constrain $\beta$ well, i.e. the power-law slope of the luminosity evolution of the intrinsic alignment amplitude, while the errors on $\eta_{\rm other}$ remain large, as expected for the short baseline in redshift. Adding in the two redshift-binned MegaZ-LRG samples greatly improves constraints on the extra redshift evolution and also narrows down the possible values of $A$. The marginalised $1\sigma$ contours for the SDSS LRG only and the SDSS + MegaZ-LRG fits are consistent, although the MegaZ-LRG data clearly prefers smaller values of $\eta_{\rm other}$. The reduced $\chi^2$ improves by $16\,\%$ when adding the MegaZ-LRG data. Incorporating in addition the SDSS Main L3 and L4 samples at low redshift and with low mean luminosity further tightens constraints on all parameters, in particular the amplitude $A$. The SDSS Main samples have small constraining power due to the large error bars on their signals. Their inclusion yields largely consistent confidence regions in combination with both SDSS LRG and MegaZ-LRG samples and decreases the reduced $\chi^2$ by another $5\,\%$.

\begin{figure}[t]
\centering
\includegraphics[scale=.36,angle=270]{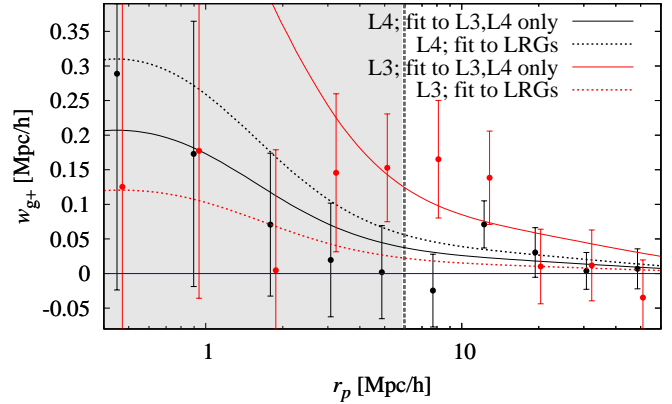}
\caption{Consistency between the combined LRG and SDSS Main samples. Shown is $w_{\rm g+}$ for the red SDSS L4 (in black) and L3 (in red) samples. The solid curves correspond to the best-fit models to the combined L3 and L4 samples when varying $A$, $\eta_{\rm other}$, and $\beta$. The corresponding best-fit models to the combined SDSS LRG and MegaZ-LRG samples are given by the dotted lines. The prior ranges $A \in \bb{0;20}$, $\eta_{\rm other} \in \bb{-10;10}$, and $\beta \in \bb{-5;5}$ were employed in both cases. Note that the red points have been slightly offset horizontally for clarity, and that the error bars are correlated.}
\label{fig:gifit_consistency}
\end{figure}

It is not a priori clear that the intrinsic alignment model determined for the LRG samples also holds for the fainter, non-LRG SDSS Main samples. The validity of this model for galaxies with luminosities around and below $L^*$ is paramount for its applicability to intrinsic alignments in cosmic shear data that has many more faint red galaxies than LRGs. To affirm consistency, we read off the combination of intrinsic alignment parameters that yield the minimum $\chi^2$ for the joint fit to the SDSS LRG and MegaZ-LRG samples. Then we compare the $\chi^2$ of this parameter combination for the fit to the combined L3 and L4 samples to the minimum $\chi^2$ obtained for fitting to this sample combination. We find a difference $\Delta \chi^2=2.01$ to the minimum $\chi^2$ of the latter samples in the parameter range $A \in \bb{0;20}$, $\eta_{\rm other} \in \bb{-10:10}$, and $\beta \in \bb{-5;5}$, corresponding to a $p$-value of 0.57. Thus the SDSS Main samples are fully consistent with the LRG data sets, but note that the L4 and L3 data are noisy and hence not very conclusive, e.g. we also compute a reduced $\chi^2$ of 1.19 (L3) and 0.89 (L4) for a fit to zero. For illustration, we have plotted both the best-fit models for a fit to L3 and L4 only, and for the fit to the SDSS LRG and MegaZ-LRG samples in Fig.$\,$\ref{fig:gifit_consistency}.

The joint fit of all considered samples clearly favours an increase of the intrinsic alignment signal with galaxy luminosity; indeed we find that $\beta < 0.5$ can be excluded at more than the $4\sigma$ level. The data are perfectly consistent with $\eta_{\rm other}=0$, i.e. with the redshift evolution inherent to the corrected NLA model as discussed in Appendix \ref{sec:larederive}. The combination of MegaZ-LRG and SDSS LRG samples is the main driver for these constraints on the redshift dependence. Using the NLA model with the redshift dependence of \citet{hirata04} instead would have shifted the constraints on $\eta_{\rm other}$ by $-2$. Consequently, our results disfavour this redshift dependence of the original NLA model, excluding $\eta_{\rm other} \geq 2$ by about $2.8\sigma$. We find an overall intrinsic alignment normalisation $A = 5.8 \pm 0.6$, which e.g. translates into an amplitude of $90\,\%$ of the standard NLA model with corrected redshift dependence and SuperCOSMOS normalisation for a typical red galaxy with $L=0.2\,L_0\approx L_*$ at redshift $z=0.5$ (see Appendix \ref{sec:LF} for a justification of these values). Using the NLA model with the redshift dependence of \citet{hirata04}, which has been employed in most weak lensing studies hitherto, yields an amplitude of about $40\,\%$ of the SuperCOSMOS normalisation.

Our findings for the marginalised parameter constraints on $\beta$ from the fits to the SDSS LRG samples are in good agreement with the results presented in \citet{hirata07}, despite a different binning in transverse separation and a fit of a pure power-law dependence on $z$ and $r_p$ instead. The latter yields a scaling with $r_p$ which is comparable to the NLA model. Due to these differences in modelling, the intrinsic alignment amplitude parameters are not directly compatible, while the \citet{hirata07} power-law index for the redshift term should roughly correspond to $\eta_{\rm other} - 1$, see (\ref{eq:GIlinalignnew}). The results for the redshift dependence are also consistent at the $1\sigma$ level, albeit with large error bars. In addition \citet{hirata07} considered joint constraints from SDSS LRG and 2SLAQ samples, but did not impose colour cuts on 2SLAQ, so that a quantitative comparison is difficult. However, we observe similar tendencies in the best-fit values of the intrinsic alignment parameters when adding the 2SLAQ and MegaZ-LRG samples, respectively. Besides, parameter errors from the joint analysis of SDSS LRG and 2SLAQ, or SDSS LRG and MegaZ-LRG samples are of the same order of magnitude.

\subsection{Dependence on $k+e$-corrections}
\label{sec:kcorrdep}

\begin{figure}[t]
\centering
\includegraphics[scale=.36,angle=270]{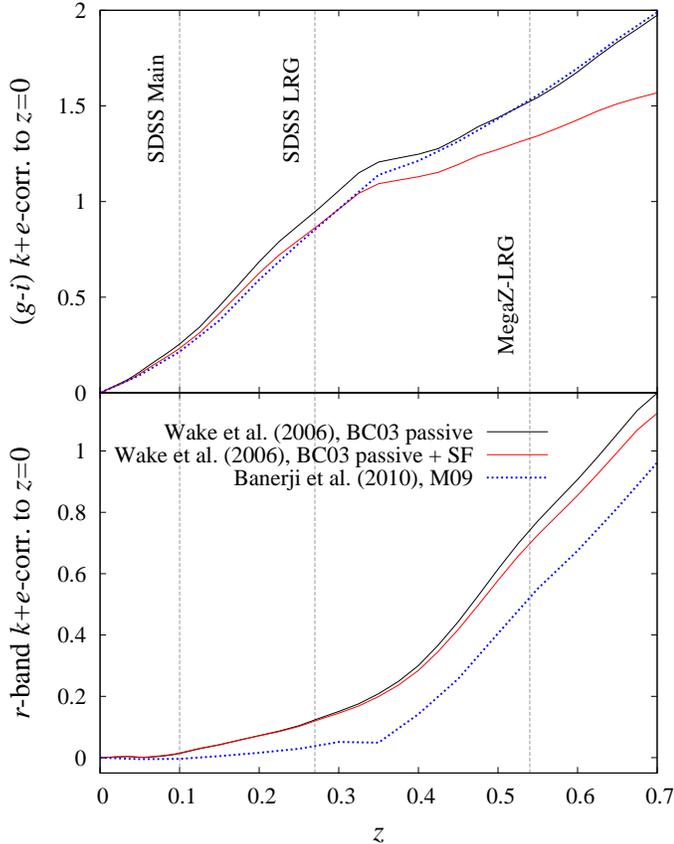}
\caption{$k+e$-corrections to $z=0$ using different sets of templates. \textit{Top panel}: $k+e$-corrections in the colour $g-i$ used to determine the compatibility in colour ranges of the galaxy samples. The black curve corresponds to the templates used in this work by \citet{wake06} which are based on the stellar population synthesis code by \citet{bruzual03} using passive evolution. If a low level of continuous star formation is added to the model, the red solid line results. In addition we show the $k+e$-corrections from \citet{banerji10} using population synthesis models by \citet{maraston09} as the blue dotted curve. \textit{Bottom panel}: $k+e$-corrections in the $r$ band in which we compute galaxy luminosities. The coding of the curves is the same as above. The grey vertical lines indicate typical redshifts of the samples under consideration.}
\label{fig:kecorr}
\end{figure}

We have specified our intrinsic alignment model in terms of the physically meaningful rest-frame $r$ band luminosity (passively evolved to $z=0$), which renders our fit results dependent on the $k+e$-corrections employed. Recently, \citet{banerji10} have utilised new corrections based on improved stellar population synthesis models \citep{maraston09} that perform better than those of \citet{wake06} at describing observed colours of LRGs. We compare these different $k+e$-corrections in the $r$ band in Fig.$\,$\ref{fig:kecorr}, finding significant differences in particular at high redshift.

The difference between the results from \citet{wake06} and \citet{banerji10} has a negligible effect on the luminosities computed for the SDSS Main samples at low redshift, but amounts to approximately $0.1\,$mag for the SDSS LRG samples and, the details depending on the population synthesis model used, to about $0.2\,$mag at the mean redshift of the MegaZ-LRG sample. Hence, switching to luminosities calculated according to \citet{banerji10} would imply a $10\,\%$ increase in the intrinsic alignment amplitude $A$ for the fits to the combined SDSS LRG samples, using the corresponding best-fit value for $\beta$. While this change is subdominant to the $1\sigma$ error, a change by $0.2\,$mag for the MegaZ-LRG sample leads to a shift by about $20\,\%$ in the term $\br{L/L_0}^\beta$, which clearly exceeds the $1\sigma$ errors e.g. of the joint fit by all galaxy samples. Besides, the effect is redshift-dependent, so that all three fit parameters $A$, $\beta$, and $\eta_{\rm other}$ would be affected in that case.

However, in principle the definition of luminosity used for (\ref{eq:iamodel}) is arbitrary, as long as it is consistent for all samples. This is the case since we employ the $k+e$-corrections according to \citet{wake06} throughout. As demonstrated above, the transformation of our intrinsic alignment model to a different convention for galaxy luminosity has a considerable effect on all parameters and needs to be executed with care.

The actual quantity of interest is not the power spectrum (\ref{eq:iamodel}) but rather the observable intrinsic alignment signals, in particular those contaminating cosmic shear surveys. Changes in the values of the intrinsic alignment fit parameters are only meaningful in how they modify these observables. The observable intrinsic alignment signals are determined by observer-frame apparent magnitude limits of a survey and can consistently be calculated from (\ref{eq:iamodel}) if the same $k+e$-corrections as used in the computation of galaxy luminosities for the model are applied. In Sect.$\,$\ref{sec:impact} we will use such a procedure to predict the contamination of a cosmic shear survey by intrinsic alignment signals derived from our best-fit model. As will be detailed in there, there are sources of uncertainty, especially concerning the choice of luminosity functions, that are likely to be more important than the effect of $k+e$-corrections.

Besides the luminosities of the galaxy samples, their $^{0.2}(g-i)$ colours used to assess the compatibility of the samples are expected to depend on the $k+e$-corrections as well. Again, we have made sure that identical $k+e$-corrections were employed to produce Fig.$\,$\ref{fig:2dtrends}, so that the MegaZ-LRG samples are consistent with the other data also in their colour cuts. Note however that, contrary to the luminosity dependence, our results cannot readily be reformulated for galaxy colours obtained from other versions of $k+e$-corrections due to the imposed colour cut which is dependent on the \citet{wake06} templates. 

As shown in Fig.$\,$\ref{fig:kecorr}, we have compared the differences between the $k+e$-corrections in the $g$ and $i$ bands obtained from \citet{wake06} and \citet[see \citealp{banerji10} for details]{maraston09} as a function of redshift. We find that the latter are in good agreement with the two variants of colour corrections derived from \citet{wake06}. The \citet{maraston09} $k+e$-corrections display a slightly steeper increase with redshift, yielding $g-i$ corrections that are 0.03 lower at the redshift of the SDSS Main samples and about 0.1 higher at the mean redshift of the full MegaZ-LRG sample compared to the mean of the \citet{wake06} models. Consequently, the MegaZ-LRG samples would on average be bluer with respect to the other samples if we had used the $k+e$-corrections based on \citet{maraston09} instead. While a difference of $0.1$ in the colour appears to be substantial compared to the typical spread of the MegaZ-LRG sample in $^{0.2}(g-i)$ of $\gtrsim 0.5$, the uncertainty in the colour cut due to the fuzziness in the lower $g-i$ limit of the SDSS LRG and Main samples is of the same order, see Fig.$\,$\ref{fig:2dtrends}. Besides, as Fig.$\,$\ref{fig:kecorr} top panel suggests, this level of uncertainty due to the differences between the templates by \citet{wake06} and \citet{maraston09} is of the same order as the uncertainty in the evolutionary model chosen for a galaxy within a given set of templates.

\subsection{Systematics tests}
\label{sec:systematics}

As in \citet{mandelbaum06}, \citet{hirata07}, and \citet{mandelbaum09} we perform several tests to ensure that our results from the MegaZ-LRG data are not contaminated by instrumental or other effects. We also repeat the systematics tests for the re-defined red SDSS Main L3 and L4 samples. No signatures of systematics were found during the previous analyses of the SDSS LRG and the original SDSS Main samples (see the aforementioned references). 

\begin{table*}[t]
\centering
\caption{Systematics tests for the MegaZ-LRG data and the re-defined SDSS L3 and L4 samples.}
\begin{tabular}[t]{llcccc}
\hline
sample & correlation &  $\chi^2_{\rm red}$(all $r_p$) & $p(>\chi^2)$ & $\chi^2_{\rm red}(r_p>6\,{\rm Mpc}/h)$ & $p(>\chi^2)$\\[0.5ex]
\hline\hline
MegaZ-LRG              & $w_{\rm g \times}$, all z, (90)         & 0.51 & 0.88 & 0.09 & 0.99 \\
MegaZ-LRG              & $w_{\rm g \times}$, $z < 0.529$, (90)   & 1.08 & 0.37 & 0.50 & 0.77 \\
MegaZ-LRG              & $w_{\rm g \times}$, $z > 0.529$, (90)   & 0.38 & 0.96 & 0.49 & 0.79 \\
MegaZ-LRG              & $w_{g+}$, all z, large $|\Pi|$          & 1.01 & 0.43 & 1.44 & 0.20 \\
MegaZ-LRG ($g-i$ cut) & $w_{\rm g \times}$, all z, (90)          & 0.72 & 0.70 & 0.77 & 0.57 \\
MegaZ-LRG ($g-i$ cut) & $w_{\rm g \times}$, $z < 0.529$, (90)    & 1.25 & 0.25 & 1.66 & 0.14 \\
MegaZ-LRG ($g-i$ cut) & $w_{\rm g \times}$, $z > 0.529$, (90)    & 0.30 & 0.98 & 0.22 & 0.96 \\
SDSS Main L3 red      & $w_{\rm g \times}$                       & 0.97 & 0.47 & 0.70 & 0.62 \\
SDSS Main L3 red      & $w_{g+}$, large $|\Pi|$                  & 0.49 & 0.90 & 0.21 & 0.96 \\
SDSS Main L4 red      & $w_{\rm g \times}$                       & 1.24 & 0.26 & 0.49 & 0.78 \\
SDSS Main L4 red      & $w_{g+}$, large $|\Pi|$                  & 1.06 & 0.39 & 1.01 & 0.41 \\
\hline
\end{tabular}
\tablefoot{We list the reduced $\chi^2$ and the corresponding $p$-values of a zero signal fit to the correlation function $w_{\rm g \times}$, and to $w_{\rm g+}$ for large values of $|\Pi|$. We use $270 < |\Pi|/[h^{-1}{\rm Mpc}] < 315$ for the photometric MegaZ-LRG data and $100 < |\Pi|/[h^{-1}{\rm Mpc}] < 150$ for the spectroscopic SDSS Main samples. We include all transverse scales available into the fit (third and fourth columns), but also restrict $r_p$ to scales larger than $6$\hmpc\ (fifth and sixth columns), i.e. the range in which the intrinsic alignment fits were performed. The numbers in parentheses indicate $\Pi_{\rm max}$ in units of \hmpc\ for the MegaZ-LRG samples.}
\label{tab:systematics}
\end{table*}

\begin{figure}[t!!!]
\centering
\includegraphics[scale=.36,angle=270]{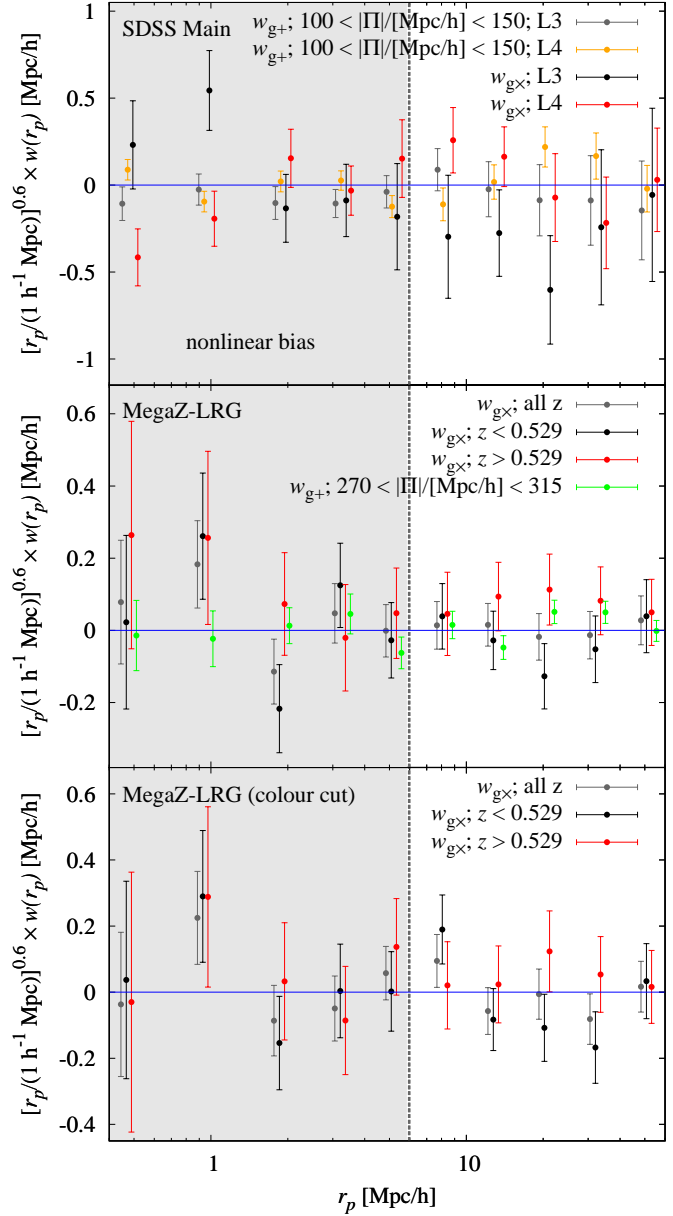}
\caption{Projected correlation functions $w_{g \times}$ and $w_{\rm g+}$ for large line-of-sight separations as a function of comoving transverse separation $r_p$. \textit{Top panel}: Shown is $w_{g \times}$ for the re-defined SDSS Main L4 (red points) and L3 (black points) samples. Moreover $w_{\rm g+}$ integrated along the line of sight for $100 < |\Pi|/[h^{-1}{\rm Mpc}] < 150$ is plotted as grey points for L3 and orange points for L4. Note that the orange, black, and red points have been slightly offset horizontally for clarity. \textit{Centre panel}: Shown is $w_{g \times}$ for the full MegaZ-LRG sample in grey, for the low-redshift sample with $z < 0.529$ in black, and for the high-redshift sample with $z > 0.529$ in red. In addition we plot the correlation function $w_{\rm g+}$ for the full MegaZ-LRG sample, integrated along the line of sight for $270 < |\Pi|/[h^{-1}{\rm Mpc}] < 315$, as green points. Note that the black, red, and green points have been slightly offset horizontally for clarity. \textit{Bottom panel}: Same as in centre panel, but for the MegaZ-LRG sample with the colour cut imposed. Note that we did not repeat the computation of $w_{\rm g+}$ for large $|\Pi|$. Error bars at different $r_p$ are correlated. Note that all correlation functions have been rescaled by $[r_p/(1\,h^{-1}{\rm Mpc})]^{0.6}$ for easier inspection. All signals are consistent with zero.}
\label{fig:sysfit}
\end{figure}

First, we compute the correlation function $w_{\rm g \times}$ using the cross-component of the shear instead of the radial component. The cross-component and thus $w_{\rm g \times}$ change sign under parity transformations, measuring a net curl of the galaxy shape distribution. Since we do not expect that galaxy formation and evolution violates parity symmetry, any non-vanishing $w_{\rm g \times}$ serves as an indicator for systematics, such as residual PSF distortions. The resulting signals for the full MegaZ-LRG sample, as well as for the two MegaZ-LRG redshift bins, are shown in the centre panel of Fig.$\,$\ref{fig:sysfit}. All signals are comfortably consistent with zero, and fits of a zero line to the data in the full range of $r_p$ yield reduced $\chi^2$ values well below unity and correspondingly large $p$-values, as shown in Table \ref{tab:systematics}. We repeat this analysis for the same MegaZ-LRG samples but with the cut in $g-i$ imposed, again finding no evidence for systematics, see the bottom panel of Fig.$\,$\ref{fig:sysfit}.

Furthermore we consider $w_{\rm g+}$ computed only for large line-of-sight separations $\Pi$ at which one does not expect astrophysical correlations anymore. A non-zero signal in this measure could for instance be caused by an artificial alignment of galaxy images in the telescope optics. Due to the photometric redshift scatter in MegaZ-LRG data, we use much larger values of $|\Pi|$ for this systematics test than preceding works, integrating the three-dimensional correlation function along the line of sight for $270 < |\Pi|/[h^{-1}{\rm Mpc}] < 315$. Still, gI correlations might not be completely negligible. We estimate the signal from the best-fit intrinsic alignment model obtained in the foregoing section, finding amplitudes below $0.004$ for $r_p>6$\hmpc\ and below $0.018$ for all $r_p$ scales considered. Thus, residual gI correlations should be negligible in this case, and indeed the resulting $w_{\rm g+}$ is consistent with zero. Note that we cannot apply this test to $w_{\rm gg}$ because the much stronger galaxy clustering signal is not negligible even in the extreme range of $\Pi$ that we have chosen.

We also compute $w_{\rm g \times}$, as well as $w_{\rm g+}$ in the range $100 < |\Pi|/[h^{-1}{\rm Mpc}] < 150$, for the SDSS L3 and L4 samples with the new colour cut to isolate red galaxies, as shown in the top panel of Fig.$\,$\ref{fig:sysfit}. Note that for these spectroscopic samples we can resort to much smaller values of $|\Pi|$ to obtain a range which is not expected to contain physical correlations anymore. All systematics tests for the SDSS Main L3 and L4 samples are consistent with zero, see Table \ref{tab:systematics}. Although some points deviate from zero clearly outside the $1\sigma$ limit, and reduced $\chi^2$ values slightly exceed unity, the $p$-values indicate that there is no significant signal in the data.

As expected, we find that error bars on $w_{\rm g \times}$ are of similar size as for $w_{\rm g+}$ when calculated for the same range in line-of-sight separation; compare e.g. the centre panel of Fig.$\,$\ref{fig:sysfit} to Fig.$\,$\ref{fig:gifit} (note the different scaling of the ordinate axes). Since the projected correlation functions are given by the sum of bins in $\Pi$, and not the mean, see e.g. (\ref{eq:wobs}), one expects larger errors on $w_{\rm g \times}$ or $w_{\rm g+}$ if the range in $\Pi$ included is broader. Therefore $w_{\rm g+}$ in the range $100 < |\Pi|/[h^{-1}{\rm Mpc}] < 150$ for the SDSS Main samples and $w_{\rm g+}$ in the range $270 < |\Pi|/[h^{-1}{\rm Mpc}] < 315$ for the MegaZ-LRG sample features smaller error bars than $w_{\rm g \times}$, calculated in the range $|\Pi|/[h^{-1}{\rm Mpc}] < 60$ and $|\Pi|/[h^{-1}{\rm Mpc}] < 90$, respectively.

One additional type of systematic is the calibration of the shear.  A
multiplicative calibration offset would manifest directly as a multiplicative factor 
in $w_{g+}$.  Aside from any impact on the best-fitting intrinsic
alignment amplitude $A$ from the combined samples, there could also be
some effect that is a function of galaxy properties (typically
apparent size and $S/N$, see \citealp{massey07step}), which would
manifest as a difference between the SDSS Main samples (very bright
apparent magnitudes and well-resolved), the spectroscopic LRGs
(moderately bright and well-resolved), and MegaZ-LRG (significantly
fainter and less well-resolved). Because the samples occupy different
places in both redshift and luminosity space, the effect of such a
bias cannot be estimated in a straightforward way.  However, the shape
measurements used for this work were subjected to significant
systematics tests in \cite{mandelbaum05}, including tests for
calibration offsets between different samples, and thus we do not
anticipate that shear calibration is a significant systematic relative
to others (such as photometric redshift error uncertainties, or
$k+e$-corrections) or relative to the size of the statistical
error bars on $w_{g+}$ measurements.

\section{Constraints on intrinsic alignment contamination of cosmic shear surveys}
\label{sec:impact}

Intrinsic alignments constitute the potentially major astrophysical source of systematic uncertainties for cosmic shear surveys. If left untreated, they can severely bias cosmological parameters estimates \citep[e.g.][]{bridle07}. If the contamination by intrinsic alignments is well known, it can ideally be incorporated into the modelling by subtracting the mean intrinsic alignment signal from the lensing term and accounting for the residual uncertainty in the systematic by introducing nuisance parameters over which one can then marginalise. 

To elucidate the implications for cosmological constraints from cosmic shear surveys by our constraints on intrinsic alignments, we will optimistically assume that the mean systematic signal is indeed given by our best-fit model. In this approach, the decisive quantity is not the mean value of the bias on cosmological parameters, which can be easily corrected for by subtracting the mean intrinsic alignment signal, but the uncertainty in the bias due to uncertainty in intrinsic alignment model parameters, which directly affects the accuracy with which the cosmological model can be constrained when taking the systematics into account.  We do not address uncertainty due to adoption of the generalised NLA model (\ref{eq:iamodel}), i.e. the possibility that the underlying intrinsic alignments model is different, because we see no tension between the model and our data. 

We assess the range of possible biases on cosmological parameters that originate from intrinsic alignment signals using the constraints obtained from the foregoing investigation. Since the SDSS Main L3 and L4 samples proved to be fully consistent with the results for the two LRG samples, we will assume in the following that our intrinsic alignment model also holds for typical, less luminous early-type galaxies predominantly found in a cosmic shear survey. We emphasise that this study requires the extrapolation of the best-fit intrinsic alignment model to combinations of galaxy redshifts and luminosities that have not been probed directly by any of the galaxy samples analysed in this work.  However, that extrapolation is less worrisome now that we have galaxy samples at $z\sim 0.6$, given that the lower redshift galaxies in a cosmic shear survey will tend to be the greatest culprits in causing GI contamination due to the scaling of the effect with redshift separations of galaxy pairs. 

By means of a Fisher matrix analysis we compute the effect on a present-day, fully tomographic (i.e. including all independent combinations of redshift bins) cosmic shear survey, roughly following CFHTLS parameters \citep{hoekstra06,semboloni06,fu07}. To calculate the matter power spectrum, we use the same cosmology, transfer function, and nonlinear correction as outlined in Sect.$\,$\ref{sec:gIcorrelationfct}. For computational simplicity we use the convergence power spectrum (the GG signal henceforth) as the observable cosmic shear two-point statistic, using the following Limber equation,
\eq{
\label{eq:limber}
C^{(ij)}_{\rm GG}(\ell) = \int_0^{\chi_{\rm hor}} \dd \chi ~\frac{q_\epsilon^{(i)}(\chi) ~q_\epsilon^{(j)}(\chi)}{\chi^2} ~P_\delta\br{\frac{\ell}{\chi},\chi}\;,
}
where $q_\epsilon^{(i)}$ again denotes the lensing weight. Instead of specifying a photometric redshift for a redshift probability distribution, we switch here to the usual notation of using an index $i$ that characterises a (broad) distribution $p^{(i)}(z)$ entering (\ref{eq:weightlensing}). The corresponding Limber equations for the GI and II signals can be readily formulated accordingly, yielding \citep[e.g.][]{hirata04}
\eqa{
\label{eq:limberII}
C_{\rm II}^{(ij)}(\ell) \!\!&=&\!\! \int^{\chi_{\rm hor}}_0 \!\!\!\dd \chi\; \frac{p_\epsilon^{(i)}(\chi)~p_\epsilon^{(j)}(\chi)}{ \chi^2}\; P_{\rm II} \br{\frac{\ell}{\chi},\chi}\;;\\ 
\label{eq:limberGI}
C_{\rm GI}^{(ij)}(\ell) \!\!&=&\!\! \int^{\chi_{\rm hor}}_0 \!\!\!\dd \chi\; \frac{p_\epsilon^{(i)}(\chi)~q_\epsilon^{(j)}(\chi) + q_\epsilon^{(i)}(\chi)~p_\epsilon^{(j)}(\chi)}{\chi^2}\; P_{\delta {\rm I}} \br{\frac{\ell}{\chi},\chi}\;,
}
where we assume that the intrinsic shear power spectrum can be described by an extension of our intrinsic alignment model analogous to (\ref{eq:iamodel}), 
\eq{
\label{eq:IImodel}
P^{\rm model}_{\rm II}(k,z,L) = A^2\; P_{\rm II}(k,z)\; \br{\frac{1+z}{1+z_0}}^{2\eta_{\rm other}} \br{\frac{L}{L_0}}^{2\beta}\;,
}
where $P_{\rm II}(k,z)$ is given by the NLA model, i.e. (\ref{eq:IIlinalign}) with the linear matter power spectrum replaced by the full, nonlinear one. Writing (\ref{eq:IImodel}) with the square of the extra redshift and luminosity terms (again similar to \citealp{kirk10}) involves the assumption that the galaxies correlated are located at similar redshifts, which is valid because the II signal is restricted to physically close pairs; see also the narrow kernel in (\ref{eq:limberII}). Then the two galaxies of a pair also underlie the same luminosity distribution, and since we average (\ref{eq:IImodel}) over this distribution, one can write $L^{2\beta}$ to good approximation although the luminosities of the galaxies in an individual pair may be largely different. Note that we have not measured intrinsic ellipticity correlations in this work, so that the computation of the II signal is entirely based on the validity of the NLA model. However, the contribution of the II signal to the bias on cosmology will be smaller than the one by GI correlations in a CFHTLS-like survey, so that this assumption does not affect our conclusions substantially. 

We employ an overall redshift distribution according to \citet{smail94},
\eq{
\label{eq:redshiftdistribution}
p_{\rm tot}(z) \propto \br{\frac{z}{z_{\rm Smail}}}^{\alpha_{\rm Smail}} \exp \bc{ -\br{\frac{z}{z_{\rm Smail}}}^{\beta_{\rm Smail}}}\;,
}
with parameters $\alpha_{\rm Smail}=0.836$, $\beta_{\rm Smail}=3.425$, and $z_{\rm Smail}=1.171$ yielding a median redshift of 0.78 \citep{benjamin07}. We slice this distribution into 10 \lq photometric\rq{} redshift bins such that every bin contains the same number of galaxies. The corresponding redshift distribution for each bin is then computed via the formalism detailed in \citet{joachimi09}, assuming a Gaussian photometric redshift scatter of width $0.05(1+z)$ around every spectroscopic redshift. We compute Gaussian covariances for the power spectra including cosmic variance and shape noise \citep[for details see][]{joachimi08}, assuming a survey size of $A_{\rm survey}=100\,{\rm deg}^2$. Shape noise is incorporated with an overall galaxy number density of $n_\Omega = 12\,{\rm arcmin}^{-2}$ and a dispersion of the absolute value of the complex intrinsic ellipticity of $0.3$ \citep{hoekstra06}.

We consider a parameter vector $\vek{p}=\bc{\Omega_{\rm m},\sigma_8,h,n_{\rm s},\Omega_{\rm b},w_0}$ for the cosmological analysis, for a flat universe with constant dark energy equation-of-state parameter $w_0$. Assuming that the covariance is not dependent on these parameters \citep[see][]{eifler08}, one obtains the Fisher matrix \citep{tegmark97}
\eq{
\label{eq:gicorr_fisher}
F_{\mu\nu} = \sum_{\ell,i \leq j,k \leq l} \frac{\partial C_{\rm GG}^{(ij)}(\ell)}{\partial p_\mu}\;  {\rm Cov}^{-1} \br{C_{\rm GG}^{(ij)}(\ell),\; C_{\rm GG}^{(kl)}(\ell)} \frac{\partial C_{\rm GG}^{(kl)}(\ell)}{\partial p_\nu}\;,
}
where we use 40 logarithmically spaced angular frequency bins between $\ell=10$ and $\ell=3000$. With the Fisher matrix, one can calculate the bias on a cosmological parameter via \citep[e.g.][]{kim04,huterer05b,huterer06,taylor07,amara08,kitching08,joachimi09}
\eqa{
\label{eq:gicorr_bias}
b(p_\mu) &=& \sum_\nu \br{F^{-1}}_{\mu \nu} \sum_{\ell,i \leq j,k \leq l} \bb{C_{\rm II}^{(ij)}(\ell) + C_{\rm GI}^{(ij)}(\ell)}\\ \nn
&& \hspace*{1cm} \times\; {\rm Cov}^{-1} \br{C_{\rm GG}^{(ij)}(\ell),\; C_{\rm GG}^{(kl)}(\ell)} \frac{\partial C_{\rm GG}^{(kl)}(\ell)}{\partial p_\nu}\;,
}
where the systematic is given by the sum of II and GI power spectra. Note that the parameter bias is independent of the survey size while the statistical errors obtained from $F_{\mu\nu}$ are proportional to $1/\sqrt{A_{\rm survey}}$.

The intrinsic alignment analysis presented above only dealt with red
galaxies, whereas a typical galaxy population in cosmic shear surveys
is dominated by blue galaxies for which \citet{mandelbaum09} reported
a null detection for a galaxy sample spanning a similar range of
redshifts as in this paper. Thus we assume that only the red fraction
$f_r$ of galaxies in the survey carries an intrinsic alignment
signal. Consequently the II power spectrum is multiplied by a factor
$f_r^2$, and the GI power spectrum by $f_r$, resulting in the same
model as used by \citet{kirk10}. Note that this approach is overly
simplistic in splitting the galaxy population into two disjoint groups
with largely different intrinsic alignment properties although one
expects the intrinsic alignment parameters to vary in a more
continuous manner with galaxy colour.

Moreover it is important to note that we ignore any uncertainty in the null measurement of blue galaxy intrinsic alignment which would add to the scatter of systematic errors on the cosmological parameters. In principle it should be feasible to take into account finite constraints on the intrinsic alignment parameters determined from blue galaxy samples, such as those studied by \citet{mandelbaum09}. However, before incorporating them into this formalism, these samples would have to undergo the same compatibility tests as performed in this work for red galaxy samples, and then be combined to yield joint fits. These steps are beyond the scope of this analysis and left to future investigation.

We expect that both $f_r$ and the distribution of luminosities of red galaxies depend on redshift and thus have different values in each photometric redshift bin of our fictitious cosmic shear survey. To estimate realistic values for these parameters, we make use of the luminosity functions provided by \citet{faber07}. They fit Schechter functions $\phi(L,z)$ jointly to samples from SDSS, 2dF, COMBO-17, and DEEP2 in redshift bins out to $z \sim 1$, considering early- and late-type galaxies individually. The criteria used by \citet{faber07} to separate red and blue galaxies differ from the ones employed in this work, but still we consider their samples as representative for differentiating between blue galaxies with negligible intrinsic alignments and red galaxies with an intrinsic alignment signal consistent with our best-fit model. We defer the technical details and also further discussion of this approach to Appendix \ref{sec:LF}.

In combination with a minimum galaxy luminosity $L_{\rm min}(z,r_{\rm lim})$, computed from the apparent magnitude limit at each redshift, a set of luminosity functions also specifies the redshift probability distribution $p_{\rm tot}(z)$. However, the luminosity functions by \citet{faber07} are unlikely to reproduce exactly the redshift probability distribution of CFHTLS because weak lensing surveys have additional galaxy angular size cuts and thus deviate from a purely flux-limited sample. Also, we must extrapolate to fainter magnitudes than the sample used to determine the luminosity function, which can be a source of significant uncertainty, particularly for blue galaxies given their steep faint-end slope.  Besides, the blue galaxy sample of \citet{faber07} is composed of several galaxy types, so that it is unclear which $k$-corrections and filter conversions are applicable. Hence, we make the assumption that the red galaxy luminosity functions from \citet{faber07} are compatible with the selection criteria of a weak lensing survey and derive the total comoving volume density of galaxies from (\ref{eq:redshiftdistribution}) by noting that $n_{\rm V} = \dd N/\dd V_{\rm com} = \dd N/\dd z\, (\dd V_{\rm com} /\dd z)^{-1}$. Then the fraction of red galaxies reads
\eq{
\label{eq:redgalfrac}
f_r(z,r_{\rm lim}) = \frac{n_{\rm V,red}(z,r_{\rm lim})}{n_{\rm V,tot}(z)} = \frac{\chi^2(z)\;\chi'(z)}{n_\Omega\; p_{\rm tot}(z)}\; \int_{L_{\rm min}(z,r_{\rm lim})}^{\infty} \!\!\!\!\!\!\!\! \dd L\; \phi(L,z)\;,
}
where the lower limit of the integration $L_{\rm min}$ is determined
by the magnitude limit of the survey which we assume to be $r_{\rm
  lim}=25$, roughly compatible with CFHTLS ($i_{\rm lim}=24.5$;
\citealp{hoekstra06}). The redshift dependence of $L_{\rm min}$ is
introduced by the conversion of $r_{\rm lim}$ to absolute magnitude
and by the $k+e$-correction. The latter is again computed in the
$r$ band by means of the templates used in \citet{wake06} for
early-type galaxies (specifically, by a $k$-correction that is between
that of the two very closely related templates used in that
paper, see also Fig.$\,$\ref{fig:kecorr}).

\begin{figure*}[t]
\centering
\includegraphics[scale=.38,angle=270]{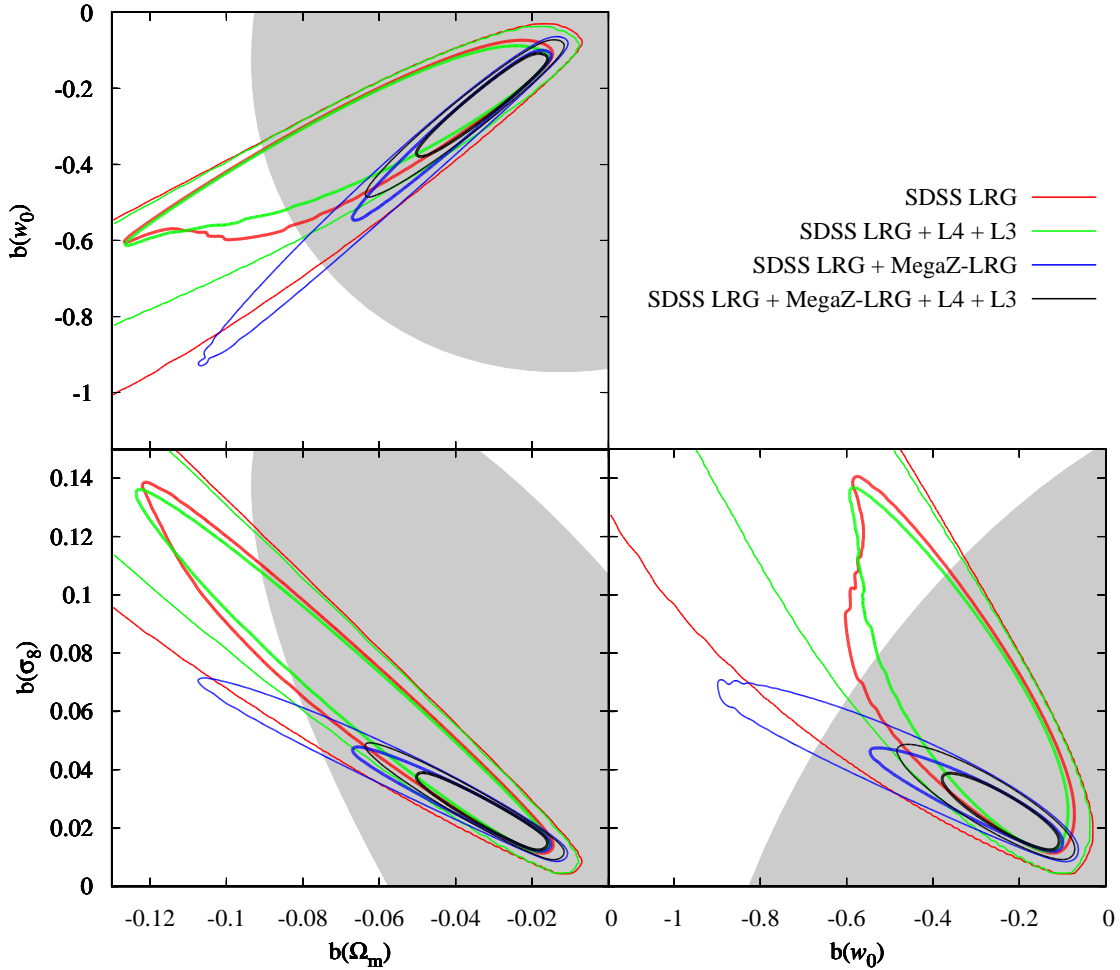}
\caption{Bias on cosmological parameters due to intrinsic alignments for a CFHTLS-like weak lensing survey. Shown are the regions in which $99\,\%$ of the possible biases on the parameters $\bc{\Omega_{\rm m},\sigma_8,w_0}$ are located when the parameters $\bc{A,\beta,\eta_{\rm other}}$ in the intrinsic alignment model are sampled from the $1\sigma$ confidence region (thick lines) and the $2\sigma$ confidence region (thin lines) of our fits. The contours resulting from the SDSS LRG constraints are shown in red, the ones from the SDSS LRG + SDSS Main (L3 and L4) constraints in green, the ones from the MegaZ-LRG + SDSS LRG constraints in blue, and the contours from the joint constraints by the MegaZ-LRG, SDSS LRG, and SDSS Main samples in black. The grey regions indicate the $1\sigma$ confidence regions of the constraints on cosmological parameters. For this analysis the 6 parameters $\bc{\Omega_{\rm m},\sigma_8,h,n_{\rm s},\Omega_{\rm b},w_0}$ were varied. The fraction $f_r$ of red galaxies and the distribution of luminosities of red galaxies were determined from the luminosity functions provided by \citet{faber07}. Note that the contours corresponding to the $2\sigma$ confidence region for the SDSS LRG-only and SDSS LRG + SDSS Main constraints extend far beyond the plot boundaries.}
\label{fig:cosmobias}
\end{figure*}

In contrast to the various SDSS samples in which we measured intrinsic alignments, the distribution of galaxy luminosities in each photometric redshift bin of our mock cosmic shear survey is wide, so that our approximation that we can replace $L$ by the mean luminosity in (\ref{eq:iamodel}) and (\ref{eq:IImodel}) is not sufficiently accurate anymore, in particular in those regions of the intrinsic alignment parameter space where $\beta$ deviates significantly from unity. Instead, we have to average (\ref{eq:iamodel}) and (\ref{eq:IImodel}) over the luminosity function, which reduces to evaluating
\eq{
\label{eq:redlum}
\frac{\ba{L^\beta}(z,r_{\rm lim})}{L_0^\beta(z)} = \frac{1}{L_0^\beta(z)}\; \frac{\int_{L_{\rm min}(z,r_{\rm lim})}^{\infty} \dd L\; L^\beta\; \phi(L,z)}{\int_{L_{\rm min}(z,r_{\rm lim})}^{\infty} \dd L\; \phi(L,z)}\;,
}
where we have taken into account that the rest-frame reference luminosity $L_0$ must be $e$-corrected back to redshift $z$, which we do via the redshift dependence of $M^*$ given in \citet{faber07}; see again Appendix \ref{sec:LF} for details. For every photometric redshift bin of the cosmic shear survey, we use the values of $\ba{(L/L_0)^\beta}$ and $f_r$ at the median redshift of the underlying redshift distributions, which is a good assumption if the redshift distributions corresponding to the photometric redshift bins are sufficiently narrow.

We further assess the accuracy and limitations of this ansatz in Appendix \ref{sec:LF} by comparing results from sets of luminosity functions other than those from \citet{faber07}. We also provide volume densities $n_{\rm V,red}$ and mean luminosities for a range of redshifts and limiting $r$ band magnitudes, which can be employed together with our best-fit intrinsic alignment model parameters to estimate the expected intrinsic alignment contamination of other surveys.

The GI and II signals are then computed via (\ref{eq:iamodel}) and (\ref{eq:IImodel}), where the free intrinsic alignment parameters $\bc{A,\beta,{\eta_{\rm other}}}$ are determined as follows. We overlay the three-dimensional $1\sigma$ and $2\sigma$ volumes of the intrinsic alignment fits to four of the sample combinations shown in Table \ref{tab:fitscombined} and Fig.$\,$\ref{fig:contours_AzL} with a square grid, containing $N$ nodes in total. For the combination of $\bc{A,\beta,{\eta_{\rm other}}}$ on each grid node we compute the projected intrinsic alignment power spectra according to (\ref{eq:limberII}) and (\ref{eq:limberGI}) and subsequently the parameter biases via (\ref{eq:gicorr_bias}). This way we obtain a bias vector $\vek{b}=\bc{b(p_1),\,..\,,b(p_{N_{\rm D}})}^\tau$, where $N_{\rm D}$ is the number of cosmological parameters under consideration, in cosmological parameter space for every grid node within the $1\sigma$ or $2\sigma$ confidence volume in intrinsic alignment parameter space.

\begin{table*}[t]
\centering
\caption{Range of possible cosmological parameter biases due to intrinsic alignments for a CFHTLS-like cosmic shear survey.}
\begin{tabular}[t]{cccccccccc}
\hline\hline
parameter & $\sigma_{\rm stat}$ & \multicolumn{2}{c}{sample set (1)} & \multicolumn{2}{c}{sample set (2)} & \multicolumn{2}{c}{sample set (3)} & \multicolumn{2}{c}{sample set (4)}\\
 & & $1\sigma$ & $2\sigma$ & $1\sigma$ & $2\sigma$ & $1\sigma$ & $2\sigma$ & $1\sigma$ & $2\sigma$ \\[0.5ex]
\hline\\[-1ex]
$\Omega_{\rm m}$ & 0.06 & 0.10 & 0.44 & 0.10 & 0.37 & 0.05 & 0.08 & 0.03 & 0.05\\
$\sigma_8$       & 0.11 & 0.11 & 0.53 & 0.11 & 0.42 & 0.03 & 0.06 & 0.02 & 0.04\\
$w_0$            & 0.62 & 0.49 & 2.84 & 0.46 & 2.40 & 0.40 & 0.75 & 0.24 & 0.37\\
\hline
\end{tabular}
\tablefoot{We have listed the $1\sigma$ statistical error $\sigma_{\rm stat}$, resulting from the Fisher matrix analysis after marginalising over all remaining parameters, in the second column, as well as the range of biases we obtained by sampling from the $1\sigma$ and $2\sigma$ confidence regions in the parameter space spanned by $\bc{A,\beta,\eta_{\rm other}}$. In the third to sixth columns results from the fits to four sets of galaxy samples, which are also shown in Fig.$\,$ \ref{fig:cosmobias}, are given. These sets are (1) the combined SDSS LRG samples, (2) SDSS LRG and SDSS Main L4 and L3 samples combined, (3) SDSS LRG and MegaZ-LRG samples combined, and (4) SDSS LRG, MegaZ-LRG, and the SDSS Main L4 and L3 samples combined. The range of biases is defined as the interval containing $99\,\%$ of the distribution of biases (\ref{eq:kde}).}
\label{tab:cosmobias}
\end{table*}

We convert the ensemble of $N$ parameter bias vectors $\bc{\vek{b}_1,\,..\,,\vek{b}_N}$ into a distribution of bias values via Gaussian kernel density estimation, i.e. we approximate this distribution by
\eq{
\label{eq:kde}
p(\vek{x}|\bc{{\vek{b}_i}}) = \frac{1}{N}\; \sum_{i=1}^N \prod_{j=1}^{N_{\rm D}}\; \frac{1}{\sqrt{2\pi}\, \Delta_j}\; \exp \bc{-\frac{\br{x_j - b_i(p_j)}^2}{2\, \Delta_j^2}}\;,
}
where we use $N_{\rm D}=2$ when considering the distribution in a two-dimensional parameter plane, and $N_{\rm D}=1$ when computing the one-dimensional distributions. The widths $\vek{\Delta}$ of the Gaussians in every dimension of cosmological parameter space are free parameters, and we choose them to take the minimum values which still produce a smooth distribution (except for small wiggles in some of the sparsely sampled regions). While we use six cosmological parameters to compute the biases on cosmology, we focus in our presentation of the uncertainty in the biases on a subset with three cosmological parameters of particular interest in cosmic shear analyses, $\bc{\Omega_{\rm m},\sigma_8,w_0}$. For the tightly constrained parameters $\Omega_{\rm m}$ and $\sigma_8$ we use $\Delta=0.001$ and in the dimension corresponding to $w_0$ we set $\Delta=0.005$. Note that we use the same widths for all sample combinations considered in order not to distort the comparison between the resulting bias distributions.

In Fig.$\,$\ref{fig:cosmobias} we show the contours comprising $99\,\%$ of the distribution (\ref{eq:kde}) in the two-dimensional parameter planes spanned by all pair combinations in the set $\bc{\Omega_{\rm m},\sigma_8,w_0}$, sampling from the posteriors of the intrinsic alignment parameters obtained for SDSS LRGs alone, SDSS LRGs and SDSS Main samples combined, SDSS LRGs and MegaZ-LRG combined, and the joint analysis of MegaZ-LRG, SDSS LRG and SDSS Main samples. In this figure we have given the parameter biases (and not the parameter values) on the axes such that in the absence of any intrinsic alignment contamination, the contours should be centred around $\br{0;0}$ in each panel.

The general direction of parameter biases, for instance along the $\Omega_{\rm m}-\sigma_8$ degeneracy, is in agreement with other predictions on biases due to intrinsic alignments \citep[see for instance][]{joachimi09}\footnote{Note that in deep cosmic shear surveys the negative GI signal dominates the contamination by intrinsic alignments, so that one may expect a negative bias on the power spectrum normalisation $\sigma_8$. However, the bias acts mainly along the $\Omega_{\rm m}-\sigma_8$ degeneracy line (apparently causing a positive systematic shift in $\sigma_8$), with only a small shift perpendicular to this line in the downward direction. Consequently, for fixed $\Omega_{\rm m}$, the bias on $\sigma_8$ is indeed negative.}. As is evident from (\ref{eq:gicorr_bias}), if the GI term dominates, which is expected for deep cosmic shear surveys, the bias is proportional to the amplitude parameter $A$ of the intrinsic alignment model. Thus the remaining uncertainty in $A$ explains the strong elongation of the contours, pointing approximately radially away from $\br{0;0}$. The large errors on intrinsic alignment parameters, in particular on $\eta_{\rm other}$, in the case of using the SDSS LRG samples alone allow for a vast region of possible parameter biases. Adding the SDSS L4 and L3 samples slightly narrows the contours, but does not reduce their radial extent.

The contours tighten dramatically when adding in the MegaZ-LRG data which allowed us to fix the redshift dependence to good accuracy. The additional information provided by the SDSS Main samples constrains the total amplitude of the intrinsic alignment signal still better, thereby further reducing e.g. the extent of the $2\sigma$ contours by about a factor of two along the degeneracy direction. Note that, since the $2\sigma$ regions of the intrinsic alignment model fits to SDSS LRGs only and SDSS LRG and MegaZ-LRG samples combined do not completely overlap (see Fig.$\,$\ref{fig:contours_AzL}), the corresponding bias distributions partly cover different regions in cosmological parameter space.

In Table \ref{tab:cosmobias} we list the $1\sigma$ marginalised statistical errors for the three cosmological parameters of interest, obtained from (\ref{eq:gicorr_fisher}) via $\sigma_{\rm stat}(p_\mu) = \sqrt{(F^{-1})_{\mu \mu}}$. Moreover we give the size of the interval that contains $99\,\%$ of the one-dimensional distribution (\ref{eq:kde}) when sampling from the $1\sigma$ and $2\sigma$ confidence volumes of the intrinsic alignment parameter fits, again as a measure for the spread of biases on cosmology. In agreement with the two-dimensional plots of Fig.$\,$\ref{fig:cosmobias} we find that adding the MegaZ-LRG samples to the SDSS LRG and SDSS Main data considerably shrinks the range of biases, e.g. by more than a factor three (seven) in the case of $\sigma_8$ when sampling from the $1\sigma$ ($2\sigma$) confidence volume. In combination with the SDSS Main L3 and L4 samples, the intervals decrease in size by roughly another $30-50\,\%$ (for the $2\sigma$ confidence volume), reaching values which are below the $1\sigma$ statistical errors. The reduction of intrinsic alignment bias to subdominant levels is also evident from the comparison with the $1\sigma$ confidence regions for constraints on the cosmological parameters plotted in Fig.$\,$\ref{fig:cosmobias}.

Hence, under the assumptions made for this prediction, and provided that the mean intrinsic alignment signal were accurately known and could be subtracted from the cosmic shear data, the uncertainty in the knowledge about the free intrinsic alignment parameters in (\ref{eq:iamodel}) and (\ref{eq:IImodel}) would be subdominant to the statistical errors in a CFHTLS-like survey, given the intrinsic alignment constraints from the joint fit to all galaxy samples considered in this work. 

It is also evident from Fig.$\,$\ref{fig:cosmobias} that the mean bias on the parameters $\Omega_{\rm m}$, $\sigma_8$, and $w_0$ due to our best-fit intrinsic alignment model is in each case appreciably smaller than the $1\sigma$ statistical errors. Consequently, it is possible that cosmology would not be significantly biased even if intrinsic alignments were simply ignored. One may ask if any of the existing cosmic shear analyses would be affected more seriously if subject to an intrinsic alignment signal that follows our best-fit model. Hitherto, weak lensing surveys have not been used in combination with photometric redshifts of individual galaxies to perform cosmic shear tomography, with the exception of the space-based COSMOS survey \citep{massey07,schrabback09}. 

Non-tomographic surveys have much lower contamination by intrinsic alignments than tomographic studies because for a wide redshift distribution the probability of having a close galaxy pair is smaller than for an auto-correlation of narrow redshift bins, thereby lowering the II signal. The probability of having a large line-of-sight separation of galaxies in turn is smaller than for cross-correlations of a low- and high-redshift tomographic bin, thereby rendering the GI contribution less important. Hence, we conclude that any intrinsic alignment signal close to our best-fit model is irrelevant for existing surveys, unless they are significantly shallower, which places a typical galaxy of the survey sample at lower redshift and higher luminosity (see e.g. Fig.$\,$\ref{fig:redgalfrac}), so that the intrinsic alignment contamination becomes stronger \citep[see also][]{kirk10}.

The COSMOS survey is deeper than the CFHTLS-like survey analysed above ($i_{814}<26.7$, \citealp{schrabback09}), so that a substantial part of the cosmic shear signal stems from high redshifts. In addition, red galaxies are likely to be less luminous on average, both effects tending to decrease the amplitude of intrinsic alignments. Besides, due to the small survey area of COSMOS, cosmological constraints are modest, further diminishing the importance of an intrinsic alignment bias. Indeed, the fully tomographic analysis by \citet{schrabback09} did not detect any effects due to intrinsic alignments by excluding auto-correlations of tomographic bins and bright red galaxies.

\section{Conclusions}
\label{sec:conclusions}

In this work we studied intrinsic alignments in the MegaZ-LRG galaxy sample, investigating for the first time an early-type galaxy sample at intermediate redshifts up to $z \sim 0.6$, and for which only photometric redshift information is available. We presented correlations between galaxy number densities and galaxy shapes as a function of the transverse comoving separation of the galaxy pairs for MegaZ-LRG and two subsamples at high and low redshift, as well as for several spectroscopic SDSS LRG and SDSS Main samples. In combination, these samples comprise wide ranges in redshift ($z \lesssim 0.7$) and luminosity ($4\,$mag) which have not been covered in a joint analysis before.

We developed the formalism to incorporate photometric redshift scatter into the modelling of the correlation function, taking special care of the large line-of-sight spread of physical correlations and the effect of contributions by galaxy-galaxy lensing and lensing magnification-shear cross-correlations, which is introduced by photometric redshift uncertainty. Our model reproduces to good accuracy the scaling of the MegaZ-LRG data with the maximum line-of-sight separation included in the computation of the correlation functions, as well as the relative contribution by galaxy-galaxy lensing when varying this maximum separation. This supports the validity of our modelling ansatz and justifies the use of the photometric versus spectroscopic relation obtained from the 2SLAQ survey to quantify the photometric redshift scatter.

Moreover we discussed a correction to the redshift dependence of the widely used linear alignment model \citep{hirata04,hirata10}. We then fitted the nonlinear version of this corrected linear alignment (NLA) model to all samples with a free overall amplitude. To allow for the transition from galaxy-intrinsic to matter-intrinsic correlations, we also determined a linear galaxy bias from galaxy clustering signals. Due to the assumption of linear biasing and the expected breakdown of the NLA model on small scales, we limited the analysis to comoving transverse separations $r_p > 6$\hmpc. We found that all samples are consistent with the scaling with $r_p$ that is inherent to the NLA model (which is identical to the scale dependence of the matter power spectrum), suggesting that the alignment of early-type galaxies is indeed determined by the local tidal gravitational field. We did not test other theories of intrinsic alignments since the linear alignment model is, at least on the largest scales, physically motivated, widespread, and reasonable for elliptical galaxies whose shapes are expected to be subjected to tidal stretching by the surrounding gravitational field.

The amplitudes of the intrinsic alignment signals that were obtained from the fits vary widely by more than an order of magnitude and are thus inconsistent with a one-parameter NLA model. We introduced additional power-law terms into the NLA model to account for an extra redshift and luminosity dependence, using in each case the power-law index as a further free parameter. With this three-parameter model, we demonstrated that the MegaZ-LRG, SDSS LRG, and SDSS Main samples under consideration are fully consistent with each other, but we add the caveat that the error bars for each sample are still quite large. We would particularly benefit from constraints from a low-luminosity sample that probes a larger volume since the statistics for the SDSS Main sample are relatively poor and thus the bright galaxies from the LRG samples dominate the fits. The joint fit to all ten samples strongly suggests an approximately linear increase of the intrinsic alignment amplitude with galaxy luminosity and is consistent with no extra redshift evolution beyond the (corrected) NLA model. Adding in the new MegaZ-LRG data is particularly beneficial in tightening the constraints on the redshift evolution of the intrinsic alignment signal due to the higher redshift of the samples. The normalisation of the NLA model is given by $0.077\,\rho_{\rm cr}^{-1}$ (combining the parameters $C_1$ and $A$), with a $1\sigma$ uncertainty of roughly $10\,\%$.

In the joint analysis of galaxy samples, special attention was payed to homogenising the samples as regards the determination of rest-frame magnitudes/luminosities and the range of galaxy colours. As a consequence, we re-defined the colour separator of the red SDSS Main L4 and L3 samples compared to \citet{hirata07} to avoid a leakage by blue-cloud galaxies. Furthermore we discarded for the joint fits about $30\,\%$ of the MegaZ-LRG galaxies with colours $^{0.2}(g-i)<1.7$, the resulting redder subsample producing slightly higher intrinsic alignment amplitudes. We also discussed the dependence on the employed $k+e$-corrections, affecting our results via the computation of luminosities and rest-frame colour cuts. 

A range of systematics tests was applied to all new and re-defined galaxy samples, finding no evidence for systematic effects in the correlation functions. Residual sources of uncertainty, e.g. in the calibration of galaxy shears or due to the statistical error on the galaxy bias, should be clearly subdominant to the uncertainty originating from the statistical errors on the correlation function measurements that were propagated into the errors on the intrinsic alignment parameters.

In the linear alignment picture, the normalisation of the intrinsic alignment signal is determined by the response of the intrinsic shape of a galaxy to the gravitational potential of that galaxy at the time of its formation. Interpreting our constraints on intrinsic alignment parameters in this framework, we obtained no evidence for an extra redshift evolution and hence a time dependence of the coupling between intrinsic galaxy shape and gravitational potential. The dependence on luminosity we found can be interpreted as an increase of this coupling with galaxy mass.  Comparing our results with \citet{mandelbaum09} who analysed blue galaxy samples out to similar redshifts, it is evident that intrinsic alignments depend strongly on the galaxy type. Whether there is a clear dichotomy between early- and late-type galaxies or a more continuous transition with galaxy colour, which the comparison between the MegaZ-LRG samples with and without colour cut may hint at, is still to be investigated.

In a Fisher matrix analysis, we predicted the bias on cosmological parameters that results from our best-fit intrinsic alignment model. To this end, we used sets of luminosity functions measured by \citet{faber07} in order to derive the fraction of early-type galaxies and their luminosity distribution at each redshift for a given apparent magnitude limit of the survey. Assuming zero intrinsic alignments for blue galaxies (without uncertainty), we then computed the expected intrinsic-ellipticity (II) and shear-intrinsic (GI) correlations, sampling the parameters of the NLA model from the confidence volume of our intrinsic alignment fits. The accuracy of this approach is limited by the substantial extrapolation of the luminosity function data to faint and high-redshift galaxies, as well as the strong scatter in luminosity function parameters obtained from different works. 

For a fully tomographic CFHTLS-like survey both the mean bias and the scatter in the bias due to the uncertainty in the intrinsic alignment model are smaller than the predicted $1\sigma$ cosmological parameter errors, and we similarly expect subdominant systematic effects by intrinsic alignments for other cosmic shear surveys performed hitherto. The MegaZ-LRG data were crucial in reducing this systematic uncertainty. However, if external priors on cosmological parameters, e.g. from the cosmic microwave background, are employed, which is likely to be the case in practice, the significance of the bias due to intrinsic alignments will be higher.

For future ambitious weak lensing surveys such as Euclid, which has roughly comparable depth to CFHTLS but considerably higher statistical power \citep{laureijs09}, the same intrinsic alignment signal would constitute a severe systematic, and marginalising over the uncertainty in intrinsic alignment parameters would significantly degrade constraints on cosmology. However, the gradually increasing precision requirements by planned cosmic shear surveys are likely to be matched by intrinsic alignment studies that continuously improve and consolidate the models. Hence, constraints on intrinsic alignment models as provided by this work and succeeding ones will prove most useful, for instance to define prior ranges in nuisance parametrisations of intrinsic alignment signals \citep[e.g.][]{bridle07}. 

A straightforward test of the intrinsic alignment model obtained in this work could be provided by a similar analysis of galaxy shape correlations from the same set of samples. This way one can verify whether the II signal agrees with the prediction by the linear alignment paradigm (based on the GI signal in this work), and whether the extra redshift and luminosity dependencies of the II signal are consistent with the present results. Furthermore, to obtain unbiased intrinsic alignment measurements for a wide range of galaxy colours is another important but challenging task because many central selection criteria such as shape measurement quality, photometric redshift scatter, or spectroscopic redshift failure rates depend strongly on the galaxy type.

Measurements of the type presented in this paper are restricted to quasi-linear scales although both cosmic shear and galaxy evolution studies have a vital interest in intrinsic alignments in the deeply nonlinear regime. A possible way forward would be the usage of a halo model approach for both galaxy bias and intrinsic alignment signals \citep[see e.g.][]{schneiderm09}. However, note that, similar to the galaxy bias, observational constraints on intrinsic alignments may ultimately be limited by an intrinsic scatter that cannot be accounted for by a deterministic model.

The generalisation of intrinsic alignment measurements to photometric redshift data has opened up a new regime of data sets which could be exploited to constrain intrinsic alignment models. For instance, all present or upcoming cosmic shear surveys with redshift information, or at least subsamples of them with low photometric redshift scatter, could be suited, thereby automatically extending the baselines in redshift and luminosity to scales most relevant for weak lensing. The higher the photometric redshift scatter, the more important become galaxy-galaxy lensing and magnification contributions to the observed correlation functions, so that e.g. at some point cosmology will have to be varied in the intrinsic alignment analysis as well. Then it might be more fruitful to instead consider simultaneous measurements of galaxy shape and number density correlations in the manner proposed by \citet{bernstein08} and investigated in \citet{joachimi10}.

\begin{acknowledgements}
We thank Peter Schneider and Gary Bernstein for useful discussions. Moreover we are grateful to Rachel Bean, Istvan Laszlo, and Chris Hirata for comparing the derivations of the redshift dependence of the linear alignment model. We would also like to thank Manda Banerji for sharing the $k+e$-corrections, and our referee for a helpful report. 

B.J. acknowledges support by the Deutsche Telekom Stiftung and the Bonn-Cologne Graduate School of Physics and Astronomy. F.B.A. acknowledges the Royal Society for a Royal Society University Research Fellowship. S.L.B. acknowledges support in the form of a Royal Society University Research Fellowship and a European Research Council Starting Grant. This work was supported by the RTN-Network \lq DUEL\rq\ of the European Commission.

Funding for the SDSS and SDSS-II has been provided by the Alfred P. Sloan Foundation, the Participating Institutions, the National Science Foundation, the U.S. Department of Energy, the National Aeronautics and Space Administration, the Japanese Monbukagakusho, the Max Planck Society, and the Higher Education Funding Council for England. The SDSS Web Site is {\slshape http://www.sdss.org/}.

The SDSS is managed by the Astrophysical Research Consortium for the Participating Institutions. The Participating Institutions are the American Museum of Natural History, Astrophysical Institute Potsdam, University of Basel, University of Cambridge, Case Western Reserve University, University of Chicago, Drexel University, Fermilab, the Institute for Advanced Study, the Japan Participation Group, Johns Hopkins University, the Joint Institute for Nuclear Astrophysics, the Kavli Institute for Particle Astrophysics and Cosmology, the Korean Scientist Group, the Chinese Academy of Sciences (LAMOST), Los Alamos National Laboratory, the Max-Planck-Institute for Astronomy (MPIA), the Max-Planck-Institute for Astrophysics (MPA), New Mexico State University, Ohio State University, University of Pittsburgh, University of Portsmouth, Princeton University, the United States Naval Observatory, and the University of Washington.

\end{acknowledgements}

\bibliographystyle{aa}

\appendix

\section{Three-dimensional number density-intrinsic shear correlation function}
\label{sec:appgicorr}

In this appendix, we derive the three-dimensional number density-intrinsic shear (gI) correlation function, detail the inclusion of photometric redshift scatter into the formalism, and establish an approximate relation between the three-dimensional correlation function and the angular power spectrum.

\subsection{Correlation function for exact redshifts}

We define the three-dimensional correlation function between the galaxy density contrast $\delta_{\rm g}$ and the radial intrinsic shear $\gamma_{\rm I,+}$ as
\eq{
\label{eq:defxigi}
\xi_{\rm gI} (r_p,\Pi,z) \equiv \ba{\delta_{\rm g} \br{0, \chi(z)-\frac{x_\parallel}{2}, z}\; \gamma_{\rm I,+} \br{\vek{x_\perp}, \chi(z)+\frac{x_\parallel}{2}, z} }\;,
}
for a given mean redshift $z$ of the galaxy pairs correlated. Here we introduced a three-dimensional comoving separation vector $\vek{x}$ which has a line-of-sight component $\Pi \equiv x_\parallel$. Its transverse components are denoted by $\vek{x_\perp}$, with modulus $r_p \equiv |\vek{x_\perp}|$. The first argument of both $\delta_{\rm g}$ and $\gamma_{\rm I,+}$ denotes the position on the sky, the second the position along the line of sight, and the third quantifies the epoch, given in terms of the redshift. Note that a line-of-sight separation $\Pi \neq 0$ implies that $\delta_{\rm g}$ and $\gamma_{\rm I,+}$ are not measured at precisely the same epoch, contrary to what we have written in (\ref{eq:defxigi}). However, as $\Pi$ is small compared to the comoving distance $\chi(z)$ to the galaxies under consideration, this approximation holds to good accuracy.

Following \citet{hirata04}, the radial component of the intrinsic shear is measured with respect to $\vek{x_\perp}$, and without loss of generality we can choose the coordinate system such that $\gamma_{\rm I,+}=\gamma_{\rm I,1}$. Note that in the majority of weak lensing studies $\gamma_+$ is defined as the tangential component of the shear. Measuring radial instead of tangential alignment implies a change of sign, so that e.g. the galaxy-galaxy lensing signal which we consider is negative. Denoting Fourier variables by a tilde, one can construct in analogy to the matter density contrast a three-dimensional intrinsic convergence $\delta_{\rm I}$\footnote{The intrinsic convergence $\delta_{\rm I}$ can best be interpreted as the convergence of the intrinsic shear field for infinitesimal slices in comoving distance. Intrinsic shear field, projected intrinsic convergence, and $\delta_{\rm I}$ have been defined in exact analogy to the cosmic shear formalism. For details we refer the reader to \citet{joachimi10} who employ the same notation except for using $\bar{\kappa}_{\rm I}$ instead of $\delta_{\rm I}$.} via
\eq{
\tilde{\gamma}_{\rm I} (\vek{k}) = \expo{2\ic \varphi}\; \tilde{\delta}_{\rm I} (\vek{k})\;,
}
where $\varphi$ is the polar angle of $\vek{k}_\perp$, i.e. the projection of the wave vector onto the plane of the sky. We will denote the line-of-sight component of $\vek{k}$ by $k_\parallel$. 

Then one can write the correlation function by Fourier transforming (\ref{eq:defxigi}) as
\eqa{
\label{eq:xigifourier}
\xi_{\rm gI} (r_p,\Pi,z) \!\!&=&\!\! \int \frac{\dd^3 k}{(2\,\pi)^3} \int \frac{\dd^3 k'}{(2\,\pi)^3} \expo{- \ic \svek{k} \cdot \svek{x} } \ba{\tilde{\delta}^*_{\rm g}(\vek{k'},z)\; \tilde{\gamma}_{\rm I,+}(\vek{k}, z) }\\ \nn
\!\!&=&\!\! \int \frac{\dd^3 k}{(2\,\pi)^3} \int \frac{\dd^3 k'}{(2\,\pi)^3} \expo{- \ic \svek{k} \cdot \svek{x} } \expo{2\ic \varphi}\; \ba{\tilde{\delta}^*_{\rm g}(\vek{k'},z)\; \tilde{\delta}_{\rm I}(\vek{k}, z) }\;.
}
Inserting the definition of the three-dimensional gI power spectrum,
\eq{
\ba{\tilde{\delta}^*_{\rm g}(\vek{k'},z)\; \tilde{\delta}_{\rm I}(\vek{k}, z) } = (2\,\pi)^3\; \delta_{\rm D}^{(3)}\br{\vek{k} - \vek{k'}}\; P_{\rm gI}(k,z)\;,
}
and subsequently integrating (\ref{eq:xigifourier}) over $\vek{k'}$ yields
\eqa{
\label{eq:xigicalc}
\xi_{\rm gI} (r_p,\Pi,z) &=& \int \frac{\dd^3 k}{(2\,\pi)^3} \expo{- \ic \svek{k} \cdot \svek{x} } \expo{2\ic \varphi}\; P_{\rm gI} (k,z)\\ \nn
&& \hspace*{-1.5cm} = \int \frac{\dd^3 k}{(2\,\pi)^3} \expo{- \ic k_\parallel \Pi } \expo{- \ic \svek{k_\perp} \cdot \svek{x_\perp} } \expo{2\ic \varphi}\; P_{\rm gI} \br{\sqrt{k_\perp^2 + k_\parallel^2},z}\\ \nn
&& \hspace*{-1.5cm} = - \int \frac{\dd k_\parallel}{2\,\pi}  \expo{- \ic k_\parallel \Pi} \int_0^\infty \frac{\dd k_\perp k_\perp}{2\,\pi}\; J_2(k_\perp r_p)\; P_{\rm gI}\br{\sqrt{k_\perp^2 + k_\parallel^2},z}\;,
}
where in order to arrive at the third equality, the definition of the second-order Bessel function of the first kind was used. In this derivation it was implicitly assumed that the intrinsic shear field does not feature B-modes, as is for instance the case for the linear alignment paradigm.

One can now integrate over the line of sight, making use of the definition of the Dirac delta-distribution, to obtain the projected gI correlation function as employed by \citet{mandelbaum06}, \citet{hirata07}, and \citet{mandelbaum09},
\eqa{
\label{eq:defcrosspower}
w_{\rm g+}(r_p,z) &\equiv& \int_{-\infty}^\infty \dd \Pi\; \xi_{\rm gI} (r_p,\Pi,z)\\ \nn
&=& - \int_0^\infty \frac{\dd k_\perp\, k_\perp}{2\,\pi}\; J_2(k_\perp r_p)\; P_{\rm gI} (k_\perp,z)\;. 
}
Real data cannot provide the correlation function for arbitrarily large line-of-sight separations, so that a truncation of the integral in (\ref{eq:defcrosspower}) is necessary. This formula is still applicable if one can stack observations for all values of $\Pi$ for which galaxy pairs carry a signal. While this can easily be achieved for spectroscopic observations, photometric redshift scatter smears the signal in $\Pi$ such that a cut-off $\Pi_{\rm max}$ needs to be taken into account explicitly in the modelling. Of course it would be possible to compute the observed correlations out to very large $\Pi_{\rm max}$, but this way many uncorrelated galaxy pairs would enter the correlation function, thereby decreasing the signal-to-noise dramatically.

Instead, we proceed from (\ref{eq:xigicalc}) by assuming that $\xi_{\rm gI}$ is a real function, and write
\eqa{
\label{eq:xigicalc2}
\xi_{\rm gI} (r_p,\Pi,z) =\\ \nn
&& \hspace*{-2cm} - \int_0^\infty \frac{\dd k_\parallel}{\pi} \int_0^\infty \frac{\dd k_\perp k_\perp}{2\,\pi}\; J_2(k_\perp r_p)\; P_{\rm gI}\br{\sqrt{k_\perp^2 + k_\parallel^2},z}\; \cos(k_\parallel \Pi)\;.
}
As can be seen from this equation, $\xi_{\rm gI}$ is an even function in both $r_p$ and $\Pi$, so that it is sufficient to compute just one quadrant. Note that by definition $r_p \geq 0$, whereas $\Pi$ can also attain negative values.

\begin{figure}[t]
\centering
\includegraphics[scale=.35,angle=270]{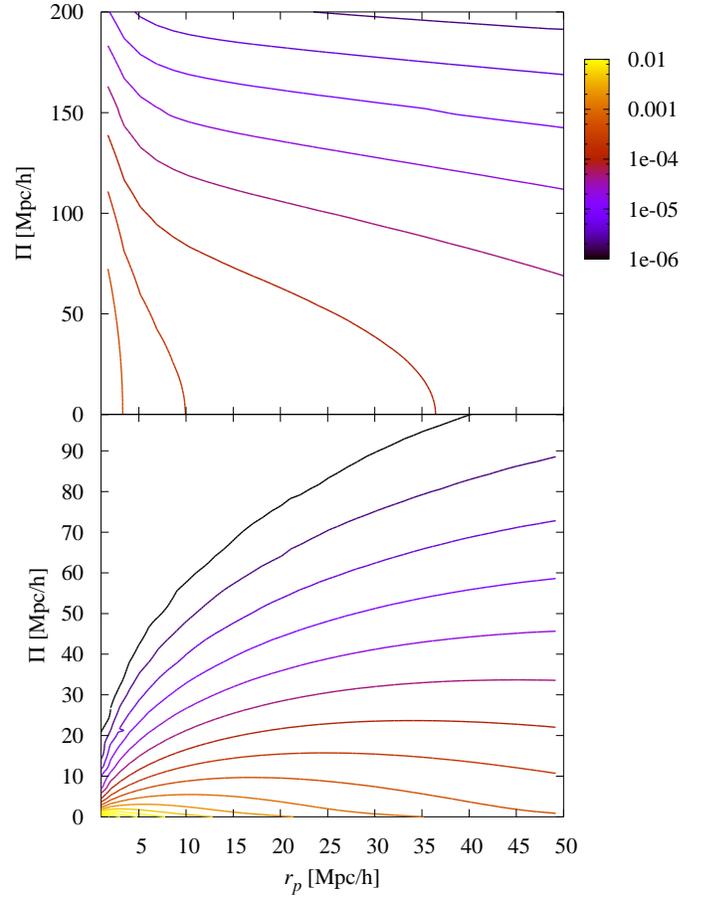}
\caption{Three-dimensional gI correlation function as a function of comoving line-of-sight separation $\Pi$ and comoving transverse separation $r_p$ at $z \approx 0.5$. Contours are logarithmically spaced between $10^{-2}$ (yellow) and $10^{-6}$ (black) with three lines per decade. \textit{Top panel}: Applying a Gaussian photometric redshift scatter of width 0.02. \textit{Bottom panel}: Assuming exact redshifts. Note the largely different scaling of the ordinate axes. The galaxy bias has been set to unity, and (\ref{eq:GIlinalign}) with SuperCOSMOS normalisation has been used to model $P_{\delta {\rm I}}$ in both cases. Redshift-space distortions have not been taken into account.}
\label{fig:gicorr_photozeffect}
\end{figure}

Equation (\ref{eq:xigicalc2}) yields the three-dimensional gI correlation function for exact or, to good approximation, spectroscopic redshifts. For the model described in Sect.$\,$\ref{sec:modelling} with SuperCOSMOS normalisation and $b_{\rm g}=1$, we plot $\xi_{\rm gI} (r_p,\Pi,z)$ for $z \approx 0.5$ in Fig.$\,$\ref{fig:gicorr_photozeffect}, bottom panel. As expected, the correlation is strongest for small separation, in particular for $|\Pi|$ close to zero. If spectroscopic data is available, essentially all information is captured when a cut-off $\Pi_{\rm max}=60$\hmpc\ is used in the integration (\ref{eq:defcrosspower}), as e.g. in \citet{mandelbaum09}. Due to the definition (\ref{eq:defxigi}), the gI correlation function measures the radial alignment of the galaxy shape with respect to the separation vector of the galaxy pair considered. Therefore the correlation function vanishes for all $\Pi$ at $r_p=0$ since then the separation vector points along the of sight. Note that the contours do not approach the $\Pi=0$-axis asymptotically, but cross this line at some value of $r_p$, as expected for a differentiable correlation function. Throughout these considerations we have not taken the effect of redshift-space distortions into account.

\subsection{Incorporating photometric redshifts}

Photometric redshift errors cause the observed correlation function to be a \lq smeared\rq\ version of (\ref{eq:xigicalc2}), introducing a spread especially along the line of sight but to a lesser extent also in transverse separation (because an uncertain redshift is used to convert angular separation to physical separation). If we denote quantities determined via photometric redshifts by a bar, the actually measured three-dimensional correlation function reads
\eqa{
\label{eq:xiphot}
\xi_{\rm gI}^{\rm phot} (\bar{r}_p,\bar{\Pi},\bar{z}_{\rm m}) =\\ \nn
&& \hspace*{-2cm} \int \dd z_{\rm m} \int \dd r_p \int \dd \Pi\; p\br{r_p,\Pi,z_{\rm m} \,|\, \bar{r}_p,\bar{\Pi},\bar{z}_{\rm m}}\; \xi_{\rm gI} (r_p,\Pi,z_{\rm m})\;,
}
where $z_{\rm m}$ denotes the mean redshift of the galaxy samples used for the number density and the shape measurement. Here, $p$ is the probability distribution of the true values of $r_p$, $\Pi$, and $z_{\rm m}$, given photometric redshift estimates of these quantities. In words, (\ref{eq:xiphot}) means that in order to obtain the observed correlation function, we integrate over $\xi_{\rm gI}$ as given in (\ref{eq:xigicalc2}), weighted by the probability that the true values for separations and redshift actually correspond to the estimates based on photometric redshifts.

The direct observables for this measurement are the redshifts of the two galaxy samples under consideration, $z_1$ and $z_2$, and their angular separation $\theta$. The sets of variables $(z_1,z_2,\theta)$ and $(r_p,\Pi,z_{\rm m})$ are related via a bijective transformation. Writing (\ref{eq:xiphot}) in terms of the other set of variables, one obtains
\eqa{
\label{eq:xiphot2}
 \xi_{\rm gI}^{\rm phot} (\bar{r}_p,\bar{\Pi},\bar{z}_{\rm m}) &=&\\ \nn
&& \hspace*{-2.5cm} \int \dd z_1 \int \dd z_2 \int \dd \theta\; p\br{z_1,z_2,\theta \,|\, \bar{z}_1 \bc{\bar{z}_{\rm m},\bar{\Pi}},\bar{z}_2 \bc{\bar{z}_{\rm m},\bar{\Pi}},\bar{\theta} \bc{\bar{z}_{\rm m},\bar{r}_p}}\\ \nn
&& \hspace*{-0.5cm} \times\; \xi_{\rm gI} \br{r_p \bc{z_1,z_2,\theta},\Pi \bc{z_1,z_2},z_{\rm m} \bc{z_1,z_2}}\\ \nn
&& \hspace*{-2.5cm} = \int \dd z_1 \int \dd z_2\; p_n\br{z_1|\bar{z}_1 \bc{\bar{z}_{\rm m},\bar{\Pi}}}\; p_\epsilon \br{z_2|\bar{z}_2 \bc{\bar{z}_{\rm m},\bar{\Pi}}}\;\\ \nn
&& \hspace*{-1cm} \times\; \xi_{\rm gI} \br{r_p\bc{z_1,z_2,\bar{\theta}(\bar{z}_{\rm m},\bar{r}_p)},\Pi\bc{z_1,z_2},z_{\rm m}\bc{z_1,z_2}}\;.
}
In the second step it was assumed that the probability distributions of $z_1$, $z_2$, and $\theta$ are mutually independent, and that $\theta$ is exactly known, i.e. $p(\theta|\bar{\theta}) = \delta_{\rm D}(\theta- \bar{\theta})$. We have introduced different redshift probability distributions for the galaxy sample with number density information $p_n$ and the one with shape information $p_\epsilon$. All quantities related to photometric redshifts have been expressed in terms of the arguments of the correlation function on the left-hand-side.

We make use of the following approximate relations between the two triples of variables,
\eqa{
\label{eq:trafo1}
z_{\rm m} &=& \frac{1}{2} \br{z_1 + z_2}\;;\\ \nn
r_p &\approx& \theta\; \chi(z_{\rm m})\;;\\ \nn
\Pi &\approx& \frac{c}{H(z_{\rm m})} \br{z_2-z_1}\;,
}
where $H(z)$ is the Hubble parameter. Note that the same transformations have been used to bin the observational data in terms of redshift, transverse and line-of-sight separation. With this equation for $\Pi$, in combination with the assignment of probability distributions in (\ref{eq:xiphot2}), we have introduced the convention that $\Pi > 0$ means that the galaxy from the density sample is at lower redshift than the galaxy from the shape sample. If and only if the distributions for the density and the shape sample are identical, which we assume throughout this work, the correlation function remains symmetric with respect to $\Pi$, i.e. $\xi_{\rm gI}^{\rm phot} (\bar{r}_p,\bar{\Pi},\bar{z}_{\rm m}) = \xi_{\rm gI}^{\rm phot} (\bar{r}_p,-\bar{\Pi},\bar{z}_{\rm m})$.

With these equations at hand, one can also write down the inverse transformation of (\ref{eq:trafo1}), which is needed to evaluate (\ref{eq:xiphot2}),
\eqa{
\label{eq:trafo2}
\theta &=& r_p\; \chi^{-1}(z_{\rm m})\;;\\ \nn
z_1 &=& z_{\rm m} - \frac{\Pi\; H(z_{\rm m})}{2c}\;;\\ \nn
z_2 &=& z_{\rm m} + \frac{\Pi\; H(z_{\rm m})}{2c}\;.
}
Then (\ref{eq:xiphot2}) can be expressed as
\eqa{
\label{eq:xiphotfinal}
\xi_{\rm gI}^{\rm phot} (\bar{r}_p,\bar{\Pi},\bar{z}_{\rm m}) &=&\\ \nn
&& \hspace*{-2cm} \int \dd z_1 \int \dd z_2\; p_n\br{z_1 ~|~ \bar{z}_{\rm m} - \frac{\bar{\Pi}\; H(\bar{z}_{\rm m})}{2c}}\; p_\epsilon \br{z_2 ~|~ \bar{z}_{\rm m} + \frac{\bar{\Pi}\; H(\bar{z}_{\rm m})}{2c}}\;\\ \nn
&& \hspace*{-1cm} \times\; \xi_{\rm gI} \br{\bar{r}_p\; \frac{\chi \br{\frac{1}{2}(z_1+z_2)}}{\chi \br{\bar{z}_{\rm m}}},\frac{c\; |z_2-z_1|}{H\br{\frac{1}{2}(z_1+z_2)}} ,\frac{1}{2}(z_1+z_2)}\;.
}
Note that the absolute value for $z_2-z_1$ has been introduced in the second argument of $\xi_{\rm gI}$, which is possible since it is an even function in this argument. The integrals in (\ref{eq:xiphotfinal}) run over the full range of spectroscopic (exact) redshifts. As a consequence, $|z_2-z_1|$ in the second argument of $\xi_{\rm gI}$ can obtain relatively large values, leading to very large $\Pi \gg 100$\hmpc. However, the spectroscopic $\xi_{\rm gI}$ becomes very small for large $\Pi$, so that the integrand in (\ref{eq:xiphotfinal}) can safely be set to zero in this case.

Still, any sizeable photometric redshift scatter leads to a considerable spread of the three-dimensional correlation function in $\Pi$, as can be seen in Fig.$\,$\ref{fig:gicorr_photozeffect}. Assuming a Gaussian photometric redshift scatter with width 0.02 around every true redshift, the strong signal concentrated at small $\Pi$ and $r_p \lesssim 10$\hmpc\ in the spectroscopic case is scattered along the line of sight, so that the values of $\xi_{\rm gI}^{\rm phot}$ at $\Pi > 200$\hmpc\ are still more than a per cent of those at $\Pi=0$ for any $r_p$. In contrast, we find that the net scatter of signal between different transverse separations is negligible. Hence, in principle the projected correlation function (\ref{eq:defcrosspower}) does not change when using photometric instead of spectroscopic redshift information as long as the complete range of $\Pi$ for which a signal is measured enters the line-of-sight integration. However, in practice the line-of-sight integral has to be truncated for reasons of a good signal-to-noise ratio, so that in the case of photometric redshifts part of the signal is lost. Therefore it is crucial to repeat the same steps applied to the data also to the model and use the same cut-off $\Pi_{\rm max}$ in (\ref{eq:defcrosspower}).

\subsection{Relation to angular power spectra}

We now derive a relation between the three-dimensional gI correlation function in the presence of photometric redshift scatter and the angular power spectrum, which proves most convenient to compute $\xi_{\rm gI}^{\rm phot}$ in practice. Inserting (\ref{eq:xigicalc}) into (\ref{eq:xiphotfinal}), one can write 
\eqa{
\xi_{\rm gI}^{\rm phot} (\bar{r}_p,\bar{\Pi},\bar{z}_{\rm m}) &=&\\ \nn
&& \hspace*{-2.5cm} - \int \dd z_1 \int \dd z_2\; p_n \br{z_1 ~|~ \bar{z}_{\rm m} - \frac{\bar{\Pi}\; H(\bar{z}_{\rm m})}{2c}}\; p_\epsilon \br{z_2 ~|~ \bar{z}_{\rm m} + \frac{\bar{\Pi}\; H(\bar{z}_{\rm m})}{2c}}\\ \nn
&& \hspace*{-2cm} \times\; \int \frac{\dd k_\parallel}{2\,\pi} \int_0^\infty \frac{\dd k_\perp k_\perp}{2\,\pi}\; J_2\br{k_\perp \bar{r}_p\; \frac{\chi\br{z_{\rm m}}}{\chi(\bar{z}_{\rm m})}}\\ \nn
&& \hspace*{-0.5cm} \times\; P_{\rm gI}\br{\sqrt{k_\perp^2 + k_\parallel^2},z_{\rm m}}\; \exp \bc{- \ic\, k_\parallel\; \frac{c\,(z_2-z_1)}{H\br{z_{\rm m}} }}\;.
}
where we employed $z_{\rm m}=(z_1+z_2)/2$ as a shorthand notation. Making use of $\theta=\bar{r}_p/\chi(\bar{z}_{\rm m})$, see (\ref{eq:trafo2}), and defining the angular frequency $\ell = k_\perp \chi\br{z_{\rm m}}$, one obtains
\eqa{
\label{eq:xitoP1}
\xi_{\rm gI}^{\rm phot} (\bar{r}_p,\bar{\Pi},\bar{z}_{\rm m}) &=&\\ \nn
&& \hspace*{-2.5cm} - \int \dd z_1 \int \dd z_2\; p_n \br{z_1 ~|~ \bar{z}_{\rm m} - \frac{\bar{\Pi}\; H(\bar{z}_{\rm m})}{2c}}\; p_\epsilon \br{z_2 ~|~ \bar{z}_{\rm m} + \frac{\bar{\Pi}\; H(\bar{z}_{\rm m})}{2c}}\\ \nn
&& \hspace*{-2cm} \times\; \int \frac{\dd k_\parallel}{2\,\pi} \int_0^\infty \frac{\dd \ell\; \ell}{2\,\pi}\; J_2\br{\ell \theta(\bar{r}_p,\bar{z}_{\rm m})}\; \chi^{-2}(z_{\rm m})\\ \nn
&& \hspace*{-1.2cm} \times\; P_{\rm gI}\br{\sqrt{\br{\frac{\ell}{\chi(z_{\rm m})}}^2 + k_\parallel^2},z_{\rm m}}\; \exp \bc{- \ic\, k_\parallel\; \frac{c\,(z_2-z_1)}{H\br{z_{\rm m}} }}\;.
}
We then transform the integration variables $\bc{z_1,z_2}$ to $\bc{z_m,\Delta z \equiv z_2 -z_1}$. Note that the determinant of the Jacobian of this transformation is unity. We apply Limber's approximation, which in this case can be written as 
\eqa{\nn
&& p_n \br{z_{\rm m} -\frac{\Delta z}{2} ~|~ \bar{z}_{\rm m} - \frac{\bar{\Pi}\; H(\bar{z}_{\rm m})}{2c}}\; p_\epsilon \br{z_{\rm m} +\frac{\Delta z}{2} ~|~ \bar{z}_{\rm m} + \frac{\bar{\Pi}\; H(\bar{z}_{\rm m})}{2c}}\\ 
&\approx& p_n \br{z_{\rm m} | \bar{z}_1(\bar{z}_{\rm m},\bar{\Pi})}\; p_\epsilon \br{z_{\rm m} | \bar{z}_2(\bar{z}_{\rm m},\bar{\Pi})}\;. 
}
Here we have assumed that the two redshift probability distributions are sufficiently broad and have similar forms, so that an evaluation at $z_{\rm m}$ instead of $z_{\rm m} \pm \Delta z/2$ does not change the results significantly. Since the photometric redshifts on which the distributions are conditional encapsulate the dependence of $\xi_{\rm gI}^{\rm phot}$ on the line-of-sight separation $\Pi$, we do not extend this approximation to the second argument. Equation (\ref{eq:xitoP1}) thereby simplifies to
\eqa{
\label{eq:xitoP2}
\xi_{\rm gI}^{\rm phot} (\bar{r}_p,\bar{\Pi},\bar{z}_{\rm m})  &\approx&\\ \nn
&& \hspace*{-2.2cm} - \int \dd z_{\rm m} \int \dd \Delta z\; \frac{p_n \br{z_{\rm m} | \bar{z}_1(\bar{z}_{\rm m},\bar{\Pi})}\; p_\epsilon \br{z_{\rm m} | \bar{z}_2(\bar{z}_{\rm m},\bar{\Pi})}}{\chi^2(z_{\rm m})}\\ \nn
&& \hspace*{-2cm} \times\; \int_0^\infty \frac{\dd \ell\; \ell}{2\,\pi}\; J_2\br{\ell \theta(\bar{r}_p,\bar{z}_{\rm m})} \int \frac{\dd k_\parallel}{2\,\pi}\; \exp \bc{- \ic\, k_\parallel\; \frac{c\,\Delta z}{H\br{z_{\rm m}} }} \\ \nn
 && \hspace*{0.2cm} \times\; P_{\rm gI}\br{\sqrt{\br{\frac{\ell}{\chi(z_{\rm m})}}^2 + k_\parallel^2},z_{\rm m}}\;\\ \nn
&& \hspace*{-2cm} = - \int \dd z_{\rm m}\; \frac{H\br{z_{\rm m}}}{c}\; \frac{p_n \br{z_{\rm m} | \bar{z}_1(\bar{z}_{\rm m},\bar{\Pi})}\; p_\epsilon \br{z_{\rm m} | \bar{z}_2(\bar{z}_{\rm m},\bar{\Pi})}}{\chi^2(z_{\rm m})}\\ \nn
&& \hspace*{-0.8cm} \times\; \int_0^\infty \frac{\dd \ell\; \ell}{2\,\pi}\; J_2\br{\ell \theta(\bar{r}_p,\bar{z}_{\rm m})}\; P_{\rm gI}\br{\frac{\ell}{\chi(z_{\rm m})},z_{\rm m}}\;,
}
where in order to arrive at the second equality, we integrated over $\Delta z$. The resulting Dirac delta-distribution renders the $k_\parallel$ integration trivial. Making use of the expressions $\dd z_{\rm m} = \dd \chi H\br{z_{\rm m}}/c$ and $p(z) = p(\chi)\, \dd \chi/\dd z$, one obtains the result
\eqa{
\label{eq:xitoP3}
\xi_{\rm gI}^{\rm phot} (\bar{r}_p,\bar{\Pi},\bar{z}_{\rm m}) &=& - \int_0^\infty \frac{\dd \ell\; \ell}{2\,\pi}\; J_2\br{\ell \theta(\bar{r}_p,\bar{z}_{\rm m})}\\ \nn
&& \hspace*{-2.5cm} \times\; \int_0^{\chi_{\rm hor}} \!\!\!\! \dd \chi\; \frac{p_n \br{\chi | \chi(\bar{z}_1(\bar{z}_{\rm m},\bar{\Pi}))} p_\epsilon \br{\chi | \chi(\bar{z}_2(\bar{z}_{\rm m},\bar{\Pi}))}}{\chi^2}\; P_{\rm gI}\br{\frac{\ell}{\chi},z(\chi)}\;\\ \nn
&&\hspace*{-2.0cm} = - \int_0^\infty \frac{\dd \ell\; \ell}{2\,\pi}\; J_2\br{\ell \theta(\bar{r}_p,\bar{z}_{\rm m})}\; C_{\rm gI} \br{\ell; \bar{z}_1(\bar{z}_{\rm m},\bar{\Pi}), \bar{z}_2(\bar{z}_{\rm m},\bar{\Pi})}\;,
}
where in the last step we implicitly defined the projected gI auto-correlation power spectrum $C_{\rm gI}$. In addition to the angular frequency, we have written the photometric redshifts $\bar{z}_1$ and $\bar{z}_2$, which characterise the redshift distributions entering $C_{\rm gI}$, explicitly as arguments. Note that Limber equations, such as (\ref{eq:gIlimber}), in general hold only approximately, the range of validity being the more limited the narrower the kernels in the line-of-sight integration \citep[e.g.][]{simon07}. 

We have verified that the calculations of the three-dimensional gI correlation function according to (\ref{eq:xiphotfinal}) and (\ref{eq:xitoP3}) agree within the numerical accuracy. The latter can be computed much more efficiently by computing the angular power spectrum via Limber's equation and then using Hankel transformations to obtain the correlation function $\xi_{\rm gI}^{\rm phot}(\bar{r}_p,\bar{\Pi},\bar{z}_{\rm m})$ via (\ref{eq:xitoP3}), employing the transformation (\ref{eq:trafo1}). One can proceed likewise to obtain analogous expressions for the gg signal. Galaxy-galaxy lensing vanishes if the density field probed by the galaxy distribution and the source galaxies on whose images the gravitational shear is measured are located at exactly the same redshift. Thus one cannot proceed with the same formalism as used to derive the gI contribution (see the assumptions underlying the definition (\ref{eq:defxigi})), but must instead incorporate redshift probability distributions from the start, again arriving at an expression analogous to (\ref{eq:xitoP3}).

\section{Redshift dependence of the linear alignment model}
\label{sec:larederive}

In this appendix we re-derive the redshift dependence of the linear alignment model, obtaining a different result than \citet{hirata04}, but being in full agreement with \citet{hirata10}. Practically all attempts at constructing a physical description for intrinsic alignments are based on the linear alignment model originally suggested by \citet{catelan01}. They assumed that the shape of the luminous part of a galaxy exactly follows the shape of its host halo, and that the ellipticity of the latter is determined by the local tidal gravitational field of the large-scale structure.

The simplest possible form allowed by the assumptions made above is a linear relation between the intrinsic shear and the gravitational field, given by \citep{catelan01}
\eqa{
\label{eq:intrinsicshearprimpot}
\gamma_{\rm I,\,1}(\vek{x}) &=& - \frac{C_1}{4 \pi G}\; \br{\frac{\partial^2}{\partial x_1^2} - \frac{\partial^2}{\partial x_2^2}} \Phi_{\rm p}(\vek{x})\;; \\ \nn
\gamma_{\rm I,\,2}(\vek{x})  &=& - \frac{C_1}{4 \pi G}\; 2\, \frac{\partial}{\partial x_1} \frac{\partial}{\partial x_2}\; \Phi_{\rm p}(\vek{x})\;,
}
where we wrote the normalisation in the notation of \citet{hirata04} in which $C_1$ is an arbitrary constant. The partial derivatives are with respect to comoving coordinates, and $\Phi_{\rm p}(\vek{x}) \equiv \Phi(\vek{x},z_{\rm p})$ denotes the \lq primordial\rq\ potential, i.e. the linear gravitational potential evaluated at the epoch of galaxy formation, at a redshift $z_{\rm p}$ well within the matter-dominated era. For ease of notation we have omitted a smoothing of the gravitational potential on galactic scales in (\ref{eq:intrinsicshearprimpot}) which can be implemented by a simple cut-off of high wavenumbers in Fourier space, see \citet{hirata04}. These authors used the relations (\ref{eq:intrinsicshearprimpot}) in their derivation of the intrinsic alignment power spectra for the linear alignment model, which we closely follow.

In a first step, the primordial gravitational potential is related to the matter density contrast via the Poisson equation 
\eqa{
\label{eq:Poisson}
\nabla_x^2 \Phi(\vek{x},z_{\rm p}) &=& \frac{3 H_0^2 \Omega_{\rm m}}{2}\;(1+z)\;\delta(\vek{x},z_{\rm p})\\ \nn
&=& 4 \pi G\; \rho_{\rm m}(z)\; (1+z)^{-2}\; \delta(\vek{x},z_{\rm p})\;,
}
where $\nabla_x^2$ denotes the comoving Laplacian. This expression is Fourier-transformed, yielding
\eq{
\label{eq:poissonfourier}
\tilde{\Phi}(\vek{k},z) = - 4 \pi G\; \rho_{\rm m}(z)\; (1+z)^{-2}\; k^{-2}\; \tilde{\delta} (\vek{k},z)\;.
}
The growth factor $D(z)$ quantifies the dependence of the matter density contrast on redshift in the limit of linear structure formation, and is normalised to $D(z)=(1+z)^{-1}$ during matter domination \citep{hirata04}\footnote{Note that this normalisation of the growth factor differs from the one used in the remainder of this work. However, this difference only affects the amplitude $C_1$ which is an arbitrary constant. Our default normalisation of the growth factor is in line with \citet{bridle07} whose value for $C_1$ we have adopted.}. Restricting (\ref{eq:poissonfourier}) to linear scales, one obtains the ratio
\eq{
\label{eq:ratiopotentials}
\frac{\tilde{\Phi}(\vek{k},z)}{\tilde{\Phi}(\vek{k},z_{\rm p})} = \frac{(1+z)\; D(z)}{(1+z_{\rm p})\; D(z_{\rm p})} = (1+z)\; D(z)\;,
}
where in the last step we made use of the fact that $z_{\rm p}$ lies in the matter-dominated era. Inserting (\ref{eq:ratiopotentials}) into (\ref{eq:poissonfourier}), and considering the linear regime, one arrives at the relation between primordial potential and linear matter density contrast,
\eq{
\label{eq:primordialpotential}
\tilde{\Phi}_{\rm p}(\vek{k}) = - \frac{ 4 \pi G\; \rho_{\rm m}(z)}{D(z)\; (1+z)^3}\; k^{-2}\; \tilde{\delta}_{\rm lin} (\vek{k},z) \;.
}
This expression differs from the result given in \citet{hirata04}, eq. (14), by an additional factor $(1+z)^{-2}$. This discrepancy was also found in other re-derivations (\citealp{hirata10}; R. Bean, I. Laszlo; priv. comm.).

\citet{hirata04} inserted their eq. (14) into (\ref{eq:intrinsicshearprimpot}) and then computed the three-dimensional intrinsic shear (II) power spectrum and the intrinsic shear-matter cross-power spectrum. Neglecting source clustering but otherwise following their steps in exact analogy, we obtain
\eqa{
\label{eq:IIlinalign}
P_{\rm II}\br{k,z} &=& C^2_1\; \rho^2_{\rm cr}\; \frac{\Omega^2_{\rm m}}{D^2(z)}\; P_{\delta, {\rm lin}} \br{k,z}\;;\\
\label{eq:GIlinalignnew}
P_{\delta {\rm I}}\br{k,z} &=& - C_1\; \rho_{\rm cr}\; \frac{\Omega_{\rm m}}{D(z)}\; P_{\delta, {\rm lin}} \br{k,z}\;,
}
where in this work we use the full nonlinear matter power spectrum on the right-hand side instead of the linear power spectrum, as written here in the original form of the linear alignment model. If \citet{hirata04}, eq. (14), were employed instead of (\ref{eq:primordialpotential}), (\ref{eq:IIlinalign}) would have an additional term $(1+z)^4$, and (\ref{eq:GIlinalignnew}) an additional term $(1+z)^2$, in the numerator. These modifications would correspond to a shift by $-2$ in $\eta_{\rm other}$ in our models (\ref{eq:iamodel}) and (\ref{eq:IImodel}).

\section{Volume density and luminosities of red galaxies}
\label{sec:LF}

To make realistic predictions for the intrinsic alignment contamination of cosmic shear surveys, we must specify, at each redshift, the distribution of galaxy luminosities that enter (\ref{eq:iamodel}). Since this intrinsic alignment model only holds for red galaxies, we additionally must estimate the fraction of early-type galaxies in the total weak lensing population as a function of redshift. In this paper, both quantities are determined using fits to the observed luminosity functions given in \citet{faber07}. In this appendix, we present technical details about these calculations, assess the sensitivity of our results to this particular luminosity function choice, and provide data that can be used to forecast the intrinsic alignment contamination of other cosmic shear surveys (with different limiting magnitudes) besides that discussed in the main text.  In all cases, we are extrapolating the luminosity functions to fainter magnitudes at a given redshift relative to the samples used to determine the luminosity function.

\begin{figure}[t]
\centering
\includegraphics[scale=.36,angle=270]{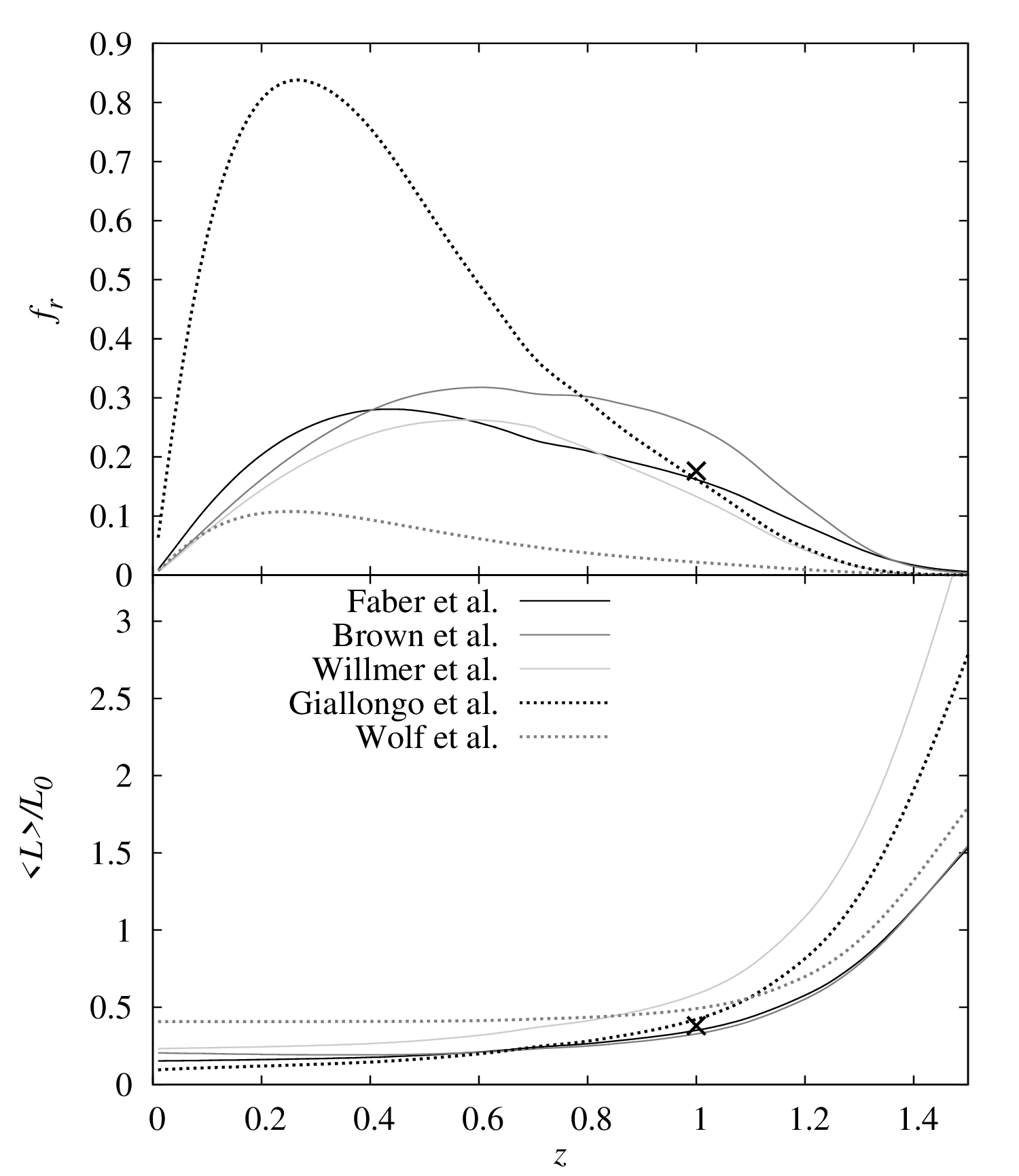}
\caption{Red galaxy fraction $f_r$ (top panel) and mean luminosity of red galaxies (bottom panel) as a function of redshift for a magnitude limit $r_{\rm lim}=25$. We compare the results for the sets of luminosity functions provided by \citet[black solid curves]{faber07}, \citet[dark grey solid curves]{brown07}, \citet[light grey solid curves]{willmer06}, \citet[black dotted curves]{giallongo05}, and \citet[grey dotted curves]{wolf03}. The results for \citet{faber07} luminosity functions used with the value of $B-r$ expected at $z=1$ are indicated by the cross in each panel.}
\label{fig:comparelf}
\end{figure}

We employ the Schechter luminosity function parameters for red galaxies from \citet{faber07}, where $\phi^*$ and $M^*$ are given as a function of redshift, and where the faint-end slope is fixed at $\alpha=-0.5$. While we consistently use magnitudes in the $r$ band, \citet{faber07} determine $M^*$ in the $B$ band. Therefore we the estimate rest-frame $B-r$ colour from the tables provided in \citet{fukugita95}, finding $B-r=1.32$ for ellipticals. This conversion from $B$ to $r$ takes into account that \citet{faber07} give $B$ band magnitudes in the Vega-based system, whereas this work uses AB magnitudes throughout. Furthermore, we have assumed $r \approx r'$, where $r'$ is the filter listed by \citet{fukugita95}. This assumption should hold to good accuracy\footnote{See {\tt http://www.sdss.org/dr6/algorithms/\\jeg\_photometric\_eq\_dr1.html\#usno2SDSS} for the transformation equation between $r$ and $r'$.} for typical colours of the galaxies in our samples, i.e. $0.2 \lesssim r-i \lesssim 0.6$. 

For early-type galaxies, $B-r$ shows little evolution between $z=0$ and $z \sim 1$ \citep{bruzual03}, so  we assume the rest-frame colour to be constant in this redshift range, which we check via the following procedure. Since the Sloan $g$ filter covers a similar wavelength range to the $B$ band (although the peaks of the transmission curves differ, see \citealp{fukugita95} for details), we use the evolution of $g-r$ as determined from the \citet{wake06} templates as an approximation for the redshift dependence of $B-r$. We find a shift of $0.15\,$mag from $z=0$ to $z=1$, which has significantly less effect on our results than employing different observational results for luminosity functions, see Fig.$\,$\ref{fig:comparelf} and the corresponding discussion below. Finally, we correct for the fact that \citet{faber07} have computed absolute magnitudes assuming a Hubble parameter $h=0.7$ while we give absolute magnitudes in terms of $h=1$.

\begin{figure*}[t]
\centering
\includegraphics[scale=.56,angle=270]{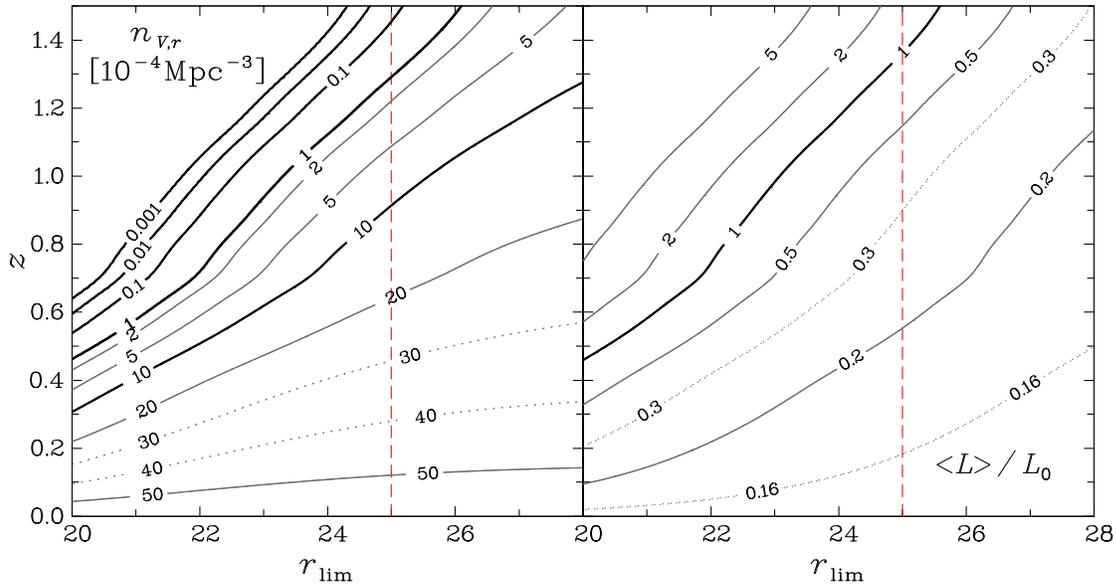}
\caption{Comoving volume number density of red galaxies $n_{\rm V,r}$ and mean luminosity $\ba{L}/L_0$ of red galaxies as a function of limiting magnitude $r_{\rm lim}$ and redshift $z$. \textit{Left panel}: Comoving volume number density of red galaxies $n_{\rm V,r}$ in units of $10^{-4}\,{\rm Mpc}^{-3}$. Contour values range between $10^{-3}$ in the upper left and 50 at the bottom. In the upper left corner of the panel $n_{\rm V,r} \approx 0$. \textit{Right panel}: Mean $r$ band luminosity. Contour values range between 0.16 in the lower part to 5 in the upper left corner. In both panels the red dashed line marks the values for $r_{\rm lim}=25$, the limiting magnitude we employ in our calculations. Decades in contour values are indicated by the black solid lines.}
\label{fig:redgalfrac}
\end{figure*}

With all of these caveats, the limiting absolute $B$ band magnitude at redshift $z$ for a given apparent magnitude limit in the $r$ band is given by
\eqa{\nn
M_{\rm min}(z,r_{\rm lim}) &=& r_{\rm lim} - \bb{ 5 \log_{10} \frac{D_L(z)}{1\,{\rm Mpc}} + 25 + k_{r,{\rm red}}(z)} + (B-r)\;,\\[-1ex]
\label{eq:mmin}
}
where $k_{r,{\rm red}}(z)$ is the $k$-correction of red galaxies for the $r$ band \citep{wake06}. In line with our convention for absolute magnitudes, the luminosity distance $D_L$ is computed with $h=1$. If absolute magnitudes are given for other values of the Hubble parameter, like e.g. in \citet{faber07}, we convert these accordingly. The limiting absolute magnitude from (\ref{eq:mmin}) can then be transformed into the minimum luminosity entering (\ref{eq:redgalfrac}) and (\ref{eq:redlum}),
\eq{
\label{eq:lmin}
\frac{L_{\rm min}(z,r_{\rm lim})}{L_0(z)} = 10^{-0.4 \br{M_{\rm min}(z,r_{\rm lim})-M_0(z)}}\;,
}
where $M_0(z)$ denotes the rest-frame absolute magnitude $-22$, evolution-corrected to redshift $z$ using the redshift dependence of $M^*$ from \citet{faber07}, which is given by $-1.2z$. Note that this dependence accounts for the redshift evolution in the $B$ band, but since $B-r$ is nearly constant as a function $z$, we can also apply the correction (to good approximation) to $r$ band magnitudes. Denoting the luminosity corresponding to $M^*$ by $L^*$, we obtain for (\ref{eq:redlum}) the expression
\eq{
\label{eq:meanlum}
\frac{\ba{L^\beta}(z,r_{\rm lim})}{L_0^\beta(z)} = \br{ \frac{L^*(z)}{L_0(z)} }^\beta\; \frac{\Gamma \br{\alpha + \beta + 1,\frac{L_{\rm min}(z,r_{\rm lim})}{L^*(z)}}}{\Gamma \br{\alpha + 1,\frac{L_{\rm min}(z,r_{\rm lim})}{L^*(z)}}}\;,
}
where the incomplete Gamma function $\Gamma(\alpha,x) = \int_x^\infty \dd y\, y^{\alpha-1} \expo{-y}$ was introduced. Analogously, we arrive at
\eq{
\label{eq:volumedensity}
n_{\rm V,red}(z,r_{\rm lim}) = \phi^*(z)\; \Gamma \br{\alpha + 1,\frac{L_{\rm min}(z,r_{\rm lim})}{L^*(z)}}\;
}
for the comoving volume density of red galaxies entering (\ref{eq:redgalfrac}).

In addition to the luminosity functions from \citet{faber07}, we also consider fitted Schechter parameters presented in \citet{giallongo05}, as well as the sets of luminosity functions published by \citet{wolf03}, \citet{willmer06}, and \citet{brown07}. We determine fit functions to the redshift dependence of both $M^*$ and $\phi^*$ for the latter three works because we have to extrapolate beyond the range of redshift analysed therein. We use linear functions for $M^*$ and various functional forms with two to three fit parameters for $\phi^*$, but note that since the fits rely on only five to six data points, the extrapolation has considerable uncertainty. All five references give $B$ band luminosity functions, but the magnitude system and the convention for $h$ vary, as well as the redshift ranges covered and the definition of red galaxies.

In Fig.$\,$\ref{fig:comparelf} the red galaxy fraction $f_r$ and the mean luminosity $\ba{L}/L_0$ for $r_{\rm lim}=25$ are plotted as a function of redshift, making use of the different luminosity functions. We find fair agreement between the results based on \citet{faber07} and \citet{brown07}, while the mean luminosities derived from \citet{willmer06} already deviate considerably at high $z$ although \citet{faber07} and \citet{willmer06} partly use the same data. The \citet{wolf03} luminosity functions produce significantly lower $f_r$ and higher $\ba{L}$ at low redshifts which is caused by the very different value for the faint end slope, $\alpha=+0.52$. We note that one of the three fields chosen by \citet{wolf03} contained two massive galaxy clusters, so that the large-scale structure in this field could strongly influence the luminosity function in particular of early-type galaxies. However, small red galaxy fractions can be compensated by higher luminosities in (\ref{eq:iamodel}), so that even the \citet{wolf03} luminosity functions may yield intrinsic alignment signals of similar magnitude to the results of e.g. \citet{faber07}.

\begin{table}[t]
\begin{minipage}[t]{\columnwidth}
\centering
\caption{Mean luminosity of red galaxies and comoving volume number density of red galaxies as a function of redshift and limiting $r$ band magnitude $r_{\rm lim}$.}
\begin{tabular}[t]{cccccccc}
\hline\hline
$r_{\rm lim}$ & \multicolumn{7}{c}{$z$}\\
              & 0 & 0.2 & 0.4 & 0.6 & 0.8 & 1.0 & 1.2\\
\hline
20 & 0.16 & 0.29 & 0.70 & 2.34 & 5.87 &       &       \\
21 & 0.15 & 0.23 & 0.41 & 1.09 & 2.50 &  5.23 &       \\
22 & 0.15 & 0.19 & 0.28 & 0.58 & 1.16 &  2.25 &  5.47 \\
23 & 0.15 & 0.18 & 0.22 & 0.36 & 0.60 &  1.05 &  2.35 \\
24 & 0.15 & 0.17 & 0.19 & 0.26 & 0.37 &  0.56 &  1.09 \\
25 & 0.15 & 0.16 & 0.18 & 0.21 & 0.26 &  0.35 &  0.58 \\
26 & 0.15 & 0.16 & 0.17 & 0.19 & 0.21 &  0.25 &  0.36 \\
27 & 0.15 & 0.16 & 0.16 & 0.17 & 0.19 &  0.21 &  0.26 \\
28 & 0.15 & 0.15 & 0.16 & 0.16 & 0.17 &  0.18 &  0.21 \\
\hline
20 & 56.64 & 22.62 &  3.32 &  0.01 &  0    &       &       \\   
21 & 57.17 & 31.02 & 10.73 &  0.63 &  0    &  0    &       \\ 
22 & 57.50 & 36.89 & 19.01 &  4.45 &  0.40 &  0.01 &  0    \\
23 & 57.71 & 40.75 & 25.62 & 11.06 &  3.22 &  0.47 &  0    \\
24 & 57.85 & 43.22 & 30.19 & 17.41 &  8.48 &  3.09 &  0.33 \\
25 & 57.93 & 44.79 & 33.19 & 22.17 & 13.70 &  7.45 &  2.34 \\
26 & 57.98 & 45.79 & 35.11 & 25.38 & 17.66 & 11.57 &  5.84 \\
27 & 58.02 & 46.41 & 36.33 & 27.47 & 20.36 & 14.63 &  9.21 \\
28 & 58.04 & 46.81 & 37.10 & 28.80 & 22.11 & 16.70 & 11.73 \\
\hline
\end{tabular}
\tablefoot{In the upper section values for $\ba{L}/L_0$ are given; in the lower section $n_{\rm V,r}$ is given in units of $10^{-4}{\rm Mpc}^{-3}$. This data is plotted in Fig.$\,$\ref{fig:redgalfrac}.}
\label{tab:redgalfrac}
\end{minipage}
\end{table}

Applying the formalism to luminosity functions from \citet{giallongo05}, we obtain very high $f_r$ at low redshift, which is clearly inconsistent with the other observations. The red galaxy sample used for the fits of \citet{giallongo05} is very small and contains only galaxies with $z>0.4$. While the resulting fit function captures the pronounced decrease in number density for high redshifts that \citet{giallongo05} observe, it can obviously not be used at $z \lesssim 0.4$. In conclusion, we find that the sets of luminosity functions by \citet{faber07} who jointly analyse galaxy samples from four different surveys produce reasonable red galaxy fractions and luminosities although the uncertainty in the values of $f_r$ and $\ba{L}$ at any given redshift is still large.

In Fig.$\,$\ref{fig:redgalfrac}, the comoving volume density of red galaxies $n_{\rm V,r}$ and the mean luminosity of red galaxies in terms of $L_0$ are plotted as a function of both $r_{\rm lim}$ and redshift, using the default set of luminosity functions from \citet{faber07}. For a fixed magnitude limit, $\ba{L}$ increases with redshift while $n_{\rm V,r}$ decreases strongly at high redshifts. Both tendencies are more pronounced if the apparent magnitude limit is brighter. At low redshift, e.g. for $z \lesssim 0.3$ at $r_{\rm lim}=25$, a large number of faint blue galaxies are above the magnitude limit and cause $f_r$ to diminish for $z$ approaching zero (see Fig.$\,$\ref{fig:comparelf}) although $n_{\rm V,r}$ continues to increase. This behaviour might change when explicitly taking into account the size cuts inherent to weak lensing surveys, but in any case, galaxies at these low redshifts constitute only a small fraction of the total survey volume and are expected to have a low luminosity and hence low intrinsic alignment signal on average, so they are unlikely to affect our results severely.

The data shown in Fig.$\,$\ref{fig:redgalfrac}, with the corresponding numbers collected in Table \ref{tab:redgalfrac}, can be used in combination with the intrinsic alignment model fits presented in Sect.$\,$\ref{sec:iafits-combined} to predict the effect of intrinsic alignments on cosmic shear surveys. The mean luminosity as a function of redshift for a given magnitude limit $r_{\rm lim}$ can be inserted into the luminosity term in (\ref{eq:iamodel}), which constitutes a fair approximation as long as values of $\beta$ close to unity (which includes our best-fit value $\beta=1.13$) are probed. Together with an overall redshift distribution $p_{\rm tot}(z)$ for the cosmic shear survey under consideration, $n_{\rm V,r}$ can be read off and used with (\ref{eq:redgalfrac}) to compute the fraction of red galaxies as a function of redshift. Again, we emphasise the limitations of this approach which relies on a substantial amount of extrapolation in luminosity, especially for fainter limiting magnitudes, and which is subject to the large intrinsic uncertainty in the different sets of luminosity functions.

\end{document}